\pdfoutput=1
\documentclass[longauth]{aa} 

\usepackage{graphicx}
\usepackage{natbib}

\begin{document}

\title{Multiwavelength study of the high-latitude cloud L1642: chain of star formation\thanks{{\it Herschel} is an ESA space observatory with science instruments provided by European-led Principal Investigator consortia and with important participation from NASA.}}

\author{
J. Malinen \inst{1} \and
M. Juvela \inst{1} \and
S. Zahorecz \inst{2} \and
A. Rivera-Ingraham \inst{3,4} \and
J. Montillaud \inst{1} \and
K. Arimatsu \inst{5,6} \and
J.-Ph. Bernard \inst{3,4} \and
Y. Doi \inst{7} \and
L. K. Haikala \inst{8,1} \and
R. Kawabe \inst{9,10,11,12} \and
G. Marton \inst{13,2} \and
P. McGehee \inst{14} \and
V.-M. Pelkonen \inst{8,1} \and
I. Ristorcelli \inst{3,4} \and
Y. Shimajiri \inst{15} \and
S. Takita\inst{5} \and
L. V. T\'oth  \inst{2} \and
T. Tsukagoshi \inst{16} \and
N. Ysard \inst{17}
}

\institute{
Department of Physics, University of Helsinki, P.O. Box 64, FI-00014 Helsinki, Finland; johanna.malinen@helsinki.fi
\and
E\"otv\"os Lor\'and University, Department of Astronomy, P\'azm\'any P. s. 1/A, H-1117 Budapest, Hungary 
\and
Universit\'e de Toulouse, UPS-OMP, IRAP, F-31028 Toulouse cedex 4, France 
\and
CNRS, IRAP, 9 Av. colonel Roche, BP 44346, F-31028 Toulouse cedex 4, France 
\and
Institute of Space and Aeronautical Science, Japan Aerospace Exploration Agency, Kanagawa 229-8510, Japan 
\and
Department of Astronomy, Graduate School of Science, The University of Tokyo, Tokyo 113-0033, Japan 
\and
Department of Earth Science and Astronomy, College of Arts and Sciences, the University of Tokyo, Tokyo 153-8902, Japan 
\and
Finnish Centre for Astronomy with ESO (FINCA), University of Turku, V\"ais\"al\"antie 20, FI-21500 Piikki\"o, Finland 
\and
National Astronomical Observatory of Japan, 2-21-1 Osawa, Mitaka, Tokyo 181-8588, Japan 
\and
Joint ALMA Observatory, Alonso de Cordova 3107 Vitacura, Santiago 763 0355, Chile 
\and
Nobeyama Radio Observatory, NAOJ, 462-2, Nobeyama, Minamisaku, Nagano, Japan 
\and
Department of Astronomy, Graduate School of Science, the Univ. of Tokyo, 7-3-1, Hongo, Bunkyo-ku, Tokyo, Japan 
\and
Konkoly Observatory, Research Center for Astronomy and Earth Sciences, Hungarian Academy of Sciences; Konkoly Thege 15-17,
H-1121 Budapest, Hungary  
\and
Infrared Processing and Analysis Center, MS 220-6, California Institute of Technology, Pasadena CA 91125 USA 
\and
Laboratoire AIM, CEA/DSM-CNRS-Universit\'e Paris Diderot, IRFU/Service d'Astrophysique, CEA Saclay, F-91191 Gif-sur-Yvette, France 
\and
College of Science, Ibaraki University, Bunkyo 2-1-1, Mito 310-8512, Japan 
\and
IAS, CNRS (UMR8617), Universit\'e Paris Sud, B\^at. 121, F-91400 Orsay, France 
}

\date{Received 11 November 2013/ Accepted 13 January 2014}

\abstract
{
L1642 is one of the two high galactic latitude ($|b| > 30^{\circ}$) clouds confirmed to have active star formation.
}
{
We examine the properties of this cloud, especially the large-scale structure, dust properties, and compact sources in different stages of star formation.
}
{
We present high-resolution far-infrared and submillimetre observations with the \emph{Herschel} and AKARI satellites and millimetre observations with the AzTEC/ASTE telescope, which we combined with archive data from near- and mid-infrared (2MASS, WISE) to millimetre wavelength observations (Planck).
}
{
The \emph{Herschel} observations, combined with other data, show a sequence of objects from a cold clump to young stellar objects at different evolutionary stages. Source B-3 (2MASS J04351455-1414468) appears to be a YSO forming inside the L1642 cloud, instead of a foreground brown dwarf, as previously classified. \emph{Herschel} data reveal striation in the diffuse dust emission around the cloud L1642. The western region shows striation towards the NE and has a steeper column density gradient on its southern side. The densest central region has a bow-shock like structure showing compression from the west and has a filamentary tail extending towards the east. The differences suggest that these may be spatially distinct structures, aligned only in projection. 
We derive values of the dust emission cross-section per H nucleon of $\sigma_e(250 \mu {\rm m}) = $ 0.5--1.5 $\times10^{-25} {\rm cm}^2/{\rm H}$ for different regions of the cloud.
Modified black-body fits to the spectral energy distribution of \emph{Herschel} and Planck data give emissivity spectral index $\beta$ values 1.8--2.0 for the different regions. The compact sources have lower $\beta$ values and show an anticorrelation between $T$ and $\beta$.
}
{
Markov chain Monte Carlo calculations demonstrate the strong anticorrelation between $\beta$ and $T$ errors and the importance of millimetre wavelength Planck data in constraining the estimates. L1642 reveals a more complex structure and sequence of star formation than previously known.
}


\maketitle

\section{Introduction} \label{sect:intro}

High galactic latitude ($|b| > 30^{\circ}$) molecular clouds have low background and foreground emission, and are usually quite nearby.
They are mostly diffuse or translucent clouds with a low density and typically no signs of star-formation, see~\citet{McGehee2008} for a review. These features make them ideal for studying the interstellar medium and interstellar radiation field, as well as low-mass objects in selected regions. LDN 1642\footnote{As a side note, a memory aid for the number: in the year 1642 Galileo Galilei died and Sir Isaac Newton was born.}~\citep{Lynds1962} (often also called L1642) is one of the two clouds at high galactic latitudes confirmed to have active star-formation, the other one is MBM 12~\citep{Luhman2001}. Studying this cloud might give constraints of the mechanisms that affect cloud evolution and trigger low-mass star formation. In this article, we use the abbreviation L1642. The cloud is also called MBM 20~\citep{Magnani1985} and G210.90-36.55~\citep{Juvela2012}.

L1642 is one of the Orion outlying clouds, see~\citet{Alcala2008} for a review of the area and earlier studies. It forms a blob in the head of a HI cloud with cometary structure, the tail of which extends over 5$^{\circ}$ in the north-east direction in equatorial coordinates, directly towards the Galactic plane. The cloud is projected on the edge of the Orion-Eridanus Bubble~\citep{Brown1995}. On the sky, L1642 is $\sim10^{\circ}$ from the filamentary reflection nebula IC 2118, or Witch Head nebula, which also contains the molecular clouds MBM 21 and 22~\citep{Kun2001,Kun2004,Alcala2008}. The different clouds are marked in Fig.~\ref{fig:Planck_Ak}.

The Galactic coordinates of L1642 are $l = 210.9^{\circ}$ and $b = -36.55^{\circ}$, and the equatorial coordinates are $\alpha_{2000} = 4^{\rm h}35^{\rm m}$ and $\delta_{2000} = -14^{\circ}15'$ ($\sim 68.75^{\circ}$ and $-14.25^{\circ}$, respectively).
\citet{Hearty2000} determined the distance of L1642 to be 112--160 pc. The X-ray observations of~\citet{Kuntz1997} suggested that the cloud is close to the edge of the Local Bubble, and not necessarily inside the Orion-Eridanus Bubble, indicating a distance of approximately 140 pc according to \citet{Sfeir1999}. See~\citet{Welsh2009} for a description of the Local Bubble. We adopted the distance of 140 pc similarly to~\citet{Russeil2003} and~\citet{Lehtinen2004}.

L1642 has been studied using several molecules, see~\citet{Liljestrom1991} for a review of the early studies.
\citet{Liljestrom1991} studied L1642 in CO, HCO$^+$ and NH$_3$ and concluded that the cloud is in virial equilibrium and older than $10^6$ years. More recently, \citet{Russeil2003} studied the morphology and kinematics of L1642 using CO observations obtained with the SEST radio telescope. Their results showed that the cloud consists of a main structure at radial velocity 0.2 km s$^{-1}$ with higher velocity components forming an incomplete ring around it, suggesting an expanding shell. Based on $^{13}$CO data, the peak column density is $N$(H$_2$) $\sim6\times10^{21}$ cm$^{-2}$ and the cloud mass $\sim$59 $M_{\odot}$.

L1642 has also been widely studied using other methods such as dust emission in mid- and far-infrared~\citep{Laureijs1987,Reach1998,Verter1998,Verter2000,Lehtinen2004,Lehtinen2007} and optical (360--600 nm) surface brightness~\citep{Laureijs1987,Mattila2012}. \citet{Mattila2007} studied scattered H{$\alpha$} light in L1642. The shadowing provided by L1642 has also been used in studies of Galactic diffuse X-ray radiation~\citep[e.g.,][]{Galeazzi2007,Gupta2009}.

The small distance, high galactic latitude with little foreground contamination, and the modest column density make L1642 a good target for
studying low-mass star formation and the effect of potential external triggering. \citet{Sandell1987} identified two IRAS sources, IRAS 04327-1419 = L1642-1 (V* EW Eri, HBC 413) and IRAS 04325-1419 = L1642-2 (HBC 410), found within L1642, to be faint nebulous binary stars with very active secondaries. The primary of L1642-1 was spectroscopically classified to be a K7IV T Tauri star.
\citet{Liljestrom1989} found a weak, bipolar outflow around the binary L1642-2.
\citet{Reipurth1990} discovered a Herbig-Haro object, HH123, also originating from L1642-2. They also concluded that the primary object of L1642-2 is a low-luminosity M0 H$\alpha$-emission star, and the secondary component is an H$\alpha$-emission star as
well.
\citet{Correia2006} observed L1642-1 in search of high-order multiplicity, but did not find a third component. \citet{Lehtinen2004} concluded that none of the other four IRAS sources projected within L1642 are likely to be young stellar objects (YSOs) inside the cloud.

The 2MASS point source catalogue object 2MASS J04351455-1414468 is situated towards the densest part of L1642 on the sky. \citet{Cruz2003} classified it as a young ($\sim$10 Myr) object. They estimated the distance to be probably within 30 pc, meaning that the object would be clearly outside the L1642 star-forming region. Later studies have made the assumption or conclusion that the object is a brown dwarf located in front of the L1642 cloud, at a
distance of 14--30 pc~\citep{Faherty2009,Antonova2013}. In this article, we re-evaluate the classification and distance estimate of this object.

The early low-resolution studies treated the L1642 cloud as a single entity.
\citet{Lehtinen2004} presented ISOPHOT far-infrared (FIR) data of L1642 and studied the cloud structure in separate components (named A1, A2, B, and C).
Neither of the IRAS point sources nor new YSO candidates were seen in their 200 $\mu$m maps.
\citet{Lehtinen2007} compared far-IR data with visual extinction and studied the dust grain emissivity properties in L1642. 
They found that FIR dust emissivity increases by a factor of two between regions with colour temperatures 19 K and 14 K. This might be caused by grain growth in the dense and cold areas.

L1642 was recently observed as part of the \emph{Herschel} open time key programme Galactic Cold Cores~\citep{Juvela2010} with unprecedented spatial resolution and sensitivity at 100, 160, 250, 350 and 500 $\mu$m. \citet{Juvela2012} presented these observations and studied the properties of the cloud on large-scale images, without separating the distinct objects. In Rivera-Ingraham et al. (in prep) we will be carrying out a detailed \emph{Herschel}-based compact source and environmental study of all high-latitude fields (including L1642) from our programme. Furthermore, we have carried out 1.1 mm dust-continuum observations toward L1642 with the AzTEC/ASTE instrument. In this paper, we use these new data, along with other, already published data, to complete the spectral energy distribution of L1642 in the long-wavelength region and advance with this (i) the study of the large-scale structure of the cloud, (ii) the investigation of general dust properties, and (iii) the characterisation of point and compact sources related to the cloud.

The contents of the article are the following: we present the observations and data processing in Sect.~\ref{sect:observations} and methods in Sect.~\ref{sect:methods}. The results based on the observations are shown in Sect.~\ref{sect:results}. We present a radiative transfer model of L1642 in the appendix. We discuss the results in Sect.~\ref{sect:discussion} and present our conclusions in Sect.~\ref{sect:conclusions}.

\section{Observations and data processing} \label{sect:observations}

\subsection{Herschel}  \label{sect:herschel}

L1642 (G210.90-36.55) was observed by the \emph{Herschel} satellite~\citep{Pilbratt2010} instruments SPIRE~\citep{Griffin2010} in March 2011 and PACS~\citep{Poglitsch2010} in July 2011 in photometric mode, using wavelengths 100 and 160 $\rm \mu m$ for PACS and 250, 350, and 500 $\rm \mu m$ for SPIRE. The observations covered a rectangular area of approximately $50\arcmin \times 50\arcmin$. The SPIRE observations were reduced with the Herschel Interactive Processing Environment HIPE v.10.0.0 using the official pipeline\footnote{HIPE is a joint development by the Herschel Science Ground Segment Consortium, consisting of ESA, the NASA Herschel Science Center, and the HIFI, PACS and SPIRE consortia.}
with the iterative destriper and the extended emission calibration options. The SPIRE maps\footnote{The reduced data will become available through the ESA web site http://herschel.esac.esa.int/UserReducedData.shtml. Further information is available on the homepage of the Cold Cores project https://wiki.helsinki.fi/display/PlanckHerschel/\\The+Cold+Cores} were then produced using the HIPE naive map-making routine, which projects the data onto the sky and then averages the time-ordered data. The PACS data were processed up to Level 1 within HIPE v10.0.0, after which Scanamorphos v20~\citep{Roussel2013} was used to create the map products\footnotemark[\value{footnote}]. The original resolution of the maps, in the order of increasing wavelength, is 7$\arcsec$, 12$\arcsec$, 18$\arcsec$, 25$\arcsec$, and 36$\arcsec$.

We made zero-point and colour corrections to the \emph{Herschel} data and corrected the effect of very small grain emission in the 100 $\mu$m data as in~\citet{Juvela2011}. In the zero-point correction, we used Planck~\citep{Tauber2010} data and the IRIS~\citep{Miville-Deschenes2005} version of IRAS data as reference. For the 100 $\mu$m map, we compared the mean of the \emph{Herschel} map with the mean of the IRIS data. For the longer wavelengths, we made a linear fit between \emph{Herschel} data and the values that were interpolated from Planck and IRIS maps using a modified black-body curve. These linear relations were extrapolated to zero Planck and IRIS values to obtain the offsets for the \emph{Herschel} data. 
The second method requires a clear correlation between the \emph{Herschel} and Planck+IRIS data. In the case of L1642, 160 $\mu$m and the SPIRE channels still give quite good correlations, and the second method can be used. An advantage of the second method is that it does not require the calibrations to be the same. Moreover, because it is insensitive to multiplicative errors, it is more reliable with respect to the uncertainties of colour corrections and interpolation of Planck and IRIS data.

The data were initially colour corrected assuming modified black-body emission with $T=15$ K and a spectral index of $\beta=1.8$. The zero-point correction and colour temperature calculations were iterated, and the final colour-correction corresponds to the temperature calculated from the final SPIRE maps. The fit used in the zero-point correction is an unweighted robust least-squares fit. The error estimates used in this procedure and later in the calculations of temperature and optical depth are 15\% and 7\% for PACS and SPIRE, respectively.

\subsection{Planck}

We used archived Planck\footnote{Based on observations obtained with Planck (http://www.esa.int/Planck), an ESA science mission with instruments and contributions directly funded by ESA Member States, NASA, and Canada.} data~\citep{Planck2013I} of the L1642 area. The data, corresponding to the first 15.5 months of Planck observations, are available through the Planck Legacy Archive\footnote{http://www.sciops.esa.int/index.php?project=planck\&\\page=Planck\_Legacy\_Archive}. We used maps at 217, 353, 545, and 857 GHz, corresponding to the wavelengths at 1380, 850, 550, and 350 $\mu$m. The angular resolutions (full width at half maximum (FWHM) beam size) of the maps are 5.01$\arcmin$, 4.86$\arcmin$, 4.84$\arcmin$, and 4.63$\arcmin$, respectively. We converted the 217 and 353 GHz maps from KCMB units to MJy/sr units using the coefficients given in Table 6 of~\citet{Planck2013IX} (483.690 and 287.450 MJy/sr/KCMB). Rotational transition lines of CO can significantly affect the signal of Planck channels 100 GHz (CO $J = 1-0$), 217 GHz (CO $J = 2-1$), and 353 GHz (CO $J = 3-2$). We performed a CO correction for the 217 GHz and 353 GHz maps using the SEST CO data described in Section~\ref{sect:CO}. The CO correction coefficients and the process are described in Section 3.2 of~\citet{Planck2013XIII}. We compare the SEST CO maps to the CO maps available from the Planck Archive in Appendix~\ref{sect:Planck_SEST_CO}. We colour-corrected all the Planck maps with the method described in~\citet{Planck2013IX} and the Planck Explanatory Supplement, using the colour-temperature map derived from \emph{Herschel} observations.

The calibration accuracy of the Planck HFI early maps is estimated to be $\sim$ 7 \% for the two highest frequencies and $\leq$ 2 \% for the lower frequencies~\citep{Planck2011VI}. We used 7 \% as an error estimate for all the four Planck maps to cover also the uncertainties in the CO correction of the 217 GHz and 353 GHz maps.

\subsection{Other infrared data}

We used the FIR AKARI survey~\citep{Murakami2007} observations of L1642 obtained with the FIS instrument in the narrow-band filters N60 and N160 (central wavelengths 65 and 160 $\mu$m) and in the wide-band filters WideS and WideL (central wavelengths 90 and 140 $\mu$m). The spatial resolutions of the AKARI maps are 37$\arcsec$, 39$\arcsec$, 58$\arcsec$, and 61$\arcsec$ for the N60, WideS, WideL, and N160 filters, respectively. We used
the AKARI/IRC Point Source Catalogue\footnote{http://www.ir.isas.jaxa.jp/AKARI/Observation/PSC/Public/\\RN/AKARI-IRC\_PSC\_V1\_RN.pdf} 9 and 18 $\mu$m data and AKARI/FIS Bright Source Catalogue\footnote{http://www.ir.isas.jaxa.jp/AKARI/Observation/PSC/Public\\/RN/AKARI-FIS\_BSC\_V1\_RN.pdf}~\citep{Yamamura2010} 65, 90, 140, and 160 $\mu$m data for the point sources.

We used the archived WISE satellite~\citep{Wright2010} mid-infrared (MIR) 3.4, 4.6, 12.0, and 22.2 $\mu$m maps and WISE All-Sky Source Catalog data for the point sources. The beam sizes are 6.1$\arcsec$, 6.4$\arcsec$, 6.5$\arcsec$, and 12.0$\arcsec$, in order of increasing wavelength.

We used the archived Two Micron All Sky Survey (2MASS)~\citep{Skrutskie2006} near-infrared (NIR) $J$, $H$, and $K_S$ band data to produce an extinction map of the L1642 field and to examine the point sources that are believed to be associated with the cloud L1642 itself. The limiting magnitudes of 2MASS are $J$ = 15.8$^{\rm m}$, $H$ = 15.1$^{\rm m}$, and $K_S$ = 14.3$^{\rm m}$.

We used the IRAS point source catalogue data (MIR-FIR 12, 25, 60, and 100 $\mu$m) and IRIS (the reprocessed IRAS) maps~\citep{Miville-Deschenes2005}. The spatial resolutions of the IRIS maps at 12, 25, 60, and 100$\mu$m are 3.8$\arcmin$, 3.8$\arcmin$, 4.0$\arcmin$, and 4.3$\arcmin$, respectively.

We compared our results with the 200 $\mu$m observations at 90$\arcsec$ resolution~\citep{Lehtinen2004}, obtained with the ISOPHOT instrument of the ISO satellite~\citep{Kessler1996}.

\subsection{AzTEC/ASTE}

From 2008 October to November, we carried out 1.1 mm dust-continuum observations towards L1642 with the AzTEC camera~\citep{Wilson2008} on the ASTE 10 m telescope~\citep{Ezawa2004, Kohno2004} located at Pampa la Bola, Chile, at an altitude of 4800 m. The AzTEC camera temporarily mounted on the ASTE telescope is a 144-element bolometric camera tuned to operate in the 1.1 mm atmospheric window, which provides a FWHM resolution of 28$\arcsec$~\citep{Wilson2008}. Observations were performed in raster-scan mode. 
The pointing uncertainty of the AzTEC map is estimated to be better than 2$\arcsec$. We observed Uranus as a flux calibrator twice per night.
We measured the flux conversion factor (FCF) from the optical loading value (in watts) to the source flux (in Jy beam$^{-1}$) for each detector element.
The principal component analysis (PCA)~\citep{Scott2008} cleaning method was applied to remove atmospheric noise. Details of the flux calibration are
described by~\citet{Wilson2008} and~\citet{Scott2008}. Since the PCA method does not preserve the extended components reliably, we applied the iterative mapping method FRUIT~\citep{Liu2010,Shimajiri2011} to recover the extended components. The noise level is 4.5 mJy beam$^{-1}$ in the central region, and the effective beam size is 35$\arcsec$ after FRUIT imaging.

\subsection{CO} \label{sect:CO}

We compared our observations with the CO observations described in~\citet{Russeil2003}, where the cloud L1642 was mapped in the $J = 1-0$ and $J = 2-1$ transitions of $^{12}$CO, $^{13}$CO and C$^{18}$O with the SEST radio telescope.
The half-power beam widths are 45$\arcsec$ ($J = 1-0$) and 23$\arcsec$ ($J = 2-1$), and the grid spacing is 3$\arcmin$. Typical noise levels in the data were $\Delta T_{\rm{rms}} = 0.06$ K (C$^{18}$O(2--1)) and $\sim$0.15 K (other transitions).

\begin{figure*}
\includegraphics[width=6.25cm]{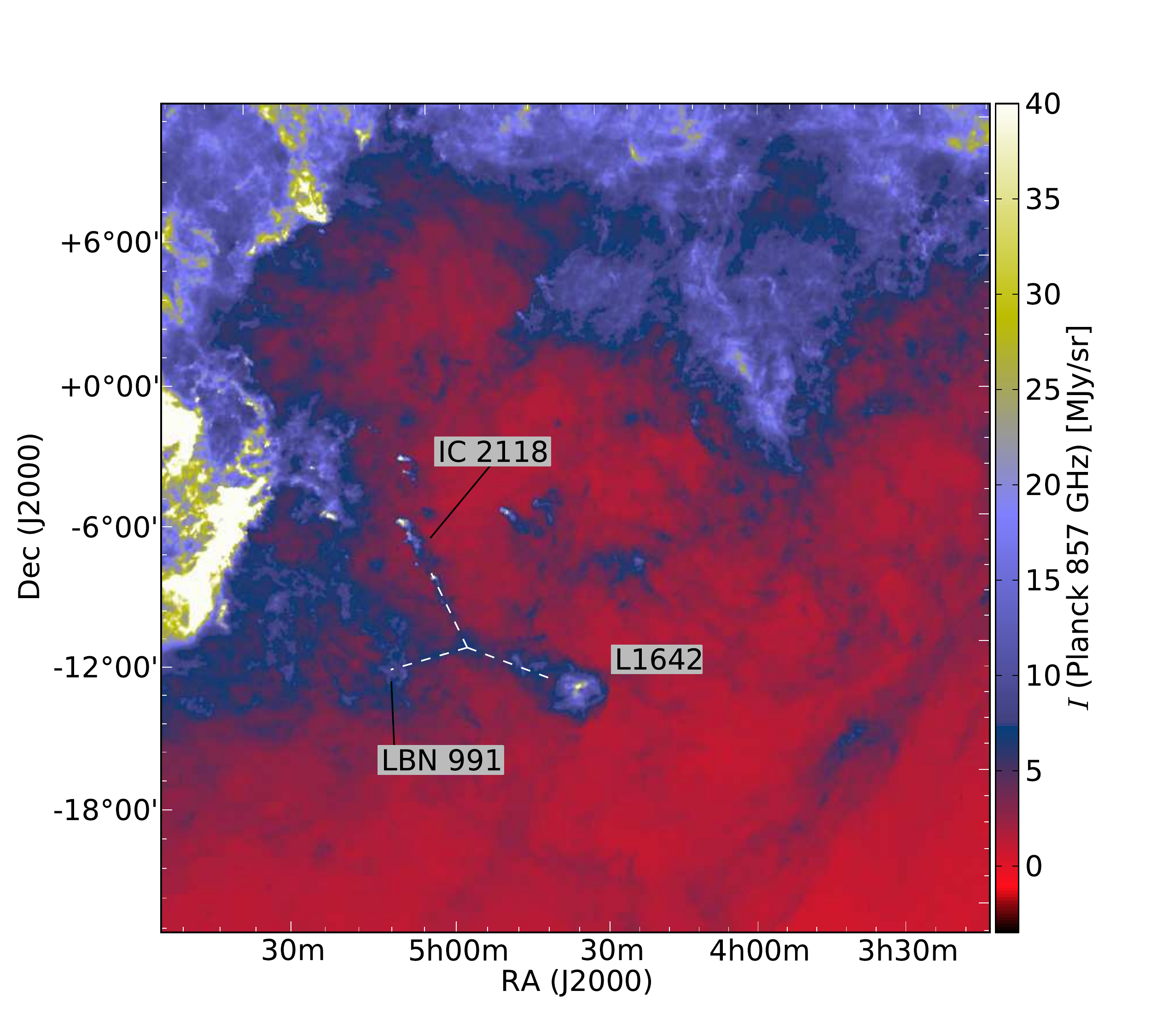}
\includegraphics[width=6.25cm]{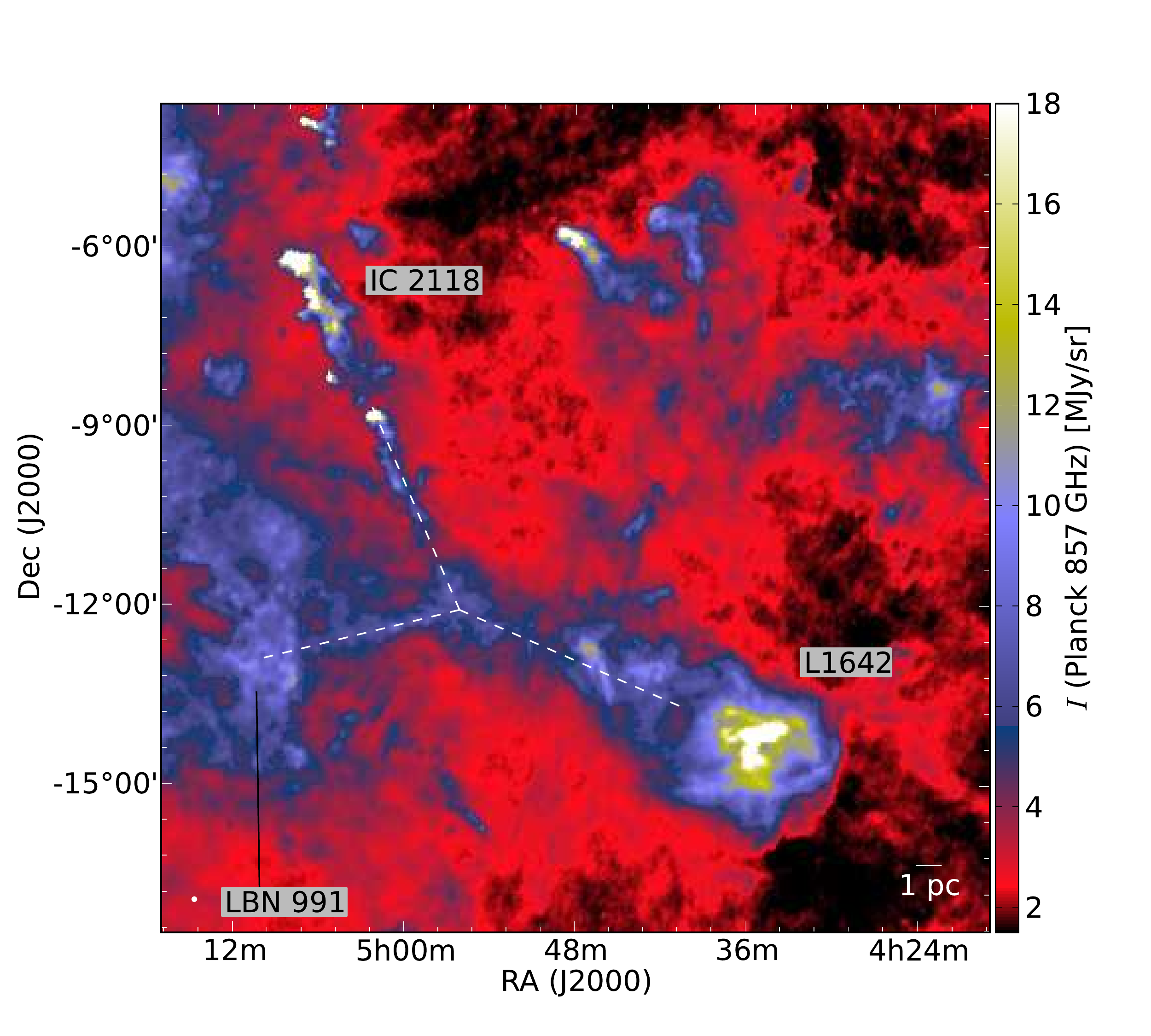}
\includegraphics[width=6.25cm]{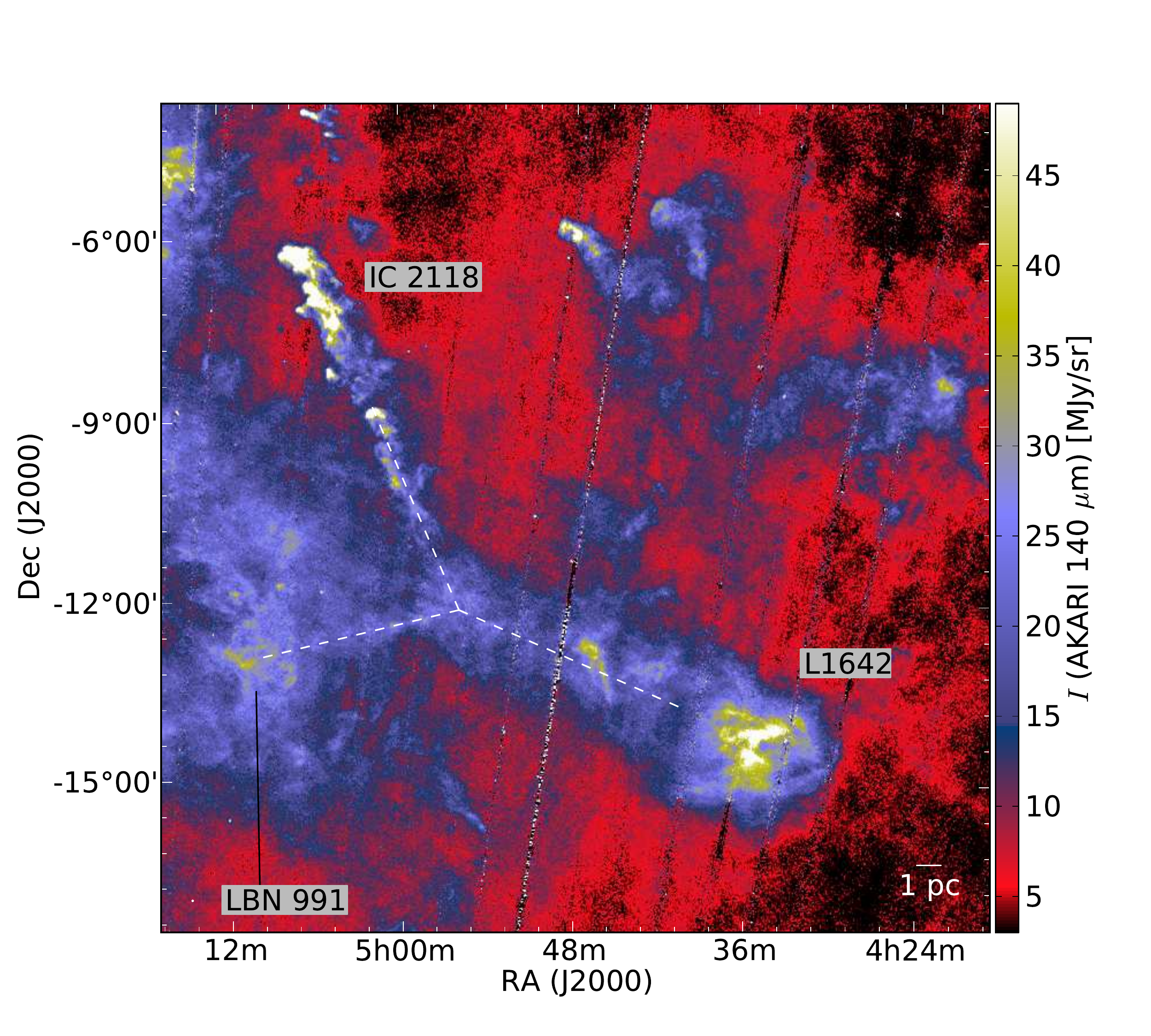}
\caption{Surrounding region of L1642. Planck 857 GHz intensity map (left) and close-ups of Planck 857 GHz (middle) and AKARI 140 $\mu$m (right) intensity maps. L1642 and other nearby clouds are marked in the maps. The 1 pc scale is marked assuming a distance of 140 pc.}
\label{fig:Planck_Ak}
\end{figure*}

\section{Methods} \label{sect:methods}

\subsection{Source extraction} \label{sect:source_extraction}

We extracted sub-millimetre clumps and calculated their fluxes from \emph{Herschel} 160-500$\mu$m maps using Getsources~\citep{Menshchikov2012}, 
a source-extraction method developed for the \emph{Herschel} Gould Belt survey~\citep{Andre2010} and \emph{Herschel} HOBYS survey~\citep{Motte2010}. The source extraction (Montillaud et al., in prep.) was made as part of the effort to catalogue the cold clumps detected in the fields observed in the Galactic Cold Cores project~\citep{Juvela2010}. The method proceeds in several steps, first extracting sources independently in each band, then combining all the data. In the extraction, we used 160, 250, 350, and 500~$\mu$m intensity maps and the optical depth at 250 $\mu$m, $\tau_{250}$, obtained from Markov chain Monte Carlo calculations (see Sect.~\ref{sect:MC_fitting}).
The optical depth $\tau_{250}$ was used in the source extraction to ensure that the extracted sources correspond to dense clumps.
All the maps were convolved to a resolution of $\sim40\arcsec$. 

We calculated the flux of the point sources in \emph{Herschel} 100$\mu$m and ASTE 1.1mm (and also in other \emph{Herschel} bands for comparison) using aperture photometry with a circular aperture. The same aperture sizes were used for all clumps and maps.
The aperture radius is 40$\arcsec$ and the annulus radii 55$\arcsec$ and 85$\arcsec$. 
We subtracted the mean value of the annulus from the aperture values.
Differences in background subtraction and in the size of the aperture can cause significant bias in the derived fluxes in the FIR regime, and it is therefore important to use the same method when comparing and fitting the values based on different data sets.

\subsection{Mass calculation}  \label{sect:mass}

We calculated the masses of a chosen region by summing the pixel values of the region using equation
\begin{equation}
M = \frac{ \mu m_{\rm H}\tau_{250}D^2\Omega_{\rm PIX}}{\sigma_{250}},
\label{eq:M}
\end{equation}
where $\mu$ is the mean weight per H atom (1.4), $m_{\rm H}$ the mass of a hydrogen atom, $\tau_{250}$ the optical depth at 250 $\mu$m, $D$ the distance, $\Omega_{\rm PIX}$ the solid angle of a pixel, and $\sigma_{250}$ the absorption cross-section per H atom at 250 $\mu$m.

The virial mass can be derived using equation
\begin{equation}
M_{\rm vir} = \frac{ k \sigma^2 R}{G},
\label{eq:Mvir}
\end{equation}
where $G$ is the gravitational constant, $R$ the cloud radius, $k$ a factor depending on the radial density distribution,
and $\sigma$ the three-dimensional velocity dispersion~\citep{MacLaren1988}. $\sigma$ is given by the equation
\begin{equation}
\sigma = \sqrt{3\Big(\frac{k_{\rm B}T_{\rm gas}}{\overline{m}} + \big(\frac{\Delta V^2}{8{\rm ln}2} - \frac{k_{\rm B}T_{\rm gas}}{m}\big)\Big)},
\label{eq:s}
\end{equation}
where $k_{\rm B}$ is the Boltzmann constant, $T_{\rm gas}$ the kinetic gas temperature,
$\overline{m}$ the mean molecular mass ($\mu m_{\rm H}$, where $\mu = 2.33$ is the mean molecular weight per free particle), $m$ the mass of the molecule used for observations ($^{13}$CO or C$^{18}$O), and $\Delta V$ the observed line width (FWHM) corrected for opacity line-broadening. 
\citet{Lehtinen2004} derived virial masses using these equations, and assuming that $k = 1.25$, a value between the density distributions $\rho(r) = r^{-1}$ and $\rho(r) = r^{-2}$. To compare our results
with this, we used the same value for $k$.

\subsection{Classifying sources}  \label{sect:class}

Young stellar objects can be classified into five classes, based on the slope of the NIR--MIR ($\sim$2--24 $\mu$m) spectral energy distribution (SED) or spectral index
\begin{equation}
\alpha = \frac{d\: \rm{log}\: \nu F_{\nu}}{d\: \rm{log}\: \nu} = \frac{d\: \rm{log}\: \lambda F_{\lambda}}{d\: \rm{log}\: \lambda},
\label{eq:alpha}
\end{equation}
where $F_{\nu}$ is the flux as a function of frequency, $\nu$, and $F_{\lambda}$ the flux as a function of wavelength, $\lambda$, (combining the work of~\citet{Lada1987}, \citet{Greene1994}, and~\citet{Andre1993}). Class 0 objects are undetectable at $\lambda < 20 \mu$m. The spectral index limits are $\alpha > 0.3$, $0.3 > \alpha > -0.3$, $-0.3 > \alpha > -1.6$, and $-1.6 > \alpha$, for Class I, Flat spectrum, Class II, and Class III sources, respectively. The classes are assumed to describe the evolutionary sequence from deeply embedded Class 0 objects through Classes I--II with NIR and MIR excess to Class III, where most of the surrounding disk has disappeared, and only little NIR excess is seen.

\citet{Robitaille2006} presented a grid of 20,000 radiation transfer models of YSOs in different evolutionary stages and from ten different viewing angles, resulting in 200,000 SEDs. These models used multiwavelength information to classify the potential YSOs, instead of just NIR--MIR fluxes used in the traditional classification system. Longer wavelengths can be useful in identifying the properties of the surrounding envelope and disk, and not just the central source. However, there are two main shortcomings in the models concerning our data. Firstly, the models cut the envelope when the dust temperature falls below 30 K. This means that the FIR spectrum maximum has to be below 100 $\mu$m. If there is a significant amount of surrounding cold dust within the aperture, the models cannot fit the \emph{Herschel} bands. Secondly, all the models are based on single objects. In our case, the two young stars in L1642 are known binary stars. However, the models allow large inner envelope and disk holes, which can correspond to the effects of binary systems. Close-by binary objects do not necessarily change the SED notably. Moreover, \citet{Robitaille2007} showed (in their Fig. 1) the SEDs of two binary stars, where the models fit the data quite well.

\section{Data analysis and results} \label{sect:results}

\subsection{Surface brightness maps and the general structure of L1642}

We present surface brightness maps of L1642 using \emph{Herschel}, Planck, AKARI, and AzTEC/ASTE data. The Planck 857 GHz map of L1642 and the surrounding area is shown in Fig.~\ref{fig:Planck_Ak} (left). We show a close-up of the same area in Fig.~\ref{fig:Planck_Ak} with Planck and AKARI data. Straight lines following the pillars from L1642, IC 2118 (Witch Head nebula) and LBN 991 are drawn to the figures, leading to the centre at $\alpha_{2000} = 4^{\rm h}56^{\rm m}$, $\delta_{2000} = -12^{\circ}10'$.

\begin{figure*}
\centering
\includegraphics[width=8.5cm]{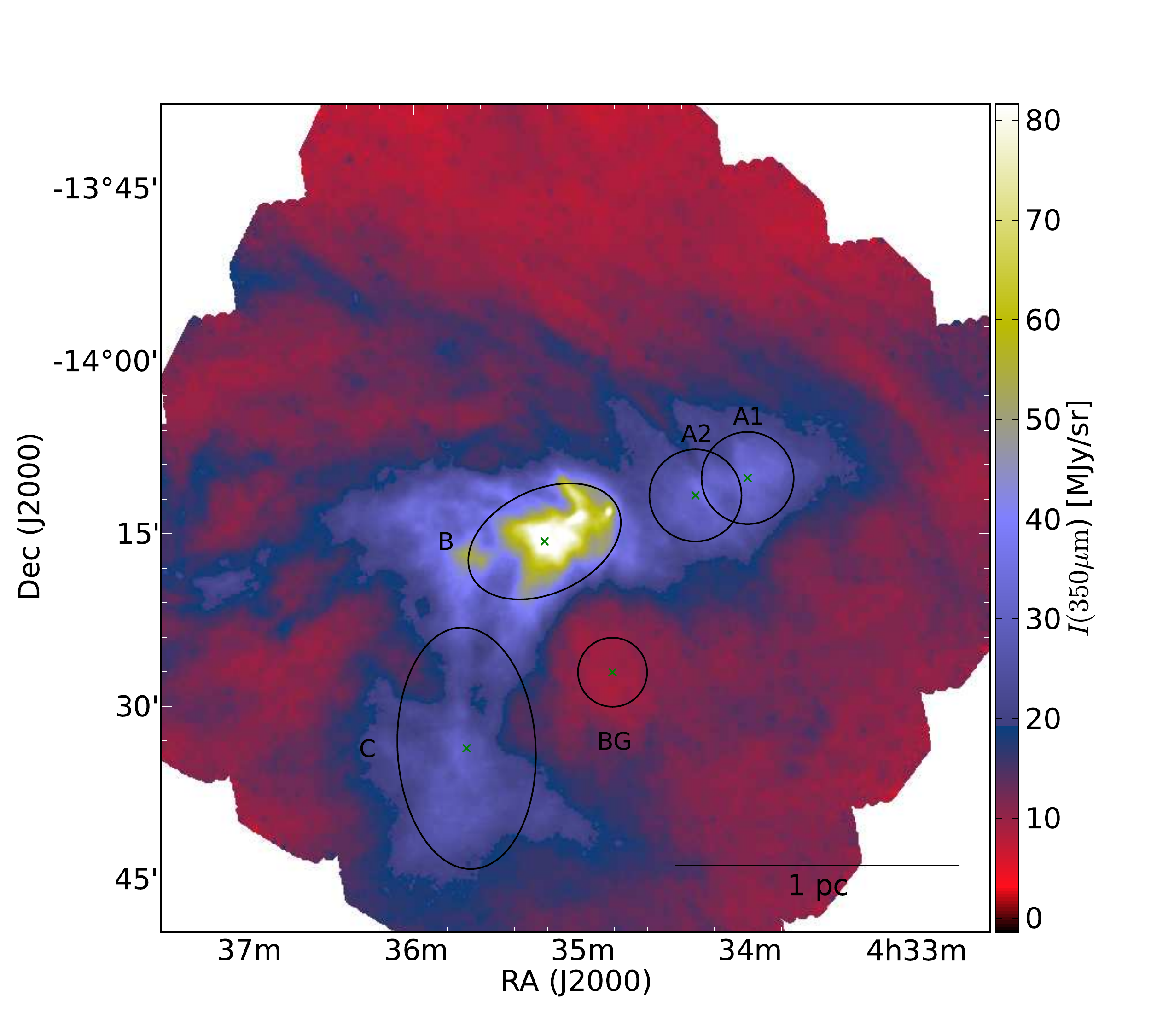}              
\includegraphics[width=8.5cm]{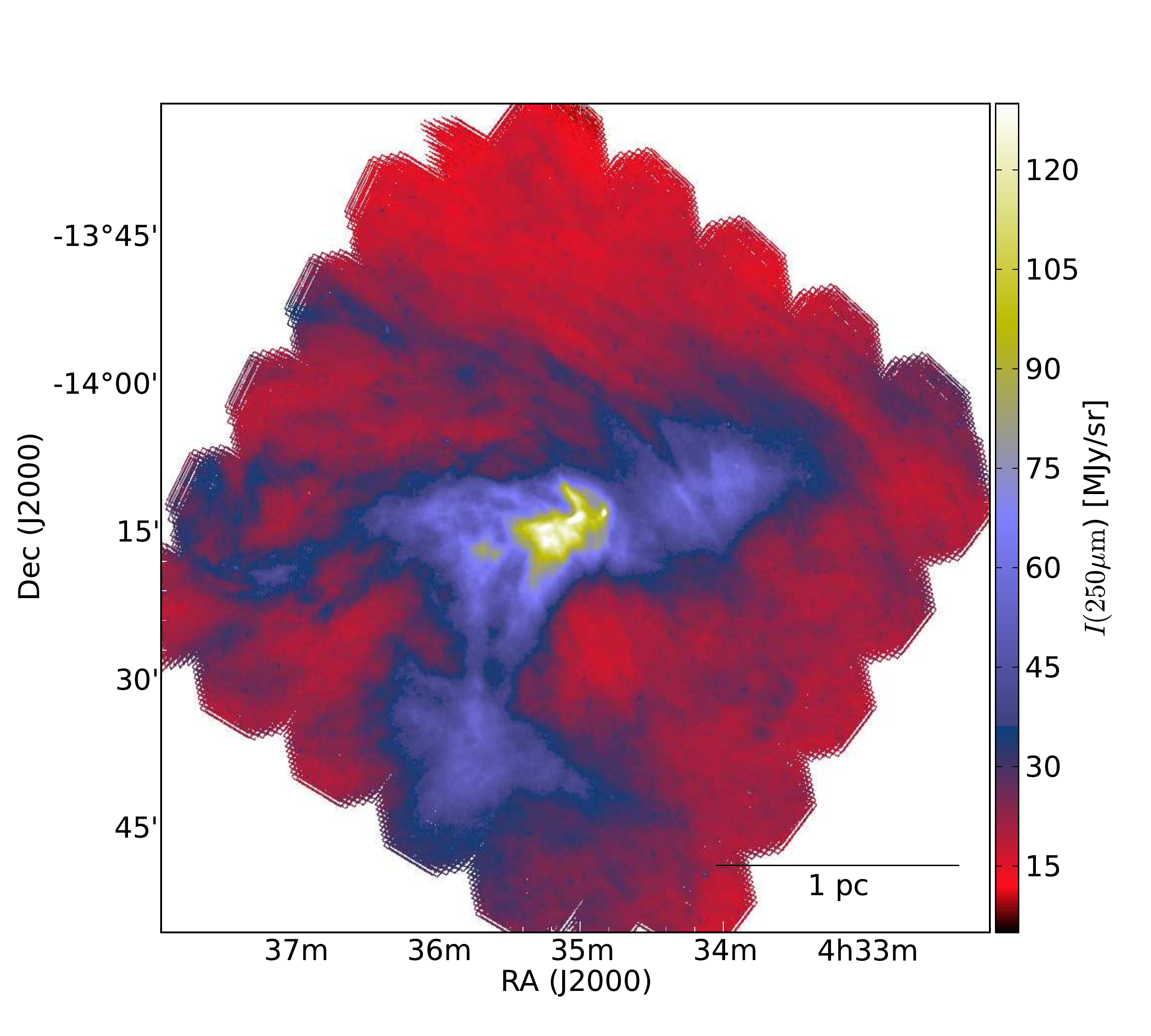}                
\includegraphics[width=8.5cm]{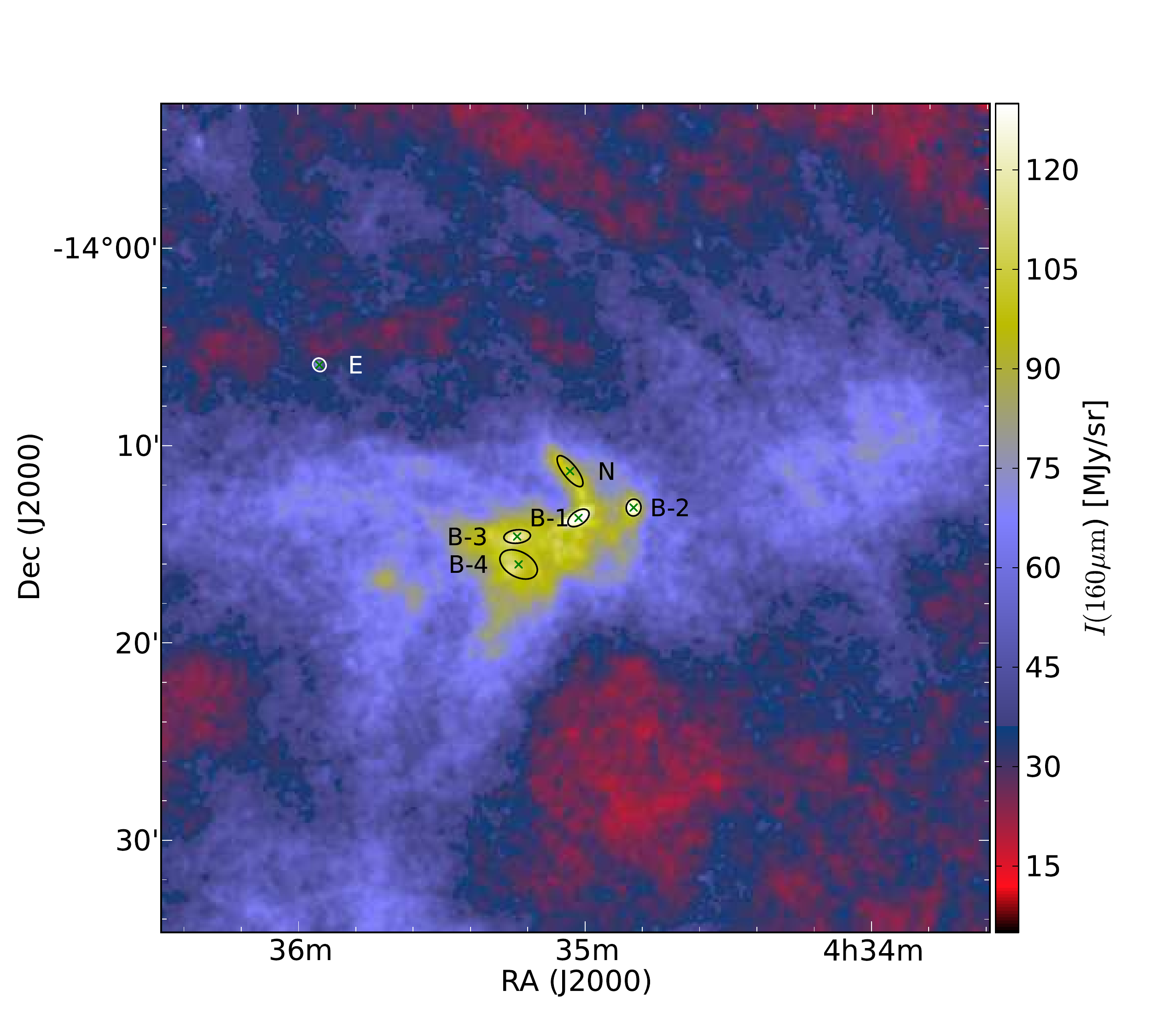}  
\includegraphics[width=8.5cm]{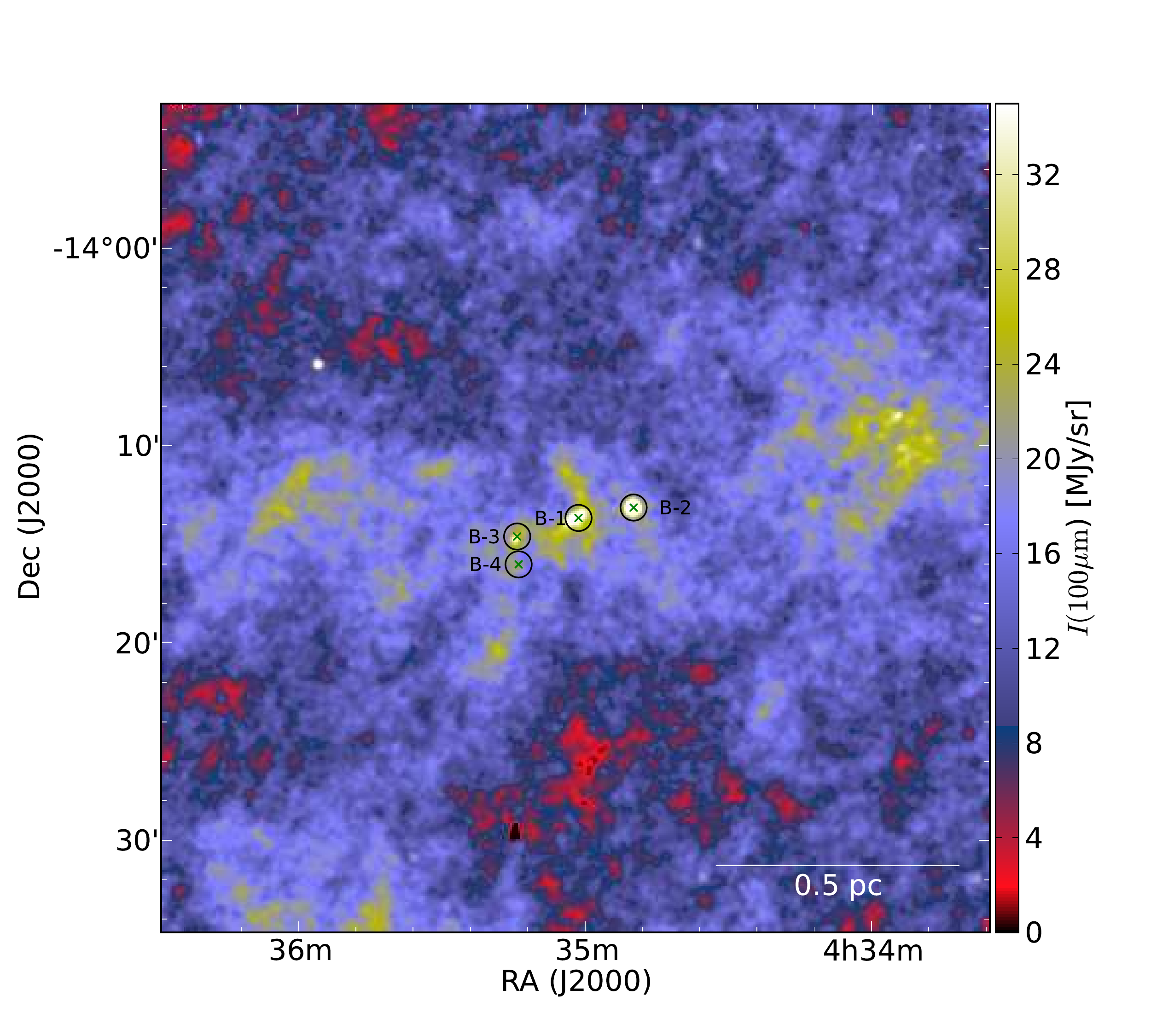}  
\caption{\emph{Herschel} intensity maps showing the areas used in the analysis.
The maps show the total intensity after zero-point correction.
(Top left) 350 $\mu$m (26.8$\arcsec$ resolution) showing the larger regions (A1, A2, B and C) and the background area (BG) used in the analysis. (Top right) 250 $\mu$m (18.3$\arcsec$ resolution) map showing the striation in L1642. (Bottom left) 160 $\mu$m (18.3$\arcsec$ resolution) showing the most prominent clumps identified by Getsources. FWHM ellipses of the clumps are drawn in the figure. (Bottom right) 100 $\mu$m (18.3$\arcsec$ resolution) showing the circular apertures (with 80$\arcsec$ diameter) used in the aperture photometry of the point sources marked in the figure. 
}
\label{fig:areas_I_maps}
\end{figure*}

L1642 consists of dense regions, surrounded by more diffuse material. \citet{Lehtinen2004} named the dense regions A1, A2, B, and C. We adopt this naming convention and show the regions in Fig.~\ref{fig:areas_I_maps} (top left frame). The high-resolution \emph{Herschel} data show considerably more details than the earlier ISO observations at 90$\arcsec$ resolution, especially in the densest region B. In the eastern part of region B, there is an intensity maximum separated from the main clump, justifying the division of region B into areas B1 and B2, similarly to region A.
Two filaments lead from these two B region maxima southwards to the C clump. The \emph{Herschel} data also show striation in the diffuse dust emission around the cloud L1642, especially at the northern part. This is a real structure and not an observational artefact.
The striation is mainly towards north-east direction from the cloud, especially near region A,
and does not follow the scanning direction.
At the northern side of the B clump there is one denser elongated structure, marked N in Fig.~\ref{fig:areas_I_maps} (bottom-left frame). This structure, like the fainter striations, was not resolved in the earlier studies~\citep{Russeil2003,Lehtinen2004}. The structure can be seen in all the \emph{Herschel} maps and the ASTE 1.1 mm map and to some extent also in the AKARI 140 and 160 $\mu$m maps.

\begin{figure*}
\centering
\includegraphics[width=8.5cm]{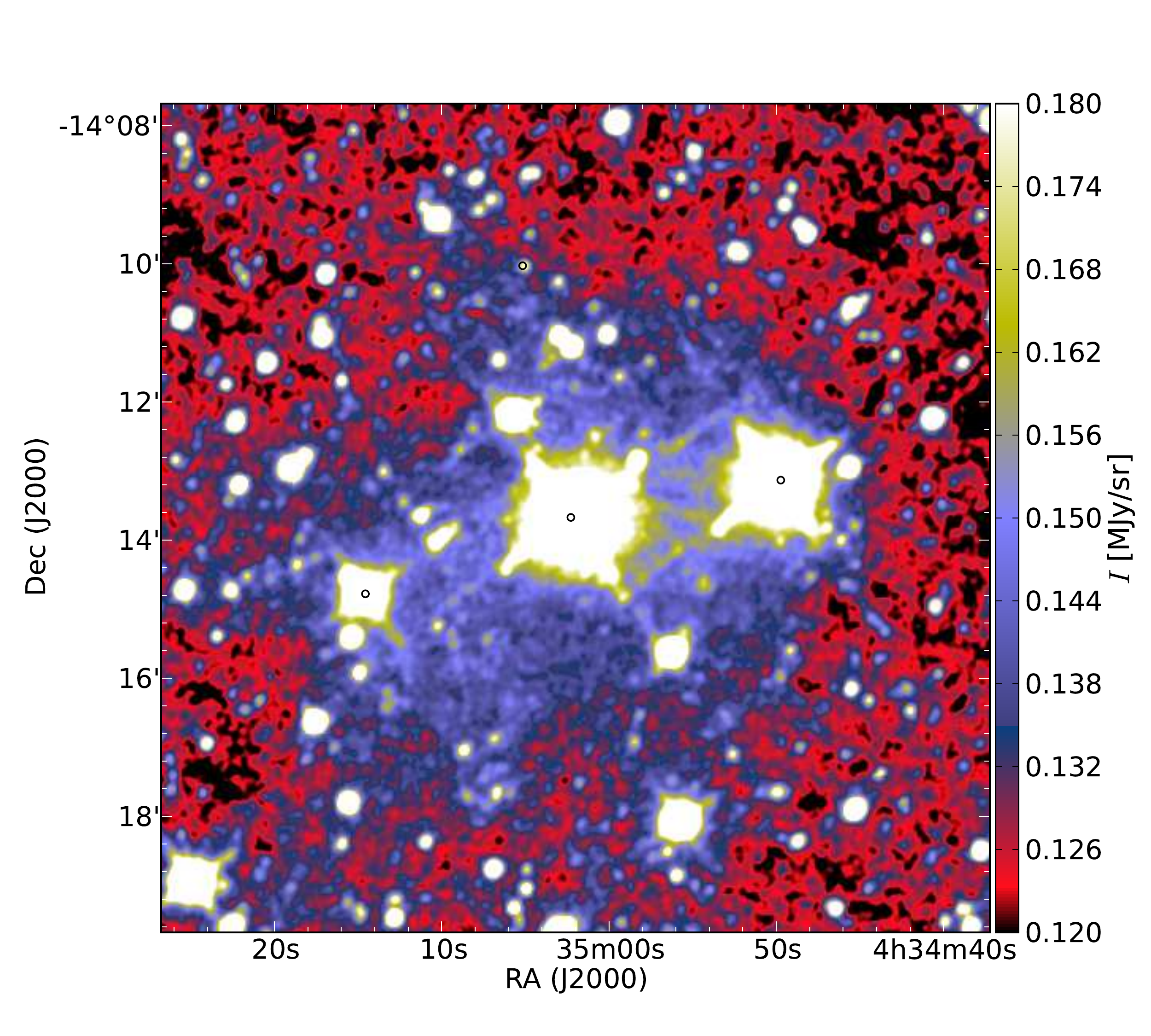}              
\includegraphics[width=8.5cm]{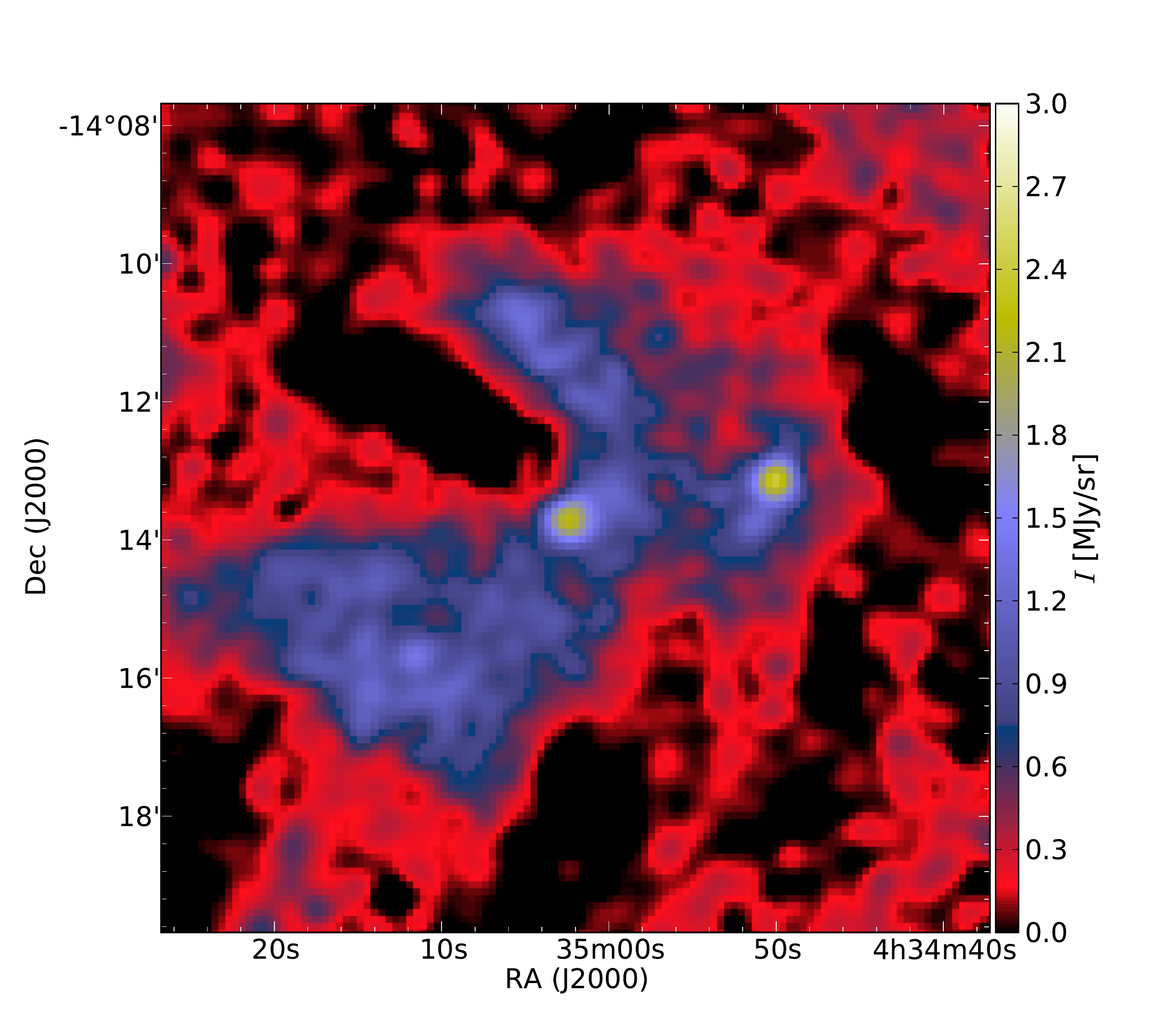}                
\includegraphics[width=8.5cm]{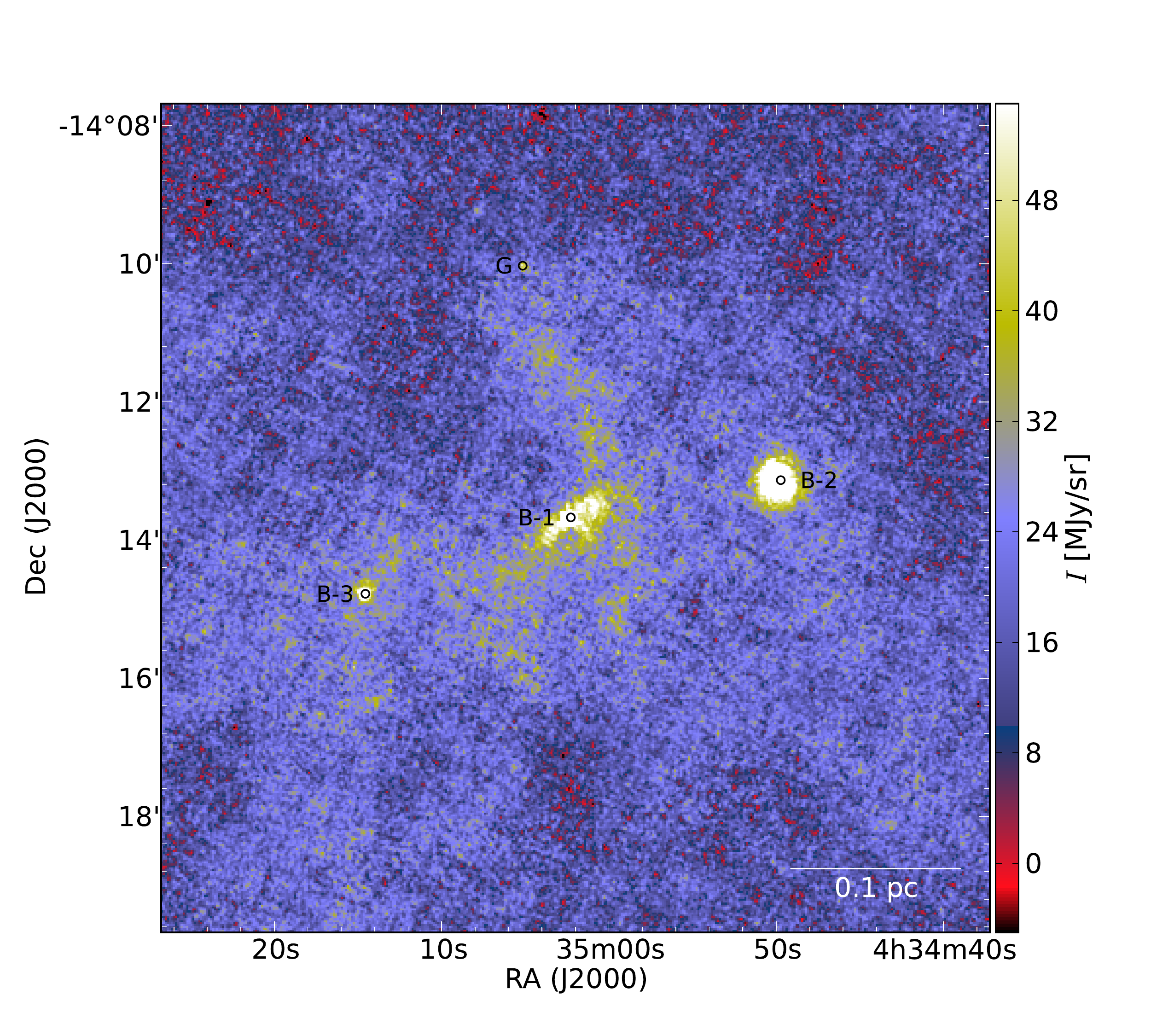}
\includegraphics[width=8.5cm]{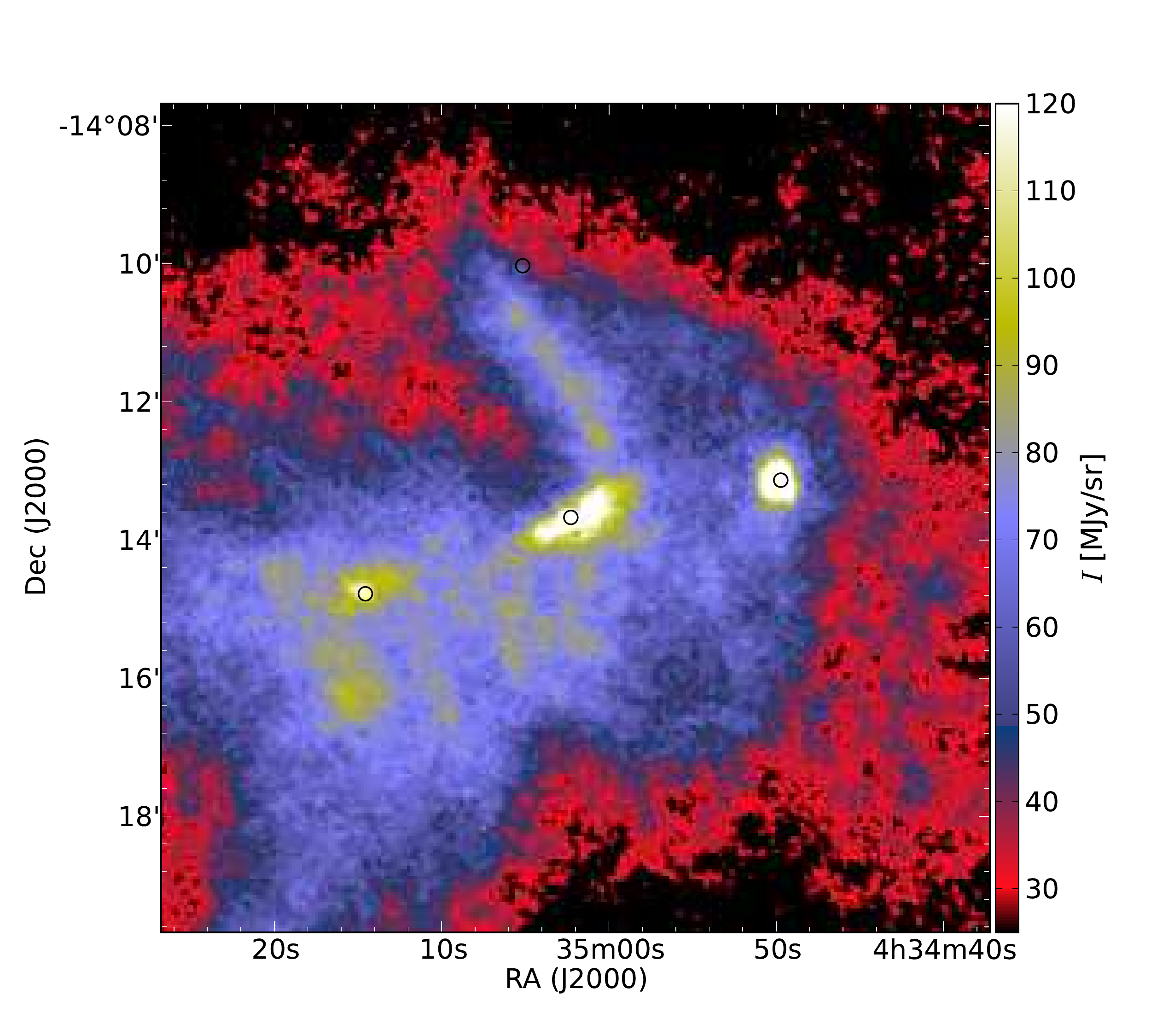}
\caption{WISE 3.4 $\mu$m and AzTEC/ASTE 1.1 mm intensity maps (top row). \emph{Herschel} 100 and 160 $\mu$m intensity maps (without zero-point correction) at the original resolution (7$\arcsec$ and 12$\arcsec$, respectively) (bottom row). The sources B-1, B-2, B-3, and G (see text) are marked on the \emph{Herschel} and WISE maps, using their 2MASS coordinates. The marker circles show the beam FWHM.}
\label{fig:I_zoom}
\end{figure*}

The point sources inside the cloud were not detected in the ISO data~\citep{Lehtinen2004}. In the \emph{Herschel} data, however, point sources can be seen even at longer wavelengths in all bands 100--500 $\mu$m. Fig.~\ref{fig:areas_I_maps} (bottom left frame) shows the most prominent clumps obtained with Getsources~\citep{Menshchikov2012} from the \emph{Herschel} data (see Sect.~\ref{sect:source_extraction}). 
We also studied sources marked B-1, B-2, B-3, and B-4 with aperture photometry. The apertures with 80$\arcsec$ diameter are marked in Fig.~\ref{fig:areas_I_maps} (bottom-right frame).
We present region B and the point sources inside it in more detail in Fig.~\ref{fig:I_zoom} with WISE, ASTE and \emph{Herschel} data. Point sources B-1, B-2, and B-3 can be seen in WISE 3.4--12.0 $\mu$m maps and in the 
\emph{Herschel} data. Sources B-1 and B-2 can also be seen even in the ASTE data at 1.1 mm. B-3 and B-4 are not as prominent in ASTE data, but they are situated near local intensity maxima.
The beam FWHM is drawn on the sources in the \emph{Herschel} maps, indicating that the point sources are extended structures in FIR wavelengths. However, in 100 $\mu$m, the FWHM of source B-3 is very close to the beam FWHM.

The object marked E in Fig.~\ref{fig:areas_I_maps} (bottom-left frame) can be clearly seen in the \emph{Herschel} 100--350 $\mu$m maps. This is the object IRAS 04336-1412 (or L 1642-3) that was mentioned, but not studied in more detail, in~\citet{Sandell1987} and~\citet{Lehtinen2004}. NED\footnote{http://ned.ipac.caltech.edu/} identifies this object as a galaxy 2MASX J04355560-1405542. There is a suitable counterpart in the WISE catalogue, WISE J043555.67-140554.1, within 2$\arcsec$.

At the northern end of the elongated structure N, a compact point source can be seen in the 100 $\mu$m map at 7$\arcsec$ resolution in Fig.~\ref{fig:I_zoom} (bottom left frame), marked G, at coordinates 4$^{\rm h}$35$^{\rm m}$05.157$^{\rm s}$ $-14^{\circ}$10$\arcmin$01.85$\arcsec$. 2MASS 04350296-1410159 and WISE J043504.98-141002.4 (within 2.7$\arcsec$) are the closest counterparts, and the object is clearly seen in the WISE maps. There are no NED objects within 1$\arcmin$, and the nature of this object is unclear. The object cannot be seen in the convolved \emph{Herschel} maps, even at 100 $\mu$m. There are several similar objects within the \emph{Herschel} map area, which are probably galaxies.

\subsection{Colour temperature, optical depth, and spectral index maps} \label{sect:MC_fitting}

Colour temperature, optical depth, and spectral index maps can be derived by fitting the observed FIR intensity maps with the modified black-body (MBB) law
\begin{equation}
I_{\nu} \sim B_{\nu}(T) \tau_{\nu} = B_{\nu}(T) \mu m_{\rm H} \kappa_{\nu} N_{\rm H} \propto B_{\nu}(T) {\nu}^{\beta},
\label{eq:MBB}
\end{equation}
where $I_{\nu}$ is the observed intensity, $B_{\nu}(T)$ is the Planck radiation law for temperature $T$ and frequency $\nu$, $\tau_{\nu}$ is the optical depth at frequency $\nu$, $\mu$ is the mean weight per H atom (1.4), $m_{\rm H}$ is the mass of H atom,
$\kappa_{\nu}$ is the mass absorption (or emission) coefficient (cm$^2$/g) relative to gas mass (often called opacity), $N_{\rm H}$ is the column density of H atoms (1/cm$^2$), and $\beta$ is the dust emissivity spectral index.
$T$ and $\beta$ can be kept as constants or free parameters in the fitting. The equation assumes optically thin emission. We use wavelength instead of frequency in the notation throughout the article, e.g., $\tau_{250}$ = $\tau_{250 \mu\rm{m}}$.

We used Markov chain Monte Carlo (MCMC) fitting~\citep{Veneziani2010, Juvela2013a} to derive colour temperature $T$, optical depth at 250 $\mu$m $\tau_{250}$, and spectral index $\beta$ maps from the \emph{Herschel} 160--500 $\mu$m intensity maps. The data used in the fitting and the resulting maps are all at 40$\arcsec$ resolution.
The calculations employed flat priors where the allowed parameter ranges were
$5\, {\rm K} < T < 35\, {\rm K}$ and
$0.3 < \beta < 4.0$. The derived $\beta$ and $T$ maps are shown in Fig.~\ref{fig:tau250_T_maps_freeb}. Within region B, $\beta$ values are mostly between 1.7 and 1.9. However, outside the densest region, noise dominates the derived $\beta$ map.

The calculations were also performed using the fixed spectral index value of $\beta=1.8$, in which case only wavelengths 250--500 $\mu$m were used.
160 $\mu$m was omitted, as the PACS channels have a smaller area, leaving the eastern tail of the cloud out of the map. PACS maps also have a lower S/N than SPIRE maps, and the shorter wavelengths might introduce more bias, which might lead to an underestimation of the optical depth~\citep[e.g.,][]{Shetty2009a,Malinen2011}. This can be more important than the increase in the statistical noise. Including 160 $\mu$m in the derivation of $\tau_{250}$ would lead to masses smaller by $< 2$ \%  for most of the regions in the analysis of Sect.~\ref{sect:regions}.

\begin{figure*}
\centering
\includegraphics[width=8.5cm]{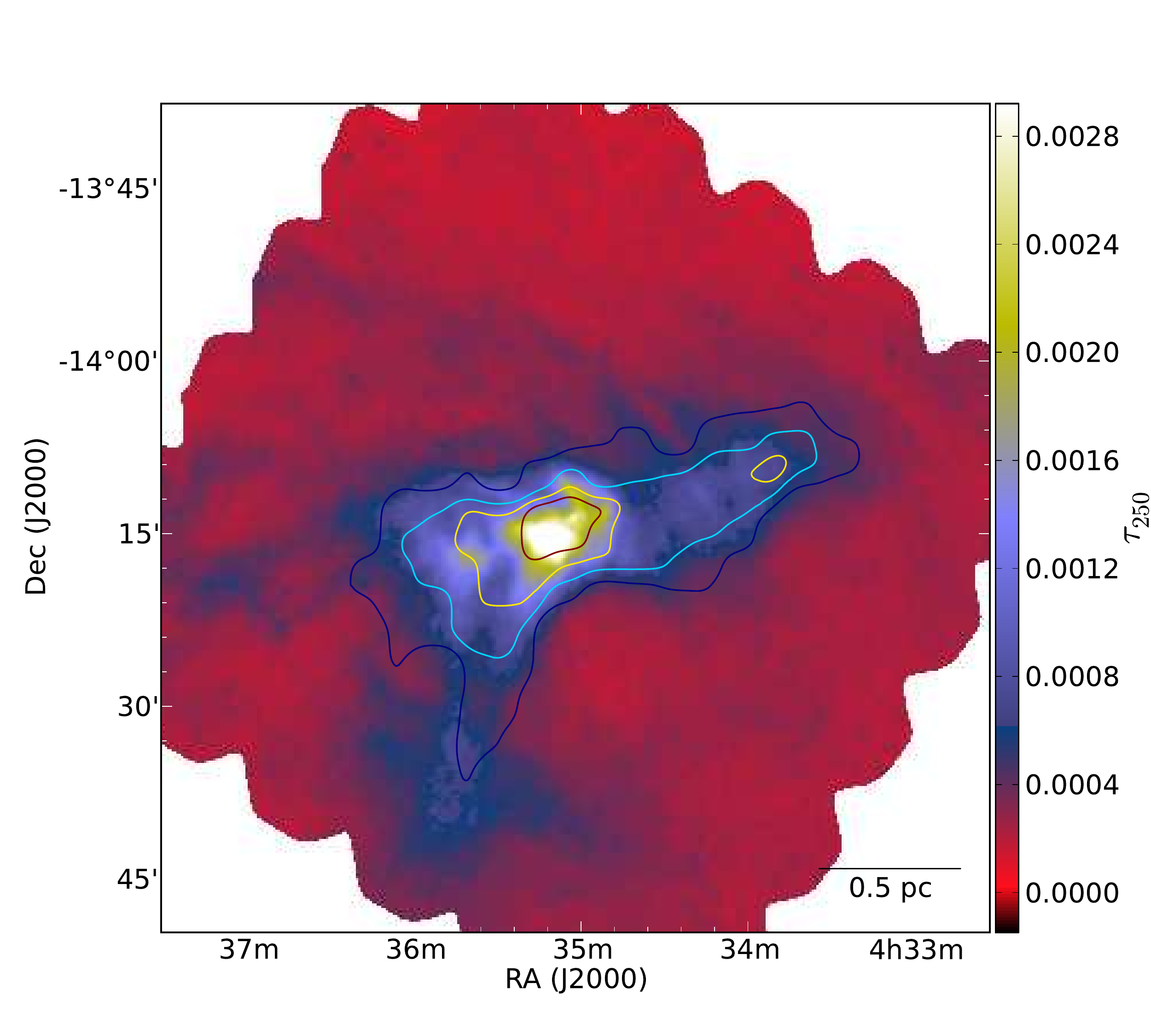}
\includegraphics[width=8.5cm]{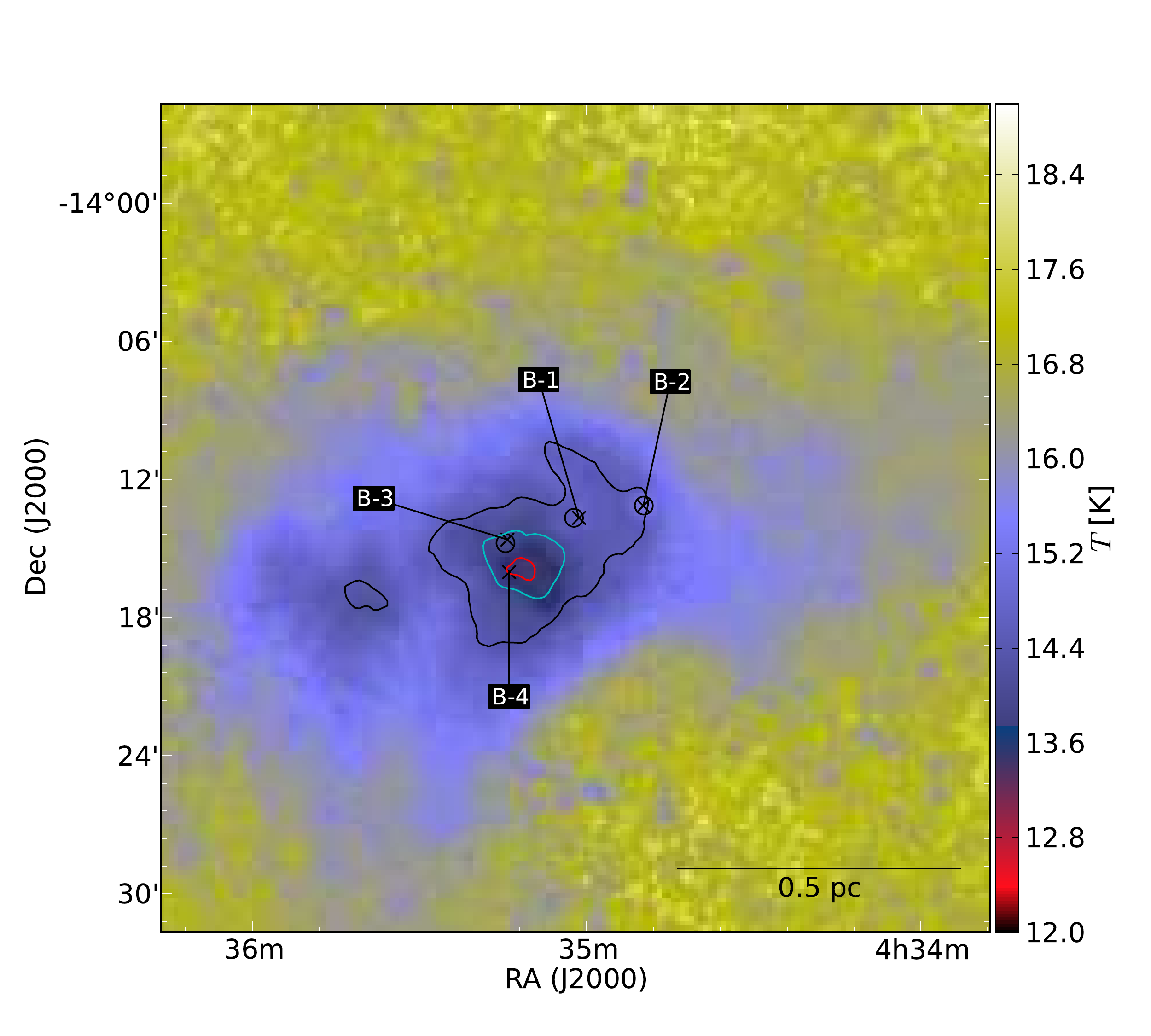}
\caption{Maps derived by MCMC fitting with a constant $\beta = 1.8$, using \emph{Herschel} 250--500 $\mu$m maps at 40$\arcsec$ resolution.
Background is not subtracted from the maps. (Left) Optical depth $\tau_{250}$ with contours of column density $N_{\rm H_2}$ derived from $^{13}$CO observations of~\citet{Russeil2003} drawn at levels 4, 3, 2, and 1 $\times 10^{21}$ cm$^{-2}$.
(Right) Colour temperature $T$ with contours of the $\tau_{250}$ map drawn at levels 0.0035, 0.0029, and 0.0017. Sources B-1, B-2, B-3, and B-4 are marked with crosses (Getsources coordinates) and circles (WISE coordinates).
}
\label{fig:tau250_T_maps}
\end{figure*}

Fitted optical depth and colour temperature maps, using $\beta = 1.8$, are shown in Fig.~\ref{fig:tau250_T_maps}. 
The contours of the column density map derived from the $^{13}$CO data of~\citet{Russeil2003} are drawn in the optical depth map in Fig.~\ref{fig:tau250_T_maps} (left). The CO-based column density map has a lower resolution because the line observations were obtained with a 3$\arcmin$ step.
In both cases, 
the local diffuse background was not subtracted and the total intensities (with the intensity zero-points derived from comparison with Planck data) were used when deriving the $\tau_{250}$, $\beta$, and $T$ maps.

For the analysis of regions A1, A2, B, and C, we derived the $\tau_{250}$ and $T$ maps with a constant value $\beta = 1.8$, using background-subtracted intensity maps. The areas used in the analysis and the reference area used for the local diffuse background subtraction are marked in the derived $\tau_{250}$ and $T$ maps in Fig.~\ref{fig:tau250_T_maps_app}.

\subsection{Dust opacity} \label{sect:correlations}
 
We studied the dust properties in different areas of the cloud by deriving the dust opacity, the ratio of FIR optical depth to column density, 
\begin{equation}
\sigma_e(\nu) = \tau_{\nu}/N_{\rm H} = \mu m_{\rm H} \kappa_{\nu} [{\rm cm}^2/{\rm H}].
\label{eq:sigma}
\end{equation}
We used $\tau_{J}$ obtained from the NIR extinction map as an independent tracer of column density. We derived an extinction map of L1642 using 2MASS stars (excluding the YSOs) and the NICER method~\citep{Lombardi2001} and converted the map to optical depth at $J$ band, $\tau_J$, shown in Fig.~\ref{fig:2mass_tauJ} (left frame). 
We used the extinction curves of~\citet{Cardelli1989}, with, for instance, $A_{\rm J}/E(J-K) = 1.67$. The original pixel size of the extinction map is 30$\arcsec$.
We compared the $\tau_J$ map based on extinction with the optical depth map at 250 $\mu$m, $\tau_{250}$, derived from \emph{Herschel} data (pixel size 12$\arcsec$).
A $\tau_{250}/\tau_J$ map at 180$\arcsec$ resolution is shown in Fig.~\ref{fig:2mass_tauJ} (right frame). The reference area used in the background subtraction of both optical depth maps is marked in the figure.
The map shows that at low resolution, the densest areas A, B, and C have a considerably higher emissivity than the surrounding diffuse areas.

\begin{figure}
\centering
\includegraphics[width=9cm]{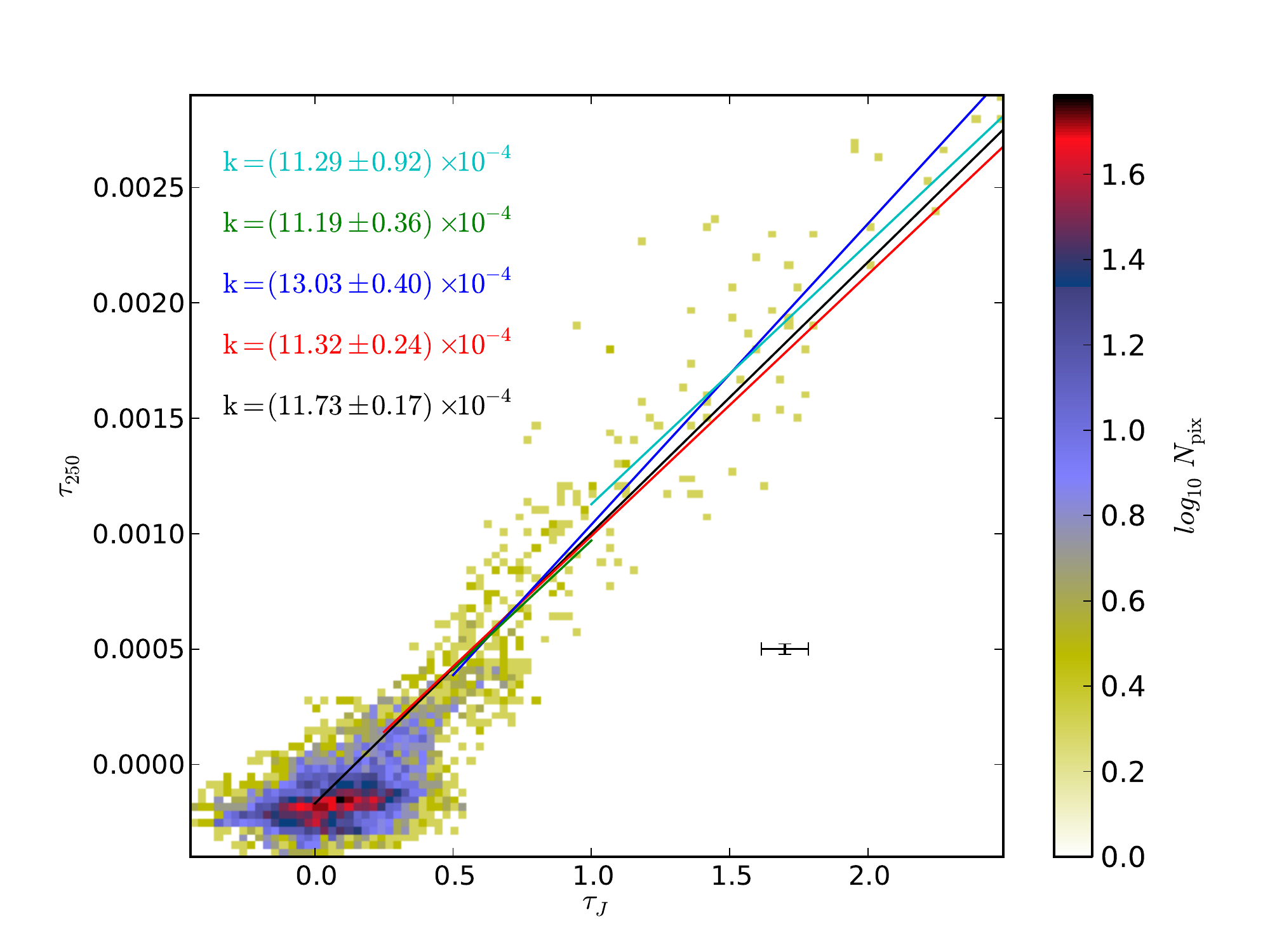}
\includegraphics[width=9cm]{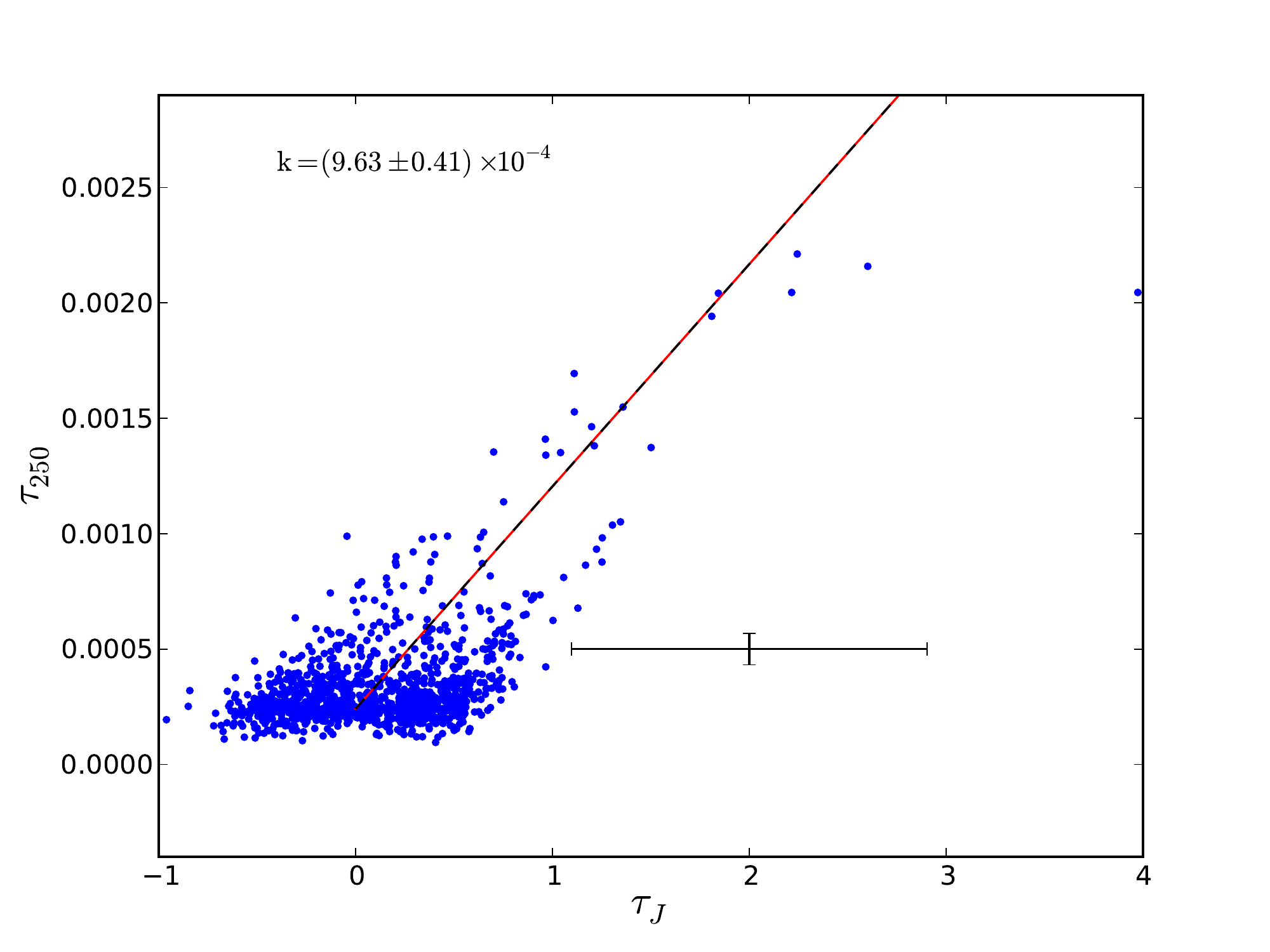}
\includegraphics[width=9cm]{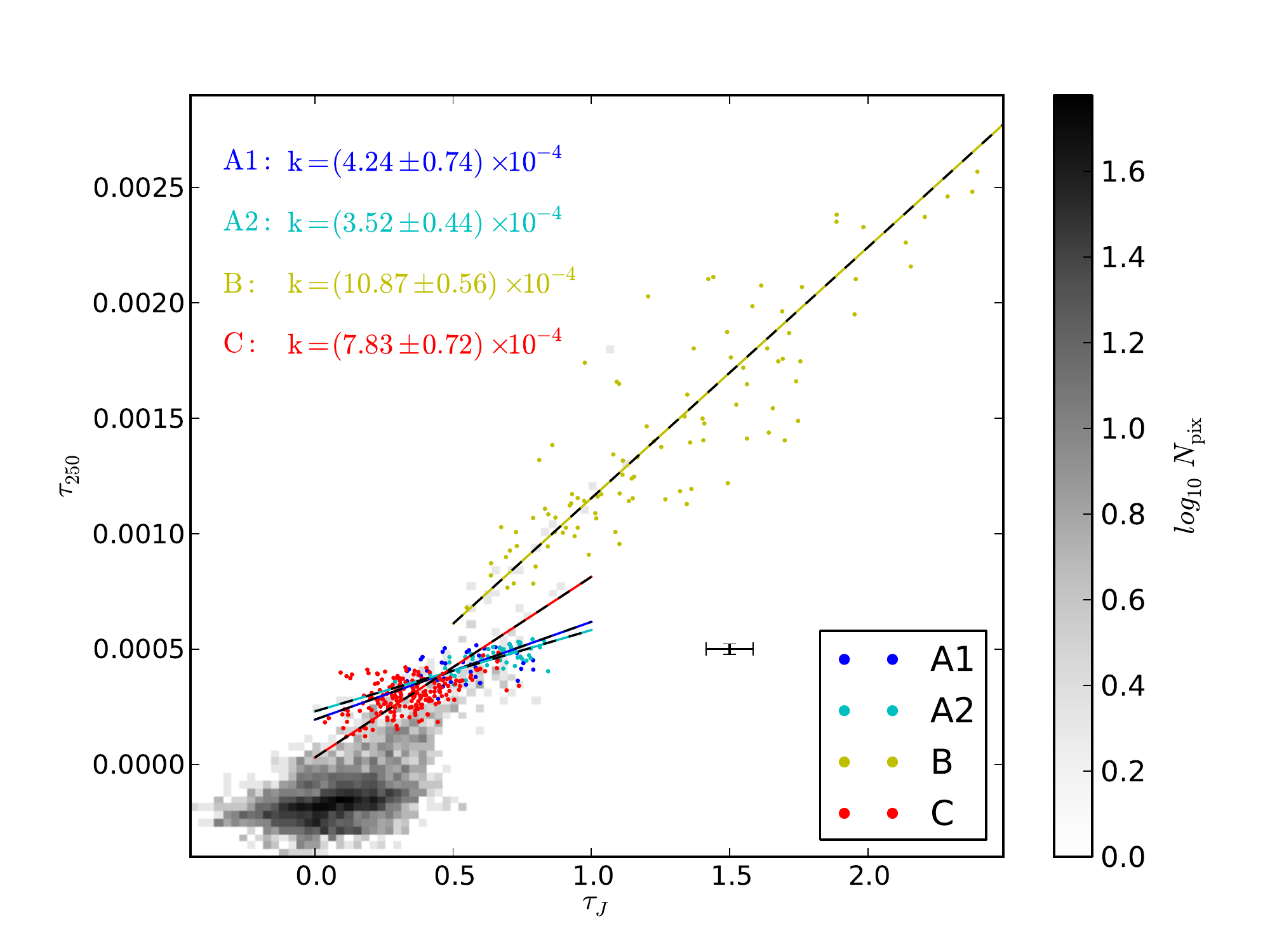}
\caption{(Top) Relation between $\tau_{250}$ and $\tau_J$. The line length shows the $\tau_J$ range used in the total least squares fitting. The slopes ($k$) of the fits are marked to the figure with the same colour as the fitted line. The used ranges are 0--2.5 (black), 0.25--2.5 (red), 0.5--2.5 (blue), 0.5--1 (green), and 1--2.5 (cyan).
(Middle) Relation between $\tau_{250}$ (with resolution 40$\arcsec$) and $\tau_J$ of individual stars, using a limit $\Delta\tau_J < 0.95$.
(Bottom) Relation between $\tau_{250}$ and $\tau_J$. The data of each region is fitted separately. The slope $k$ and the mean error bars are marked to the figures. The colour scales in the top and bottom frames correspond to the logarithmic density of points.
}
\label{fig:tau250_tauJ_hist}
\end{figure}

The relation between $\tau_{250}$ and $\tau_J$ is shown in Fig.~\ref{fig:tau250_tauJ_hist} (top frame).
We made linear fits of $\tau_{250}$ vs. $\tau_{J}$ using the total least-squares method, which takes into account the error estimates in both variables. For $\tau_{J}$ the error estimates were calculated by the NICER method. 
The uncertainties of $\tau_{250}$ were estimated as part of the MCMC calculations. Taking into account the spatial
averaging to 180$\arcsec$ resolution, the average formal uncertainty becomes $\sim 0.15\times 10^{-4}$. The
estimates are probably very optimistic because imperfections in map making result in surface brightness errors that are correlated at
scales larger than the beam size. To take this partly into account, we increased the $\tau_{250}$ error estimates
ad hoc by 50 \%. The larger error estimates also have a systematic effect of lowering the slope values by $\sim$ 10 \%, the effect rising
to $\sim$ 20 \% for regions A and C. This shows that in these regions the results are quite sensitive also to the accuracy of error
estimates. Line-of-sight (LOS) temperature variations might also cause bias and lead to an underestimation of the optical depth and of the dust opacity~\citep[e.g.,][]{Shetty2009a,Malinen2011}. The samples were taken in 1/3 beam steps.

In Fig.~\ref{fig:tau250_tauJ_hist} (top frame), we show the slopes that were obtained in different $\tau_{J}$ ranges. If the data are cut simply at fixed values of $\tau_{J}$, one can bias the slope estimates. Therefore, we first scaled the $\tau_{250}$ values so that the median uncertainties were equal along both axes, which causes the confidence regions of typical measurement points to become spherical. The data selection was then made using two delimiting lines that are perpendicular to the line fitted to all data points;  these lines cross the fitted line at the selected $\tau_{J}$ positions. Because the error distributions at this point are roughly spherical (on average, although not for every measurement point) and because the cut is made perpendicular to the expected direction of the fitted line, the selection procedure probably does not bias the
slope of the linear fit to the selected data significantly.
The fitted slope values and their error estimates are marked in the figure. The slopes differ between values $(11.2-13.0)\times 10^{-4}$ when using different $\tau_{J}$ ranges.

The correlation between $\tau_{250}$ (with resolution 40$\arcsec$) and $\tau_J$ of individual stars is shown in Fig.~\ref{fig:tau250_tauJ_hist} (middle frame). The scatter of the points below $\tau_{250} = 0.005$ shows the large errors in the $\tau_J$ values.
The $\tau_J$ values of individual stars are very noisy, but they correspond to extremely narrow beams through the cloud. The spatial averaging of $\tau_J$ values might bias the $\tau_J$ map because fewer stars are visible through regions with higher column densities.
If this were a strong effect,
the $\tau_{250}/\tau_J$ slope would be fainter when using $\tau_J$ of individual stars than when using a $\tau_J$ map at 180$\arcsec$ resolution.
Fig.~\ref{fig:tau250_tauJ_hist} (middle frame) shows no sign of such a bias. In the following analysis, we use 
$\tau_{250}/\tau_J$ data at 180$\arcsec$ resolution.

To compare with the earlier results of~\citet[][Fig. 3]{Lehtinen2007}, we derived total least-squares fits of $\tau_{250}$ vs. $\tau_{J}$ also separately for different regions. These are shown in Fig.~\ref{fig:tau250_tauJ_hist} (bottom frame), with the fitted slope values and their error estimates. The slope values show much more variation, $(3.5-10.9)\times 10^{-4}$, than when simply cutting the data based on the $\tau_{J}$ range.

The $\tau_{250}/\tau_J$ slopes can be converted into opacity or dust emission cross-section per H nucleon, $\sigma_e(250)$ using the relation $\sigma_e(250) = 1.33 \times 10^{-22} \tau_{250}/\tau_{J}$ cm$^2$/H~\citep[see the derivation in][]{Malinen2013}. 
This relation assumes the ratio of hydrogen column density and NIR colour excess $N({\rm HI} + {\rm H}_2)/E(B-V) = 5.8 \times 10^{21}$ cm$^{-2}$/mag appropriate for diffuse medium~\citep{Bohlin1978}. $E(B-V)$ is converted to $\tau_J$ using the~\citet{Cardelli1989} extinction curve. The relation is derived assuming $R_{\rm V} = 3.1$. With the assumption $R_{\rm V} = 4.0$, the derived value would be $\sim$ 40 \% higher. As the density within the cloud varies, the choice of $R_{\rm V}$ is not obvious, and the value might be even higher in the densest B region.

For $\tau_{250}/\tau_J = 10.0\times10^{-4}$, the previous relation (assuming $R_{\rm V} = 3.1$) leads to a value $\sigma_e(250) = 1.33 \times 10^{-25} {\rm cm}^2/{\rm H}$. The obtained $\sigma_e(250)$ values for regions A1, A2, B, and C are
0.56, 0.47, 1.45, and 1.04 $\times10^{-25} {\rm cm}^2/{\rm H}$.

\citet[][Table 1]{Lehtinen2007} showed their results as $\sigma_e(200)$. These can be converted into $\sigma_e(250)$ with the equation 
$\sigma_e(250) = \sigma_e(200) \times (200.0/250.0)^{\beta}$. We used value $\beta = 1.8$. Using this conversion, the $\sigma_e(250)$ values of~\citet{Lehtinen2007} are $0.27-0.47\times10^{-25}$, $0.67-0.80\times10^{-25}$, $0.40\times10^{-25}$, and $0.80-0.87\times10^{-25} {\rm cm}^2/{\rm H}$ for A (A1+A2), B, C, and L1642 interclump dust, respectively. We derived up to 2--3 times higher values than~\citet{Lehtinen2007}. This difference is mainly due to the intensity in the ISO maps, which is lower
by a factor of 2-3, see also Sect.~\ref{sect:regions}.
In~\citet{Lehtinen2007}, the change in $\sigma_e$ is $\sim$2--3 between different regions. In our data, the relative change between regions is of the same order.

\subsection{Properties of large regions}  \label{sect:regions}

We calculated the average properties of regions A1, A2, B, and C using the elliptical regions shown in Figs.~\ref{fig:areas_I_maps} (top left) and~\ref{fig:tau250_T_maps_app}. Background was subtracted from the intensity maps before deriving the optical depth and $T$ maps, using the area marked in the figures. All the maps used are at 40$\arcsec$ resolution. We used the method described in Sect.~\ref{sect:mass} to calculate the masses of the regions.

\citet{Lehtinen2004} used the value $\sigma_{200}=1.5\times10^{-25}$ cm$^2$/H for the absorption cross-section, corresponding to $\sigma_{250} \sim 1.0\times10^{-25}$ cm$^2$/H. In Section~\ref{sect:correlations} we derived the $\sigma_{250}$ values for different regions. As there are significant uncertainties in the derivation of the $\sigma_{250}$ values, and the values change between regions, we calculated masses using two different constant values, $\sigma_{250} =  1.5\times10^{-25} {\rm cm}^2/{\rm H}$ and $1.0\times10^{-25} {\rm cm}^2/{\rm H}$, in addition to the individual values for each region.

The derived properties are shown in Table~\ref{tab:result_areas}, using $\sigma_{250} =  1.5\times10^{-25} {\rm cm}^2/{\rm H}$. For comparison, we derived the masses also from the data of~\citet{Lehtinen2004} (ISO data convolved to IRAS resolution 4.5$\arcmin$) using the same method and regions. These are also shown in Table~\ref{tab:result_areas}. Using \emph{Herschel} data, we obtain $\sim$3-6\% lower temperatures and $\sim$2--3 times  higher masses than from the earlier ISO+IRAS data.
The difference in the masses is mainly caused by differences in the intensity maps, because the \emph{Herschel} map (after converting 250 $\mu$m to 200 $\mu$m) gives intensities higher
by 2--3 times  than the ISO map.

With $\sigma_{250} =  1.0\times10^{-25} {\rm cm}^2/{\rm H}$, the masses are 4.4, 4.5, 28.2, and 11.5 $M_{\sun}$ (using \emph{Herschel} data) or 
1.5, 1.7, 10.9, and 5.4 $M_{\sun}$ (using ISO+IRAS data) for the regions A1, A2, B, and C, respectively. 
\citet{Lehtinen2004} derived masses 0.8, 0.8, 13, and 2.9 $M_{\sun}$ and mean colour temperatures 16.1, 16.3, 14.2, and 16.1 K for the regions A1, A2, B, and C, respectively. The differences in the estimates based on ISO+IRAS data are mainly due to different methods of background subtraction, different $\beta$ (they had 2.0, we used 1.8) and opacity.
They calculated the background separately for each clump in the vicinity of the clump, which led to lower mass estimates. We used a common background area for the whole map.

Using the individual opacities for each region, shown in Fig.~\ref{fig:tau250_tauJ_hist} (bottom frame), we derive masses 7.8, 9.6, 19.6, and 11.1 $M_{\sun}$ for A1, A2, B, and C, respectively.
\citet{Lehtinen2004} derived virial masses for the regions using the method described in Sect.~\ref{sect:mass}.
They used values for gas temperature $T_{\rm gas}$ (10 K) and $\Delta V$ derived from the CO observations of~\citet{Russeil2003}.
They derived virial masses 10, 12, 15, and 33 $M_{\sun}$, for regions A1, A2, B, and C, respectively, concluding that only region B is close to the virial mass and therefore potentially gravitationally bound.
Our data give further evidence that region B is more massive than the approximated virial mass and confirm that it is gravitationally bound.
The other regions A1, A2, and C seem to be below their virial masses, even though we derive higher masses than previous studies
did.

\begin{table*}
\centering
\caption{Obtained values describing the regions A1, A2, B, C, and the whole cloud L1642. The columns are the region, central coordinates, FWHM size (one value for a circle or two values for an ellipse), position angle (from east to north) of the main axis of an ellipse, mean colour temperature, mass, mean optical depth $\tau_{250}$, and mean 250 $\mu$m intensity (derived from MCMC calculations). The mean colour temperature ($T_L$) and mass ($M_L$) derived from the ISO+IRAS data of~\citet{Lehtinen2004}, using our method, are also shown for comparison. The absorption cross-section is $\sigma_{250} = 1.5\times10^{-25}$ cm$^2$/H.}
\label{tab:result_areas}
\begin{tabular}{lllllllllll}
\hline \hline
Region & Central position & ~ & $FWHM$ & angle (E$\rightarrow$N) & $\langle T\rangle$ & $M$ & $\langle \tau_{250}\rangle$ & $\langle I_{250}\rangle$ & $\langle T_L\rangle$ & $M_L$\\
~      & $\alpha(2000)$ & $\delta(2000)$ & ($\arcmin$) & ($^{\circ}$) & (K) & ($M_{\sun}$) & ($10^{-4}$) & (MJy/sr) & (K) & ($M_{\sun}$)\\
\hline
A1 & 4h34m00.3s & $-$14$^{\circ}$10$'$12$''$ & 8 & ... & 16.1 & 2.9 & 4.7 & 33.6 & 17.1 & 1.0\\
A2 & 4h34m19.0s & $-$14$^{\circ}$11$'$43$''$ & 8 & ... & 15.8 & 3.0 & 4.8 & 31.3 & 16.5 & 1.1\\
B & 4h35m13.1s & $-$14$^{\circ}$15$'$43$''$ & 9$\times$14 & 25 & 14.1 & 18.9 & 15.4 & 63.8 & 14.8 & 7.3\\
C & 4h35m41.1s & $-$14$^{\circ}$33$'$41$''$ & 12$\times$21 & 93 & 16.3 & 7.7 & 3.1 & 23.0 & 16.9 & 3.6\\
L1642 & 4h35m15s & $-$14$^{\circ}$15$'$0$''$ & 60 & ... & 17.8 & 72.1 & 2.1 & 13.5 & ... & ...\\
\hline
\end{tabular}
\end{table*}

\subsection{Properties of compact sources within the main cloud of L1642}   \label{sect:point_sources}

We study the character and properties of the compact sources B-1, B-2, B-3, and B-4 found within the cloud L1642 and marked in Fig.~\ref{fig:areas_I_maps}. B-1 and B-2 are well-known binary stars, also known as L1642-1 and L1642-2~\citep{Sandell1987}, respectively. B-3 has been classified as a dwarf at a distance of $\sim$30 pc~\citep{Cruz2003}, but we study the possibility, that it is instead a YSO inside the L1642 cloud. B-4 is a cold clump seen as a temperature minimum and optical depth maximum.
The secondary component of B-2, L1642-2B, appears as a separate object in the 2MASS catalogue. There is no counterpart for this in the WISE catalogue. The secondary of B-1, L1642-1B, does not have a separate entry in the catalogues. B-4 does not appear in any of the point source catalogues.

\begin{figure}
\centering
\includegraphics[width=8.5cm]{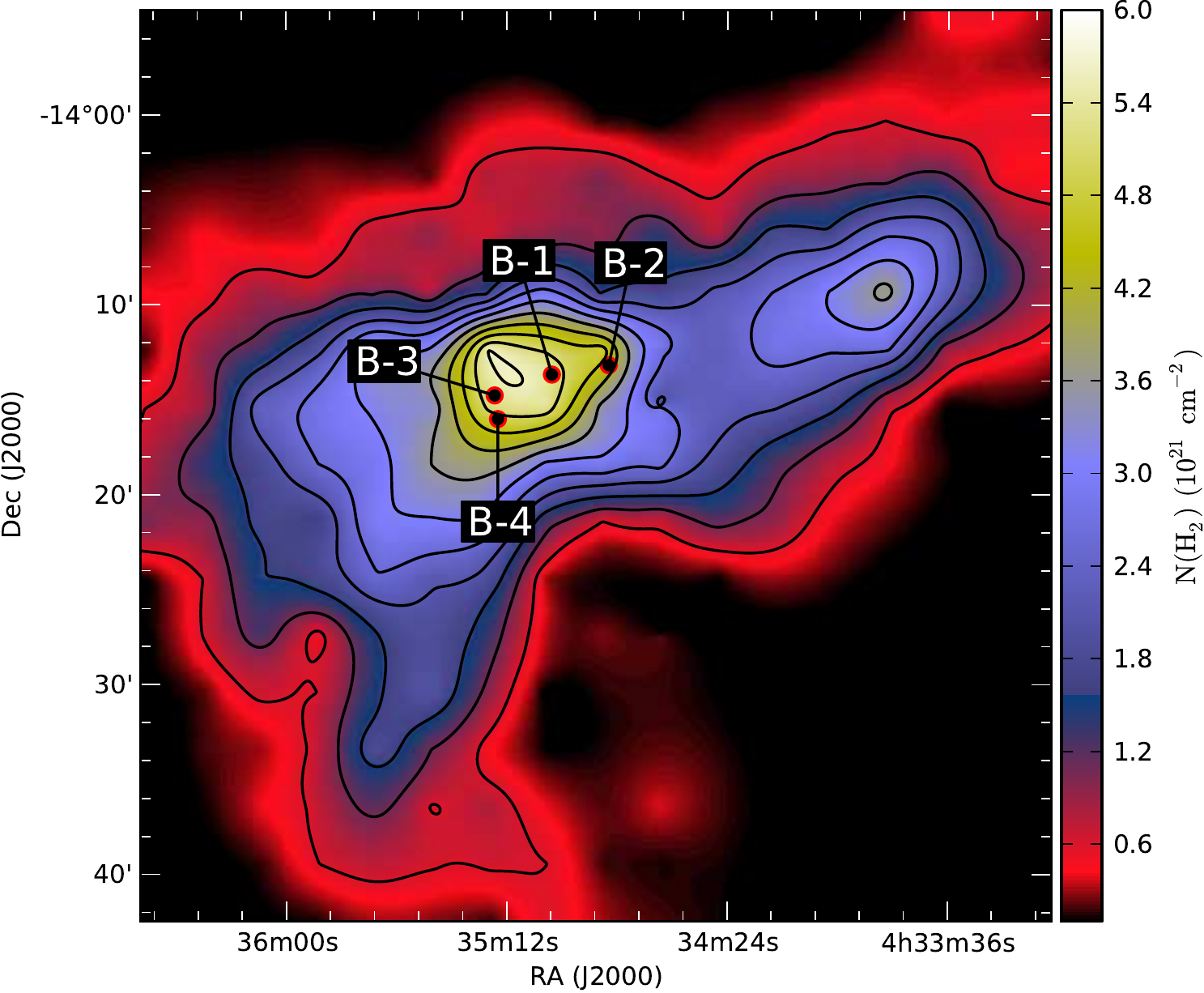}
\caption{H$_2$ column density map derived from $^{13}$CO, reproduced from the data of~\citet{Russeil2003}. Contours start from 1.0$\times10^{20}$ cm$^{-2}$ and grow by 5.0$\times10^{20}$ cm$^{-2}$ up till 6.1$\times10^{21}$ cm$^{-2}$ . The positions of the four targets are marked. The starless core, target B-4, remains invisible in the data.}
\label{fig:NCO}
\end{figure}

We calculated the \emph{Herschel} 160-500 $\mu$m fluxes for the sources using Getsources~\citep{Menshchikov2012}, as described in Sect.~\ref{sect:source_extraction}. All the maps were convolved to $\sim40\arcsec$ resolution. The most prominent clumps obtained with Getsources are marked in Fig.~\ref{fig:areas_I_maps} (bottom left frame).
We calculated the flux of the sources in \emph{Herschel} 100 $\mu$m and ASTE 1.1 mm maps (and also in other \emph{Herschel} bands for comparison) using aperture photometry with a circular aperture, with 40$\arcsec$ aperture radius and 55$\arcsec$ and 85$\arcsec$ annulus radii, as described in Sect.~\ref{sect:source_extraction}. These data are all convolved to $\sim40\arcsec$ resolution. The aperture size was chosen to be small, to avoid flux coming from the surroundings of the central source. However, also part of the flux coming from the source can therefore be lost. We use these fluxes only to make a comparison between the \emph{Herschel} and ASTE data levels, and do not use these in the other fits.
The circular apertures are shown in Fig.~\ref{fig:areas_I_maps} (bottom right frame).
The sources are also marked on the column density $N$(H$_2$) map of~\citet{Russeil2003} in Fig.~\ref{fig:NCO}.

\begin{figure}
\centering
\includegraphics[width=9cm]{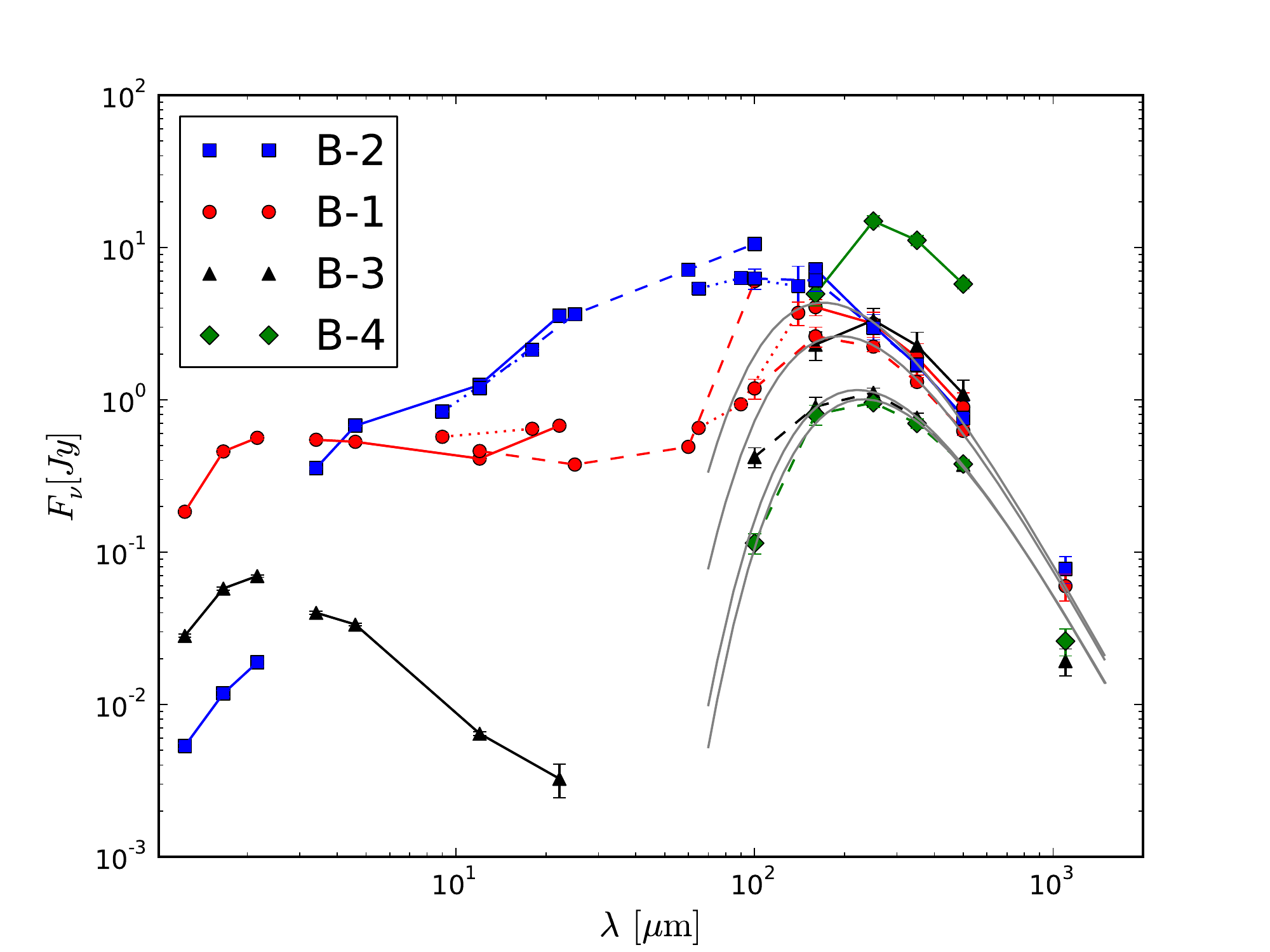}
\caption{SEDs of the sources B-1 (red circle), B-2 (blue square), B-3 (black triangle), and B-4 (green diamond). 2MASS ($J$, $H$, $Ks$), WISE (3.4, 4.6, 12.0, 22.2 $\mu$m), \emph{Herschel} (Getsources: 160, 250, 350, 500 $\mu$m) marked with solid line, AKARI (9 and 18 $\mu$m and 65, 90, 140, and 160 $\mu$m) marked with dotted line and IRAS (12, 25, 60, and 100 $\mu$m) marked with dashed line. Also fluxes obtained with aperture photometry (with 40$\arcsec$ aperture radius and 55$\arcsec$ and 85$\arcsec$ annulus radii) are shown for \emph{Herschel} 100, 160, 250, 350, 500 $\mu$m data (dashed line) and AzTEC/ASTE 1.1mm data. \emph{Herschel} and ASTE data are based on maps at 40$\arcsec$ resolution, the others are catalog values. Gray solid lines show a MBB fit, with $\beta = 1.8$, to the 160--500 $\mu$m \emph{Herschel} aperture photometry fluxes.}
\label{fig:SED_PS}
\end{figure}

SEDs of these sources are shown in Fig.~\ref{fig:SED_PS}, using values from 2MASS, WISE, IRAS and AKARI catalogs. We also show \emph{Herschel} data obtained with Getsources and by aperture photometry, and ASTE data obtained by aperture photometry, both convolved to $\sim40\arcsec$ resolution. For B-1 and B-2 ,we get rather similar values for the fluxes using these two methods. However, for B-3 and B-4, the aperture photometry gives notably smaller values, indicating the differences in the size and shape of the used area to measure flux. We tested the consistency of the ASTE fluxes by fitting a MBB model, with $\beta = 1.8$, to the 160--500 $\mu$m \emph{Herschel} fluxes obtained with aperture photometry, and compared the fit to the ASTE flux. Fig.~\ref{fig:SED_PS} shows, that the ASTE fluxes are within $\sim$ 40 \% of the values predicted by the MBB fit, and notably closer to the line in the case of source B-1.

We used the YSO classification system based on the slope of the NIR--MIR SEDs or spectral index $\alpha$, as described in Sect.~\ref{sect:class}.
Using the slope between wavelengths 3.4 and 22 $\mu$m (WISE channels 1 and 4), we derived $\alpha$ values $-$0.89, 0.23, and $-$2.34, leading to Class II, Flat spectrum, and Class III, for B-1, B-2, and B-3, respectively. Using the slope between 3.4 and 12 $\mu$m would lead to the same classes. This implies that the sources are in the order of the age, starting from the youngest, B-2, on the west side, and going towards the oldest, B-3, to east.

\begin{figure*}
\centering
\includegraphics[width=8cm]{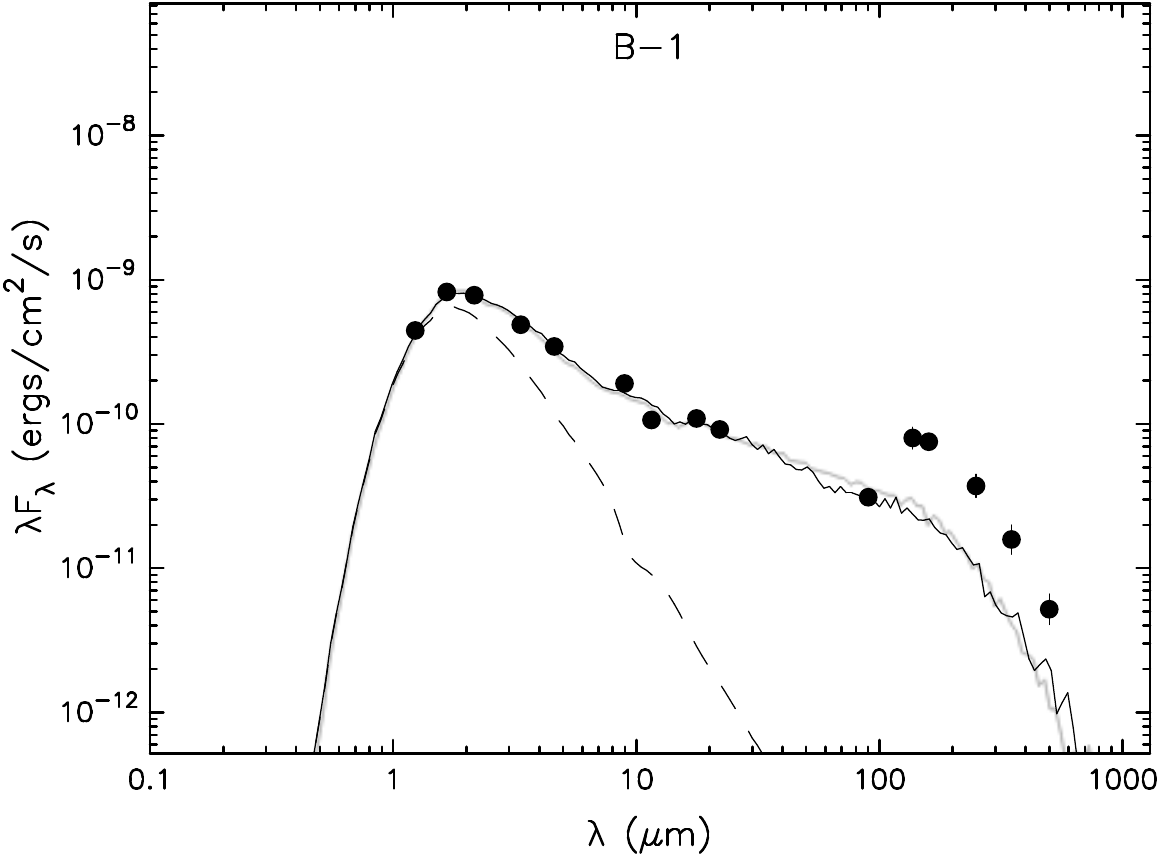}
\includegraphics[width=8cm]{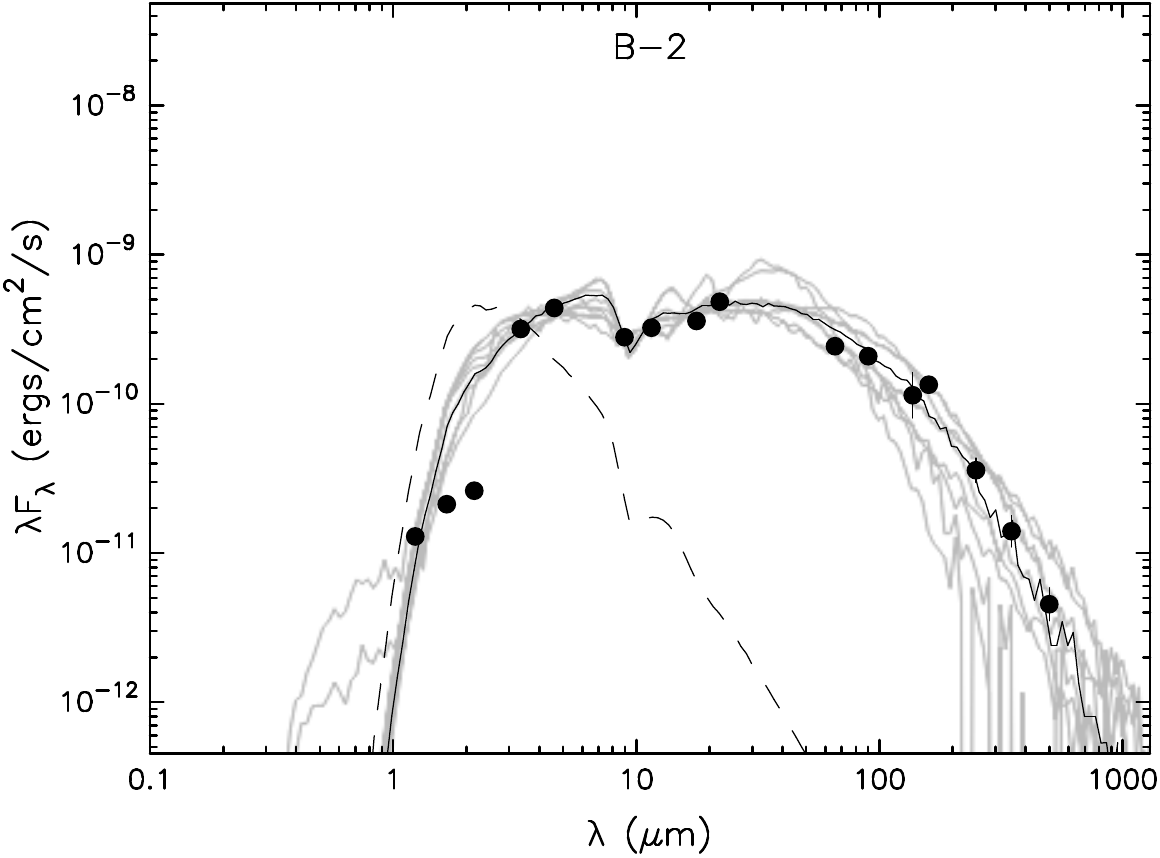}
\includegraphics[width=8cm]{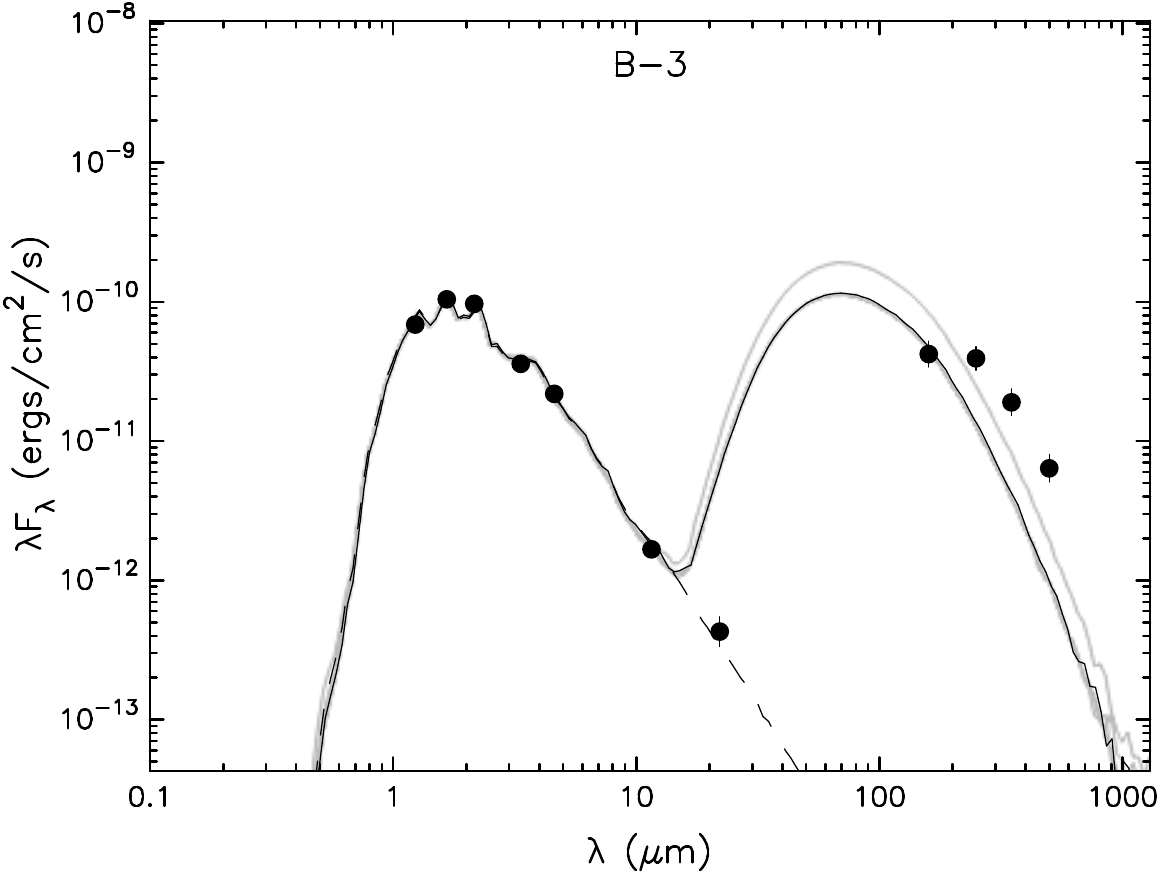}
\includegraphics[width=8cm]{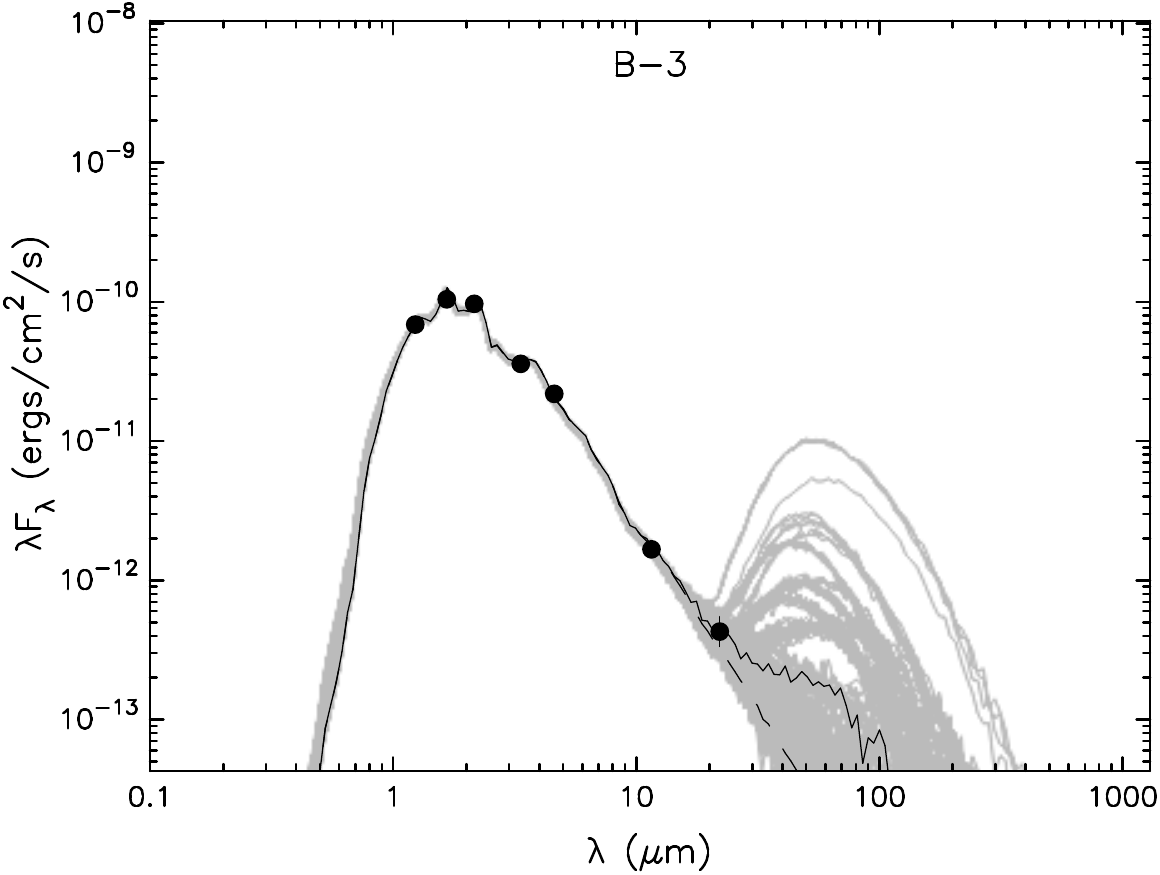}
\caption{SEDs of the sources B-1, B-2, and B-3, fitted with the YSO models of~\citet{Robitaille2007}.
Filled circles show the fitted 2MASS ($J$, $H$, $Ks$), WISE (3.4, 4.6, 12.0, 22.2$\mu$m), AKARI (only for B-1 and B-2) 
and \emph{Herschel} (Getsources: 160, 250, 350, 500$\mu$m) fluxes shown in Table~\ref{tab:Robitaille_values}. Error bars are shown, but they are often smaller than the data points. The solid black line shows the best-fitting model, and the possible grey lines show the models that fit the criteria $\chi^2 - \chi^2_{\rm{best}} < 3$, where $\chi^2_{\rm{best}}$ is the $\chi^2$ value per data point of the best-fitting model (for B-1 and B-3), or the ten best models (B-2). The dashed line shows the stellar photosphere of the best-fitting model.
(Bottom right) B-3 fitted with only NIR--MIR data and using a wide distance range (20--800 pc).
}
\label{fig:Robitaille}
\end{figure*}

We fitted the SEDs of the sources B-1, B-2, and B-3 with the online SED fitting tool of~\citet{Robitaille2007}\footnote{http://caravan.astro.wisc.edu/protostars/}. We used 2MASS, WISE, AKARI (only for B-1 and B-2), and \emph{Herschel} (Getsources) fluxes in the SEDs. The photometric data and aperture sizes used in the fitting are shown in Table~\ref{tab:Robitaille_values}. Using the different catalogue values at their original resolution should be reliable in the NIR-MIR range, and the SED fitter takes into account the different aperture sizes. %
We used a free $A_V$ range, 0.01--20 mag, in the fitting. For all the sources, we fitted the data both with a strict distance limit, with $\sim$10 \% error of the assumed 140 pc distance, leading to the range 126--154 pc, and a free distance range 20--800 pc. When the distance was free, the ten best fits to both B-1 and B-2 showed much variation of the distance estimates and often very low values ($\sim$ 40-70 pc), therefore we do not show these results.
The ten best fits of B-3 with a free distance range gave distance estimates 145--209 pc. B-2 is well fitted, except for the NIR part. Both B-1 and B-3 show a double peak in the SED, with a well-fitting NIR--MIR part and a FIR peak above 100 $\mu$m. Because the Robitaille models cannot handle FIR peaks above 100 $\mu$m (see Sect.~\ref{sect:class}),  the FIR points are poorly fitted. However, one might fit the SEDs with a model with a colder component in the envelope.
We also fitted the B-3 data using only NIR--MIR points. In that case, the ten best fits with a free distance range gave very well limited distance estimates of 174--182 pc.
The best obtained fits are shown in Fig.~\ref{fig:Robitaille}. The parameter limits for stellar mass $M$, stellar temperature $T$, and luminosity $L$ from the ten best-fitting models are shown in Table~\ref{tab:Robitaille_fits_range}. The highest obtained value for stellar $T$ and $L$ are very high compared with the best-fit values, and probably reflect the uncertainties in the fitting. B-1 and B-2 are known binary stars, and the NIR fluxes only contain the primary star, whereas the longer-wavelength fluxes with larger apertures contain both objects.

We calculated a virial mass for source B-4 using equations~\ref{eq:Mvir} and~\ref{eq:s} and a value of $T_{\rm gas} = 10$ K. \citet{Lehtinen2004} stated, based on the data of~\citet{Russeil2003}, that in region B, the mean value for the observed line width from the C$^{18}$O line is $\Delta V ({\rm C^{18}O}) = 0.52$ km s$^{-1}$. Using this value, we derived the value 1.93 $M_{\sun}$ for the virial mass of B-4, with a 40$\arcsec$ radius. We calculated the mass of B-4 as in the previous section, and obtained values of 0.83 $M_{\sun}$ with $\sigma_{250} =  1.0\times10^{-25} {\rm cm}^2/{\rm H}$ and 0.57 $M_{\sun}$ with $\sigma_{250} = 1.5\times10^{-25} {\rm cm}^2/{\rm H}$. We also tested the effect of changing the central coordinates from the coordinates given by Getsources to the $T$ minimum and to the $\tau_{250}$ maximum, which are both close by, but not exactly at the same point. These changes did not significantly affect the derived masses. The derived masses of B-4 are clearly below the derived virial mass, indicating that the object is not gravitationally bound.

In the appendix, we study the other potential YSOs located within L1642 or nearby.

\begin{table*}
\centering
\caption{Photometric data (flux $F$ and flux error $\Delta F$) from source catalogues and using Getsources for \emph{Herschel} data, and aperture sizes for the Robitaille fits of sources B-1, B-2, and B-3.}
\label{tab:Robitaille_values}
\begin{tabular}{llllllll}
\hline \hline
Band & B-1 & ~ & B-2 & ~ & B-3 & ~ & Aperture\\
~ & $F$ (mJy) & $\Delta F$ (mJy) & $F$ (mJy) & $\Delta F$ (mJy) & $F$ (mJy) & $\Delta F$ (mJy) & ($\arcsec$)\\
\hline
2MASS $J$ (1.24 $\mu$m) & 184.2 & 7.2 & 5.3 & 0.2 & 28.2 & 0.7 & 3\\
2MASS $H$ (1.66 $\mu$m) & 458.2 & 14.5 & 11.8 & 0.6 & 57.7 & 1.4 & 3\\
2MASS $K_{\rm{S}}$ (2.16 $\mu$m) & 563.3 & 9.8 & 18.9 & 1.0 & 69.7 & 1.2 & 3\\
WISE 1 (3.4 $\mu$m) & 546.9 & 16.9 & 356.4 & 4.9 & 40.0 & 0.9 & 5\\
WISE 2 (4.6 $\mu$m) & 530.3 & 10.6 & 676.3 & 8.7 & 33.5 & 0.6 & 5\\
WISE 3 (12.0 $\mu$m) & 411.4 & 4.9  & 1251.6 & 13.8 & 6.4 & 0.2 & 5\\
WISE 4 (22.2 $\mu$m) & 676.4 & 11.1 & 3569.7 & 36.0 & 3.3 & 0.8 & 5\\
IRAS 12 $\mu$m & 461.1 & ... & 1195.0 & ... & ...\tablefootmark{a} & ... & ...\\
IRAS 25 $\mu$m & 375.9 & ... & 3650.0 & ... & ... & ... & ...\\
IRAS 60 $\mu$m & 491.3 & ... & 7137.0 & ... & ... & ... & ...\\
IRAS 100 $\mu$m & 6007.0 & ... & 10570.0 & ... & ... & ... & ...\\
AKARI 9 $\mu$m & 571.9 & 28.7 & 836.1 & 2.38 & ... & ... & 60\\
AKARI 18 $\mu$m & 645.6 & 4.39 & 2132.0 & 43.9 & ... & ... & 60\\
AKARI 65 $\mu$m & ...\tablefootmark{b} & ... & 5369.0 & 356.0 & ... & ... & 120\\
AKARI 90 $\mu$m & 934.9 & 25.3 & 6308.0 & 478.0 & ... & ... & 120\\
AKARI 140 $\mu$m & 3724.0 & 648.0 & 5577.0 & 1960.0 & ... & ... & 120\\
\emph{Herschel} 160 $\mu$m & 4062.0 & 495.6 & 7208.0 & 564.1 & 2303.0 & 494.0 & 125\\
\emph{Herschel} 250 $\mu$m & 3166.0 & 591.3 & 3055.0 & 581.0 & 3335.0 & 656.0 & 125\\
\emph{Herschel} 350 $\mu$m & 1898.0 & 441.2 & 1690.0 & 411.0 & 2272.0 & 503.8 & 125\\
\emph{Herschel} 500 $\mu$m & 893.3 & 216.5 & 781.9 & 198.9 & 1093.0 & 252.5 & 125\\
\hline
\end{tabular}
\tablefoot{
\tablefoottext{a }{Source B-3 is not included in the IRAS or AKARI catalogues.}
\tablefoottext{b }{Flagged AKARI catalogue fluxes (B-1: 65 and 160~$\mu$m, B-2: 160 $\mu$m) are not shown.}
}
\end{table*}

\begin{table*}
\centering
\caption{Parameter limits from the ten best Robitaille fits for sources B-1, B-2, and B-3.}
\label{tab:Robitaille_fits_range}
\begin{tabular}{llllllllll}
\hline \hline
Source & Stellar $M$ ($M_{\sun}$) & ~ & ~ & Stellar $T$ (K) & ~ & ~ & $L$ ($L_{\sun}$) & ~ & ~ \\
~ & Min & Best & Max & Min & Best & Max & Min & Best & Max\\
\hline
B-1 & 1.61 & 1.71 & 1.94 & 4581 & 5612 & 5612 & 4.72 & 5.08 & 6.46\\
B-2 & 0.80 & 3.07 & 3.99 & 4091 & 5101 & 11582 & 7.02 & 18.7 & 121\\
B-3 & 0.10 & 0.13 & 0.13 & 2835 & 2954 & 2960 & 0.12 & 0.20 & 0.24\\
B-3 (NIR--MIR) free $D$ & 0.23 & 0.23 & 0.23 & 3168 & 3168 & 3181 & 0.35 & 0.38 & 0.39\\
\hline
\end{tabular}
\tablefoot{B-1, B-2 and B-3 are fitted with a strict distance limit (126--154 pc). B-3 (NIR--MIR) free $D$ is fitted with only NIR--MIR points, with a free distance range 20--800 pc, leading to distance estimates of 174--182 pc.}
\end{table*}

\subsection{Dust emissivity spectral index}  \label{sect:MCMC}

We studied the dust emission in the submillimetre spectral range, especially the dust emissivity spectral index $\beta$, both in the large-scale regions as defined in Sect.~\ref{sect:regions} and in the compact sources studied in Sect.~\ref{sect:point_sources}.

\begin{figure*}
\centering
\includegraphics[width=8cm]{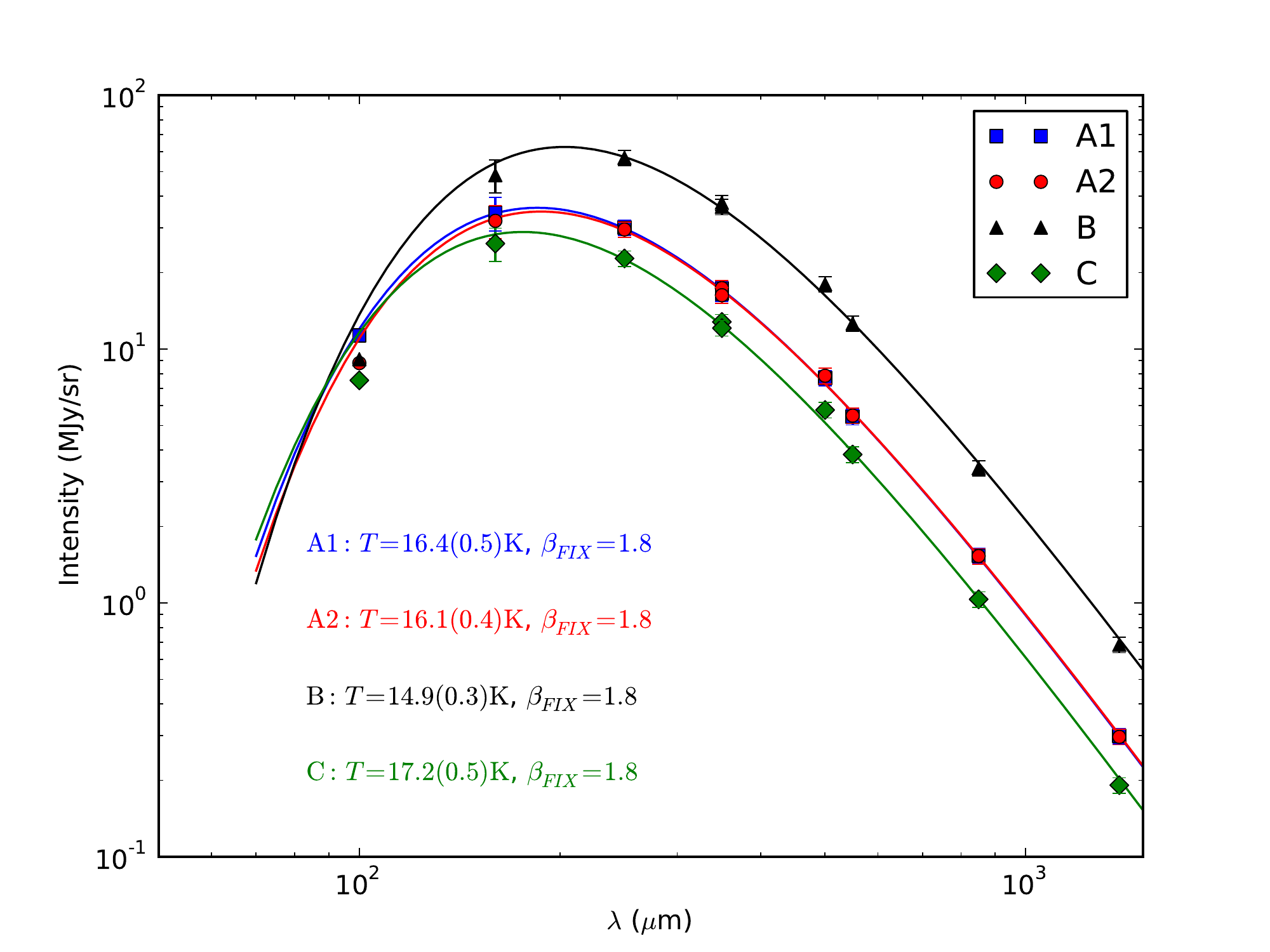}
\includegraphics[width=8cm]{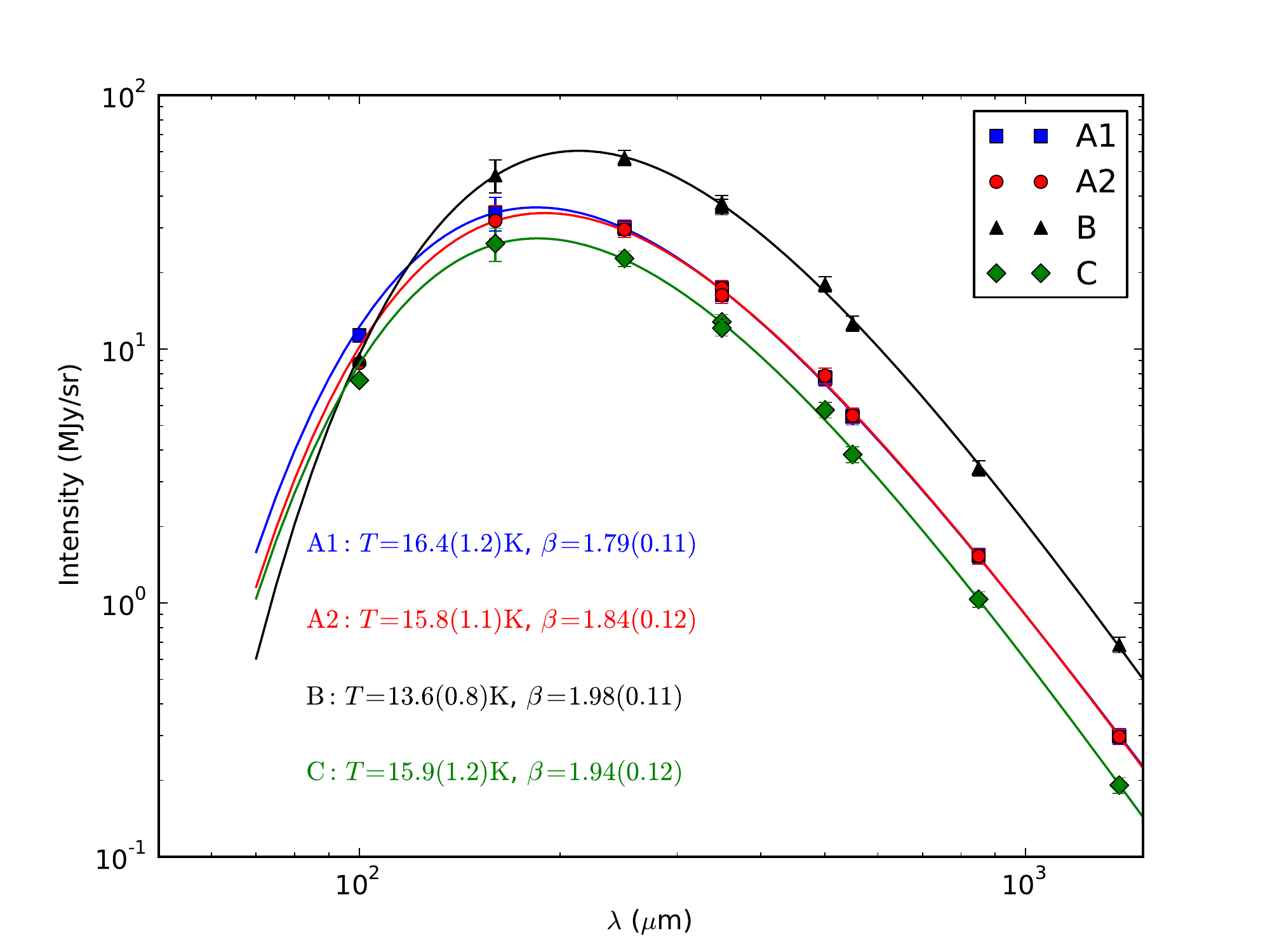}
\caption{MBB fits to the SEDs of regions A1, A2, B, and C. The values are based on background-subtracted \emph{Herschel} 100--500 $\mu$m and Planck 350--1382 $\mu$m intensity maps. 100 $\mu$m data are not included in the fitting. The fitted values of colour temperature $T$ and spectral index $\beta$ are marked in the figure. (Left) Fixed $\beta=1.8$. (Right) $\beta$ as a free parameter.}
\label{fig:KLareas_SED}
\end{figure*}

We calculated the mean intensity of the regions marked in Fig.~\ref{fig:areas_I_maps} (top left frame) using background-subtracted \emph{Herschel} 100--500 $\mu$m and Planck 350--1382 $\mu$m intensity maps. All data were convolved to the lowest resolution of the Planck maps, 5.01$\arcmin$. We fitted the SED with an MBB model, Eq.~\ref{eq:MBB}, shown in Fig.~\ref{fig:KLareas_SED}. 
Here, we used a direct least-squares fit with $\chi^2$ estimates. The flux at 100 $\mu$m can contain emission from small, transiently heated dust particles, and it is sensitive to warmer dust. The assumed effect of the small grains was corrected for in the data processing, but the data might still be biased. Therefore, the 100 $\mu$m data were not included in the fitting, but the value is shown in the figure for reference. The MBB model fits the data well, both with a fixed value of $\beta = 1.8$ and with a free $\beta$.
When $\beta$ is kept as a free parameter, the obtained $\beta$ values are $\sim$1.8 for both A1 and A2, $\sim$1.9 for C and $\sim$2.0 for the densest area, B. Temperature estimates for region B are 13.6 K with free $\beta$ and 14.9 K with a fixed value $\beta = 1.8$.

\begin{figure*}
\centering
\includegraphics[width=8cm]{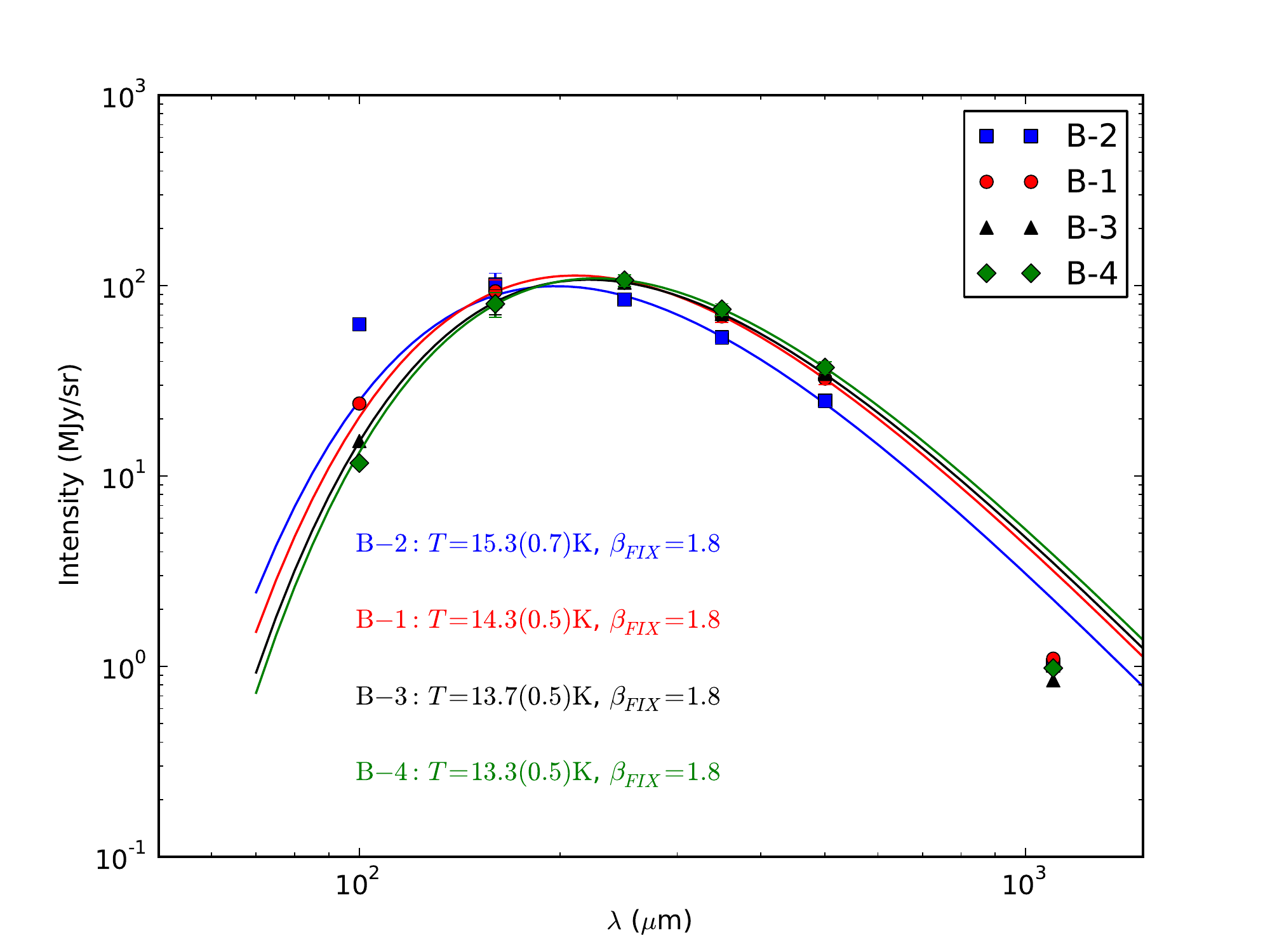}
\includegraphics[width=8cm]{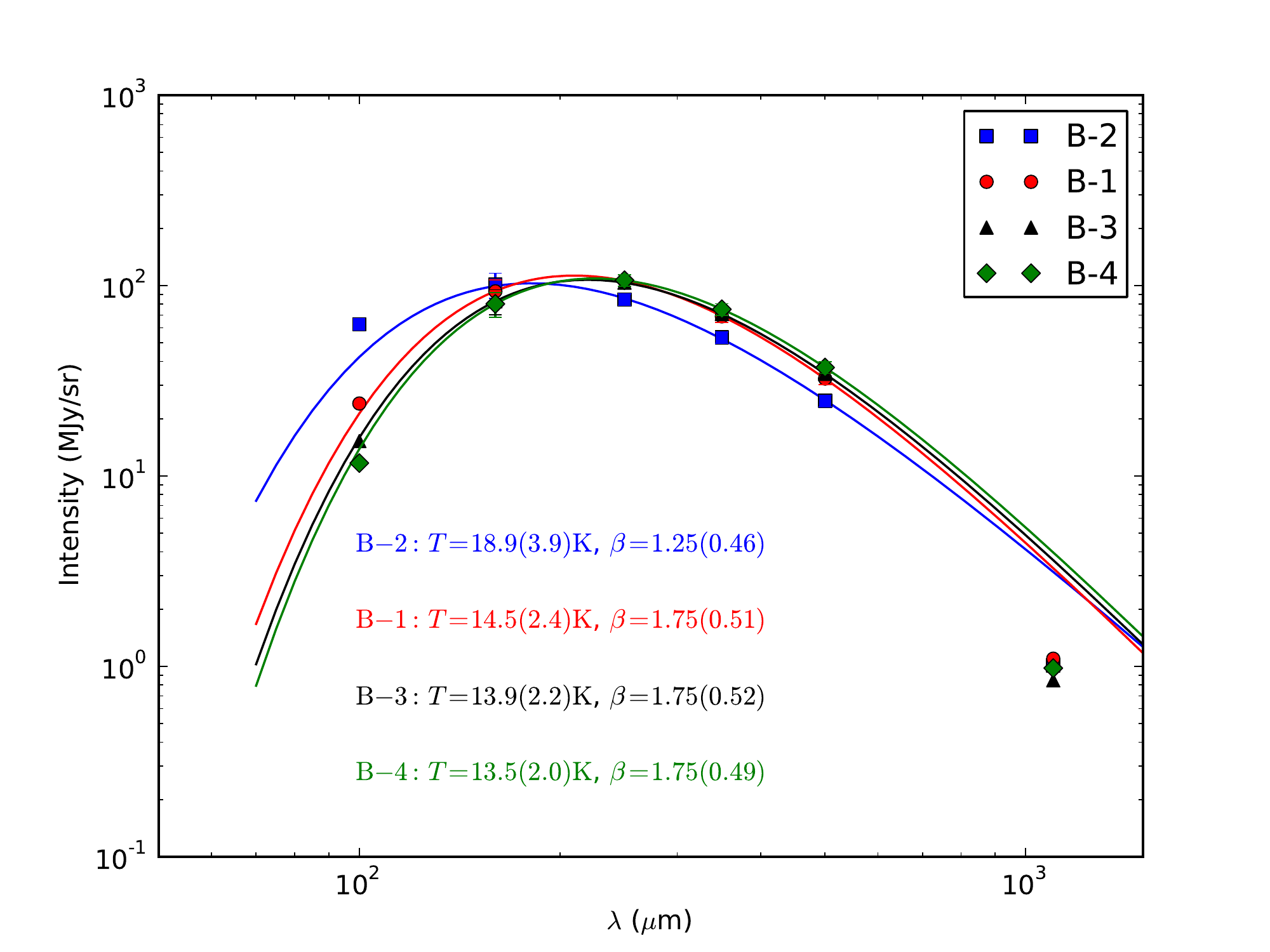}
\caption{MBB fits to the SEDs of sources B-1, B-2, B-3 and B-4, with 80$\arcsec$ aperture. The values are based on background-subtracted \emph{Herschel} 100--500 $\mu$m and ASTE 1.1 mm maps. 100 $\mu$m and ASTE data are not included in the fitting. The fitted values of colour temperature $T$ and spectral index $\beta$ are marked in the figure. 
The fits were performed both with a fixed value of $\beta=1.8$ (left frame) and with free $\beta$ (right frame).}
\label{fig:PSareas_SED}
\end{figure*}

Similar to the larger regions, we calculated the mean intensity in circular 80$\arcsec$ point-source apertures marked in Fig.~\ref{fig:areas_I_maps} (bottom-right frame) using background-subtracted \emph{Herschel} 100--500 $\mu$m and ASTE 1.1 mm intensity maps. All data were convolved to 40$\arcsec$. We fitted the obtained SED with the MBB function shown in Fig.~\ref{fig:PSareas_SED}. Extended structures are filtered in the AzTEC/ASTE data reduction, and therefore we considered the ASTE intensity only as a lower limit for 1.1 mm.  \emph{Herschel} 100 $\mu$m and ASTE data were not included in the fitting, but the values are shown in the figure for reference. Here, the MBB model also fits the \emph{Herschel} data remarkably well, except for the 100 $\mu$m data in B-2, both with fixed and free $\beta$. The ASTE data are, however, significantly lower
than those of the MBB model. The filtering of the extended emission appears to also have affected the scale of the compact sources, 80$\arcsec$. The free $\beta$ fit gives value $\beta = 1.75$ for sources B-1, B-3 and B-4. For B-2, the obtained value is much lower, $\beta = 1.25$. Here, the 100 $\mu$m values are mostly close to the line, except for B-1, in which the value is higher
than the line.

The colour temperature $T$ and emissivity spectral index $\beta$ were
also estimated with MCMC calculations,
using the same intensity values and uncertainties as in the least-squares fit
and using flat priors for the $T$ and $\beta$
parameters. Compared with the $\chi^2$ estimates, the main advantage of
the MCMC method is that it maps the full probability distribution of
the parameters~\citep{Veneziani2010, Juvela2013a}. This provides
a much better picture of the actual uncertainties than the mere
1-$\sigma$ error estimates.

\begin{figure}
\centering
\includegraphics[width=9cm]{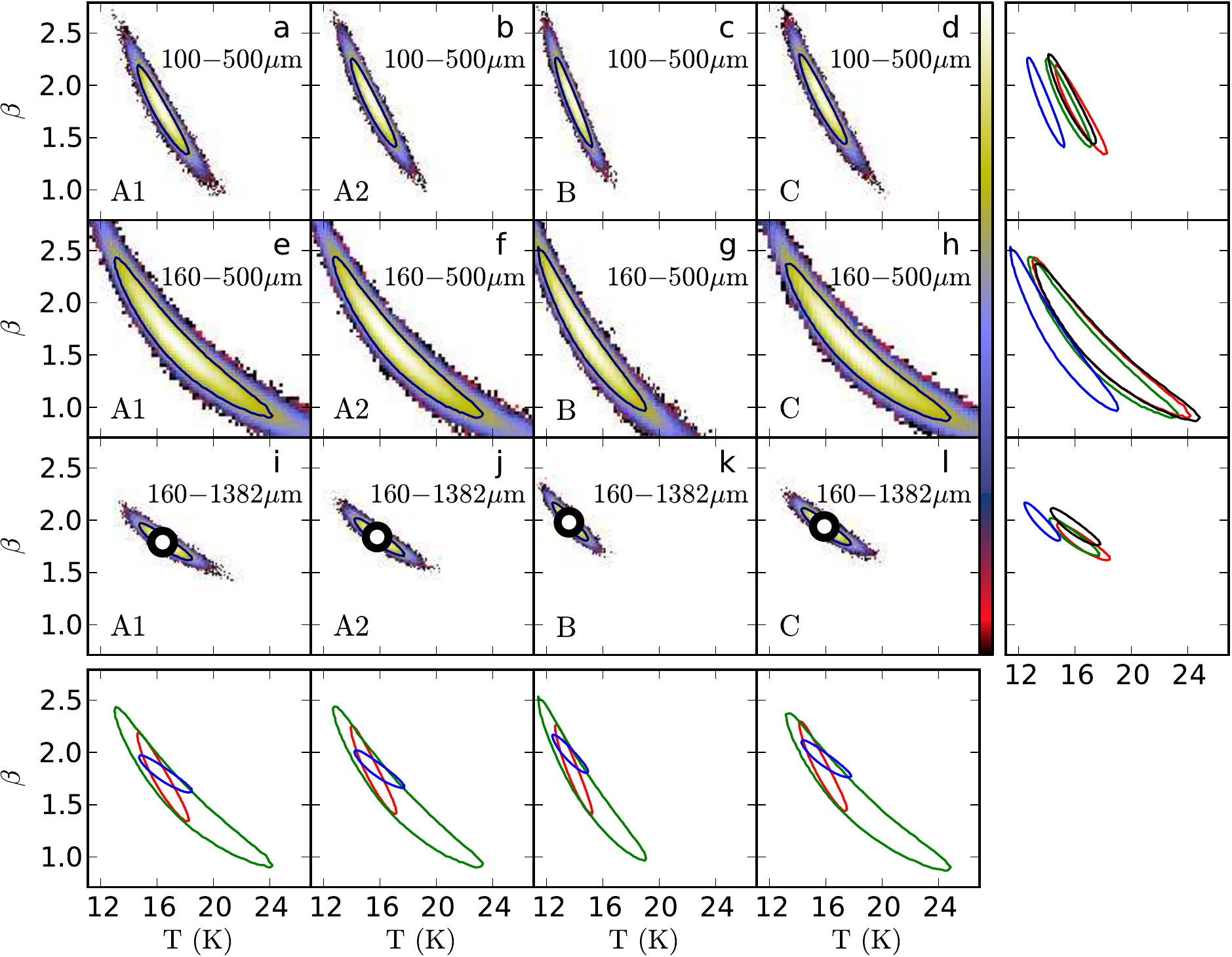}
\caption{
(T, $\beta$) probability distributions calculated for areas
A1, A2, B, and C with the MCMC method. Each column corresponds to one
of the areas and the rows to different wavelength ranges used in the
MBB fits. The contours correspond to the 90 \%
confidence regions. The maximum-likelihood solutions obtained
by combining \emph{Herschel} and Planck observations are shown as black circles.
The 90 \% contours for all cases in the same column or row are shown in the last column or row, respectively. The contour colours are red, green, blue, and black, from left to right or from top to bottom frame.
}
\label{fig:MCMC_areas}
\end{figure}

\begin{figure}
\centering
\includegraphics[width=9cm]{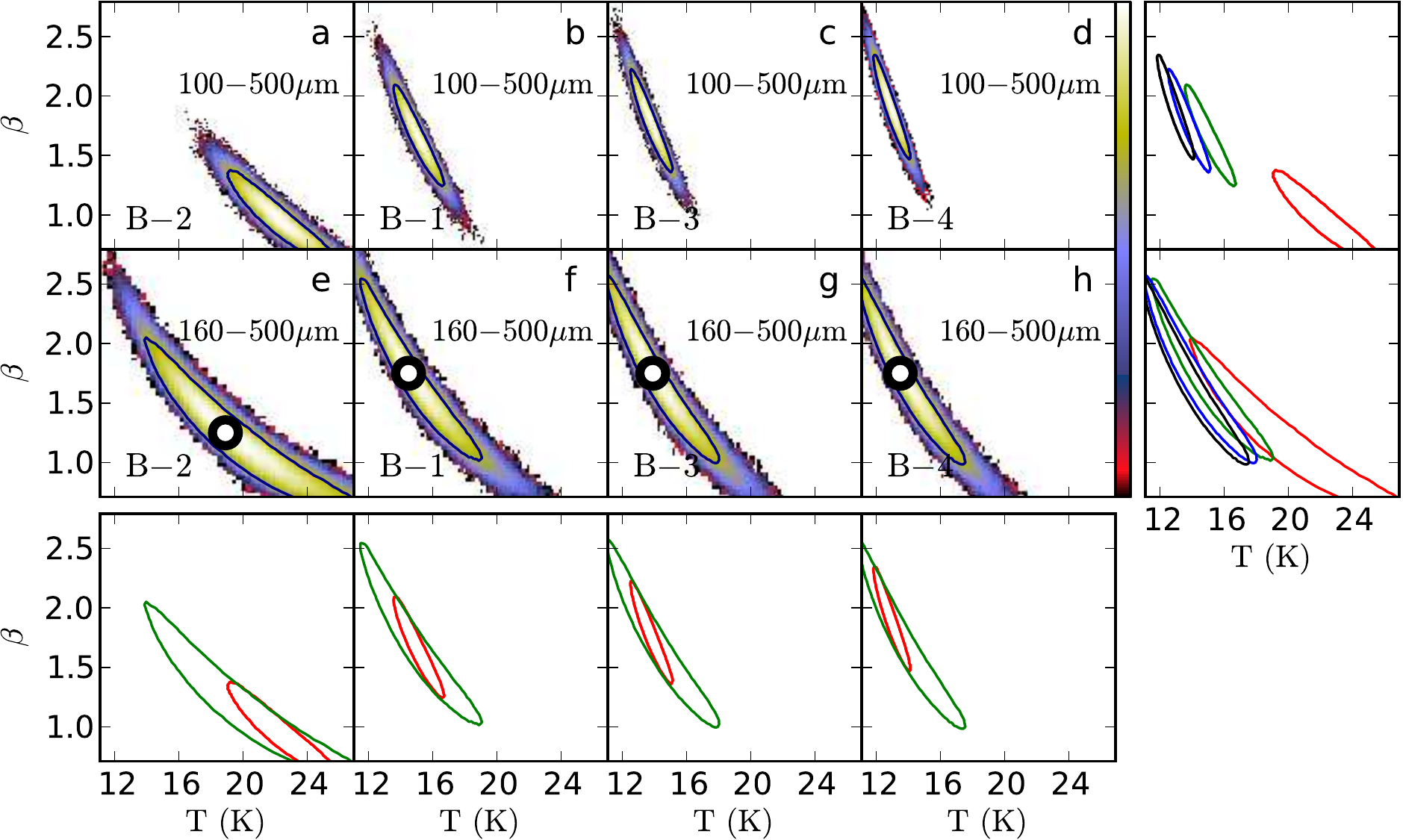}
\caption{
The ($T$, $\beta$) probability distributions calculated for
sources B-1, B-2, B-3, and B-4 using the MCMC method. The contours
correspond to 90 \% confidence regions. The black circles indicate the
maximum-likelihood solutions that were obtained with \emph{Herschel} data in
the range 160-500\,$\mu$m. The contours of all cases are shown in the last frames, similarly as in Fig.~\ref{fig:MCMC_areas}.
}
\label{fig:MCMC_ps}
\end{figure}

The obtained probability distributions are shown in
Figs.~\ref{fig:MCMC_areas} and \ref{fig:MCMC_ps} as a function of $T$ and
$\beta$. These are based on several million MCMC steps that were
registered after an initial burn-in phase, in which the initial samples were discarded.
For regions A1, A2, B, and C, the results are shown separately for the \emph{Herschel} data in the
wavelength intervals 100-500\,$\mu$m and 160-500\,$\mu$m and for the
combination of the \emph{Herschel} and Planck data 160--1380\,$\mu$m.
Figure~\ref{fig:MCMC_ps} shows the results for the compact sources
B-1, B-2, B-3, and B-4 in the wavelength ranges 100-500\,$\mu$m and
160-500\,$\mu$m.

The figures show the strong anticorrelation between the $T$ and
$\beta$ errors. The results obtained with different wavelength combinations are consistent. Including 100 $\mu$m in these fits does not produce noticeable bias, but can instead be used to constrain the values. Figure~\ref{fig:MCMC_areas} also demonstrates the
importance of the Planck data in constraining $\beta$ and, consequently,
also the temperature estimates. The temperatures in the larger regions differ slightly, and especially region B is much colder than the others. The $\beta$ values are similar in all regions, and the data do not show a clear correlation or anticorrelation between $T$ and $\beta$. Source B-2 is warmer than the other sources and has a smaller $\beta$. Using the wavelength range 100-500\,$\mu$m, the sources show a clear $T$-$\beta$ anticorrelation. However, with the range 160-500\,$\mu$m, the difference between the sources is only marginal.

\section{Discussion} \label{sect:discussion}

We have used a wide range of data from NIR to mm to study the high-latitude cloud L1642 in more detail, especially the large-scale structure, the general dust properties, and the star formation within the cloud. We discuss these main points in the following subsections.

\subsection{Structure and evolution of L1642}

L1642 forms the head of a large cometary structure whose tail extends across 5$^{\circ}$ towards the Galactic plane. 
Fig.~\ref{fig:Planck_Ak} shows that the clouds L1642, IC 2118, and LBN 991 can be connected with straight lines following the pillars leading from the clouds to the centre at approximately $\alpha_{2000} = 4^{\rm h}56^{\rm m}$, $\delta_{2000} = -12^{\circ}10'$. The Planck 817 GHz map shows the pillar to LBN 991 more clearly, but the AKARI 140 $\mu$m map reveals the warmer material and shows that the different clouds appear to be connected. L1642 is estimated to be closer than IC 2118, which is situated at $\sim$ 210 pc~\citep{Kun2001}, and it is possible that the pillar of L1642 is directed towards the viewer. These pillars can be seen already in the earlier IRAS data. These clouds have typically been studied separately, and especially LBN 991 has not been much investigated~\citep[see][]{Alcala2008}. Studying the possible connection between these clouds might help to constrain the distance estimates and birth mechanisms of the clouds.

High-resolution \emph{Herschel} data reveal new details in the structure of L1642, including small-scale elongated structures around the main cloud and striations in the diffuse dust, especially in the northern part. Whether star formation in L1642 is triggered, and if so, by which source, is still an open question. In Rivera-Ingraham et al. (in prep) we study the properties and relevance of this type of structures, the local environment, and possible triggering effects at high-latitude fields, including L1642.

We investigated the velocity structure of L1642 and its surroundings using HI data from the LAB survey~\citep{Kalberla2005}. 
The HI spectrum towards L1642 shows three radial velocity components,
broad features at $\sim$ $-$13\,km\,s$^{-1}$ and
$\sim$ +14\,km\,s$^{-1}$, and a narrow feature around 0\,km\,s$^{-1}$
that corresponds to the radial velocity of the CO lines. A detailed
study of the kinematics of the larger environment of L1642 is not
possible because we lack molecular line observations. The
filament that connects L1642 to the clouds closer to the Galactic
plane does not appear to have a clear continuous velocity gradient
in HI data,
although the analysis is complicated by the presence of several
velocity components. However, even the radial velocity of IC2118 is
only +4\,km\,s$^{-1}$ and thus the structures that appear to be
connected in the projection may be continuous structures also in real
space, in spite of the wide range of estimated distances, 140\,pc for L1642 and
210\,pc for IC2118 at the other end of the filamentary structures. 
IC2118 is likely to be inside the Orion-Eridanus bubble and is directly affected by the
massive star towards equatorial east~\citep{Kun2001}. 
In contrast, L1642 appears to be
beyond the influence of the Orion region and is mainly affected by a
flow from the west or is itself travelling west through the
ambient medium.

The velocity field within a few degrees of L1642 was analysed
by~\citet{Taylor1982} using HI observations made with the
Parkes radio telescope ($\sim 15\arcmin$ beam width). The narrow HI
line component associated with the cloud has a peak in declination one
degree south of the molecular cloud.
The HI velocity is lowest on
the eastern side of the molecular cloud but increases in all
directions. Especially towards SE both the radial velocity and the
line width increase. This is interpreted as a shearing flow that is
caused by material flowing around the cloud. The velocity gradient
5\,km\,s$^{-1}$ per degree corresponds to a radial velocity gradient
of $\sim$2\,km\,s$^{-1}$\,pc$^{-1}$. The overall kinematics is
explained by a cometary structure where the low-velocity gradients on
the western side may be a result of a geometry, where on that side of
the cloud the gas is flowing mainly in the plane of the sky. \citet{Taylor1982} 
predicted that the shearing flow will enhance the density towards
the head of the cometary cloud, increase turbulence in the outer
parts, and can lead to the formation of hydrodynamical instabilities.

This can be contrasted with the new data on the dense molecular part
of L1642. There is some compression from the west that is reflected in the general structure of the cloud (consistent with the interpretation of the HI data; see Fig.~\ref{fig:ysos}), in the striation associated especially with region A, and in the bow-shock-like structure on the western border of region B, close to source B-2 (see Fig.~\ref{fig:areas_I_maps}).
However, in the dense regions of the molecular cloud the effect of a possible slow shock remains weak.
With a steeper column density gradient on its southern edge and with striation extending towards NE, region A has the appearance of being compressed by a force from SW. In region B,
the morphology suggests a force acting more directly from the west. Region B is also associated with a filamentary structure, a possible tail, that extends directly eastward, up to the
edge of the \emph{Herschel} map, and then curves north.
The CO data indicate only a moderate level of turbulence, and there
are no signs of instabilities (e.g., classical Kelvin-Helmholtz)
within the area mapped by \emph{Herschel}. 
The difference in morphology suggests that regions A and B may be physically separate structures, aligned only in projection.
One can also conjecture that the
striation may be linked to a magnetic field geometry that is
moderating the effects of the shearing flow in the low-density
envelope. The striations look very similar to the findings of~\citet{Goldsmith2008} and~\citet{Palmeirim2013} in Taurus. These authors concluded that striations in the diffuse regions tend to follow the magnetic field direction.
This question will be clarified when Planck measurements of
dust polarisation become available and the main direction of the
magnetic field can be measured.

\subsection{Dust properties}

We derived values of the dust emission cross-section per H nucleon of $\sigma_e(250) = $ 0.56, 0.47, 1.45, and 1.04 $\times10^{-25} {\rm cm}^2/{\rm H}$ for regions A1, A2, B, and C with the assumption that $R_{\rm V} = 3.1$ (diffuse cloud).
In high-latitude, diffuse areas, the dust opacity is estimated to be $\sigma_e(250\mu \rm{m})\sim1.0\times10^{-25} {\rm cm}^2/{\rm H}$ \citep[see, e.g.,][]{Boulanger1996,Planck2011b}. In dense areas, the opacity can be 2--4 times higher~\citep[see, e.g.,][]{Juvela2011,Planck2011b,Martin2012,Roy2013}. 
\citet{Malinen2013} derived $\sigma_e(250 \mu \rm{m})$ values $\sim1.7-2.4$ $\times10^{-25} {\rm cm}^2/{\rm H}$ for a filament in Taurus at a similar distance of $\sim$ 140 pc. The newly derived opacities for the different regions of L1642 correspond to the opacities found in diffuse areas, except for region B, where the opacity is higher, at the lower limit of denser areas.

The WISE 3.4 $\mu$m map in Fig.~\ref{fig:I_zoom} shows some extended emission from the densest part of L1642. This could be caused by scattered MIR light (the core-shine phenomenon) that is associated with grain growth~\citep{Steinacker2010}.
However, this can also be explained by the PSF tails of the bright point sources and by the additional scattering (and potential dust emission) associated with the embedded sources.

MBB fits to the SED of \emph{Herschel} and Planck data give $\beta$ values 1.8--2.0 for the different regions of L1642. 
MCMC fits of the SEDs show a strong anticorrelation between $\beta$ and $T$ errors and the importance of Planck data in constraining $\beta$ and $T$ estimates. For all the regions, the recovered values of the emissivity spectral index are consistent with those found in molecular clouds at large scales~\citep{Planck2011b}, but are somewhat lower than the values reported by~\citet{Paradis2010} and~\citet{Planck2011XXIII} for regions of similarly low temperature.
We find no clear correlation or anticorrelation between $T$ and $\beta$ in the different regions. However, the compact sources show a $T$-$\beta$ anticorrelation when all \emph{Herschel} wavelengths 100-500\,$\mu$m are used. This corresponds to the results of~\citet{Malinen2011}, showing that internal heating sources produce an anticorrelation between $T$ and $\beta$.

The fits to the SED of compact sources B-1, B-3, and B-4, using only \emph{Herschel} data, give slightly lower values, $\beta \sim 1.75$. For source B-2, we obtain a much lower value, $\beta = 1.25$. Radiative transfer modelling (in Appendix~\ref{sect:models}) suggests that low $\beta$ values towards the sources can be caused by the LOS temperature variations, similarly to the results of, e.g.,~\citet{Malinen2011}. 
On the other hand, away from the sources, the optical depth estimates are probably not affected by the LOS temperature variations by more than $\sim$ 10 \%.

\subsection{Star formation in L1642}

The view of L1642 has changed from a ''small, somewhat insignificant cloud''~\citep{Taylor1982} to an interesting object in several aspects, as shown by the various studies already mentioned.
So far, B-1 and B-2 are the only known stars associated to L1642. We have questioned this view and studied the possibility of other signs of star formation within L1642.

\subsubsection{Source B-3: YSO in L1642 or a foreground dwarf} \label{sect:disc_B3}

\citet{Cruz2003} classified 2MASS J04351455-1414468 to be a young ($\sim$ 10 Myr) object. They estimated the distance to be probably within 30 pc, meaning that the object would be clearly in front of the L1642 star-forming region. Their Fig. 9 shows no clear sign of lithium (Li I) absorption at 6708\AA, which would be a sign of a probable brown dwarf (although see the caution of~\citet{Kirkpatrick2006} of the usability of this test).
\citet{Faherty2009} marked the object as a red photometric outlier, based on the red NIR colour. They also gave a very low tangential velocity for the object, $v_{tan} = 1 \pm 1$ km s$^{-1}$. They classified the object as a low surface gravity dwarf with spectral class M8.

\citet{Chauvin2012} used deep NIR $Ks$ observations to search for companions around this object. No companions were detected. They used the distance estimate of \citet{Cruz2003}, but stated that the object is close to the cloud L1642 and that the red colour is probably due to extinction caused by surrounding material and not to the chemical properties of the object itself. Based on the spectra, they assigned a tentative spectral type M6$\delta \pm 1$, where $\delta$ means a very young (Taurus-like) object, following the spectral classification system of \citet{Kirkpatrick2005} and \citet{Kirkpatrick2006}. Also, \citet{Antonova2013} treated B-3 as a low-mass dwarf in front of the L1642 cloud. \citet{Caballero2007} used the brown dwarf list of~\citet{Cruz2003} to study the proper motions of low-mass objects. The obtained low proper motion of B-3 is consistent with the one expected for the L1642 cloud.

The \emph{Herschel} data show that the point source B-3 (using 2MASS or WISE coordinates) precisely coincides with an intensity maximum in the centre of a local clump projected on L1642 on the sky, visible at all wavelengths 100--500 $\mu$m. The MBB fits to this source are very similar to the other sources known to reside in the cloud. Moreover, the models of~\citet{Robitaille2007} support the longer distance, and give a relatively good, even though not perfect, fit to the data. 
Furthermore, CO data show no other kinematic components that would suggest the presence of a second cloud on this line-of-sight.

We studied the possibility of classifying B-3 as a dwarf using WISE and 2MASS data and the method shown in~\citet{Kirkpatrick2011}.
WISE (Vega) magnitudes from PSF fitting are 9.711, 9.268, 9.136, and 8.514.
The derived colours are W1-W2 = 0.443, W2-W3 = 0.132, and W3-W4 = 0.623. 2MASS $J$, $H$, and $Ks$ magnitudes are 11.879, 10.622, and 9.951, respectively.
These values correspond to basically stellar colours with some IR excess, meaning that they are consistent with a YSO.
The colours are, however, not inconsistent with a mid-L dwarf. We tested this object with the brown dwarf search criteria of~\citet{Kirkpatrick2011}. B-3 would fail the first criteria for the coldest brown dwarfs with type T5 or bigger (W1-W2 $>$ 1.5). It would pass the colour criteria for bright, nearby L and T dwarfs (W1-W2 $>$ 0.4). However, as the object is clearly seen in 2MASS (and also faintly in DSS optical data), it would be dropped from the brown dwarf list. We also compared the colours of B-3 to the colour-colour and colour vs. spectral type plots of~\citet[][Figures 1--10]{Kirkpatrick2011}. Based on these comparisons, we conclude that B-3 colours are marginally too red in W1-W2 and clearly too red in $J$-$H$ to match~\citet{Faherty2009} spectroscopic classification of M8. However, the colours of B-3 fit the colours of a mid-L type dwarf
quite well, except in W2-W3, where it is slightly too blue.

\citet{Koenig2012} have adapted the photometric YSO classification method designed for \emph{Spitzer}~\citep[e.g.,][]{Gutermuth2008,Gutermuth2009} to WISE bands. Their method is designed to classify only YSOs with excess emission at NIR--MIR wavelengths, that is, Class I--II objects. This system cannot classify diskless Class III objects. However,~\citet{Koenig2012} also show the previously found diskless YSOs in their colour-colour plots. We compared the WISE colours of B-3 with these plots (their Figs. A.7--11) and conclude that B-3 fits the group of diskless YSOs well.

The perfect positional coincidence of the NIR/MIR and FIR source positions strongly suggests that the point source in B-3 is located within the cloud L1642. The fact that the object can be clearly seen in the 100 $\mu$m map, showing mainly dust emission, similarly to only YSOs inside the cloud and a single galaxy, suggests that the object is surrounded by dust. The 160 $\mu$m map already shows a clearly extended object, with an internal, warming source. These data support the view that B-3 is a YSO situated within the cloud L1642.

\subsubsection{Chain of star formation}

The point sources B-1 and B-2 are well known binary T Tauri stars. We discussed the characteristics of B-3 in more detail in Sect.~\ref{sect:disc_B3}. Our results show that B-3 is more likely to be a YSO within the cloud L1642 than a dwarf in front of the cloud, contrary to previous studies. 
The SEDs clearly show the difference between the evolutionary stages of these objects. Using the NIR-MIR slope, we derived YSO classes II, a flat spectrum, and III for B-1, B-2, and B-3, respectively. This indicates that the objects are situated in the age order from the youngest B-1 in west to the oldest B-3 in the east.

B-4 can be seen as a clear temperature minimum and optical depth maximum in the maps, south of the chain of the other sources. However, we derived a mass for this object using an 80$\arcsec$ aperture, and concluded that the obtained mass is smaller than the virial mass. 
It appears that B-4 is a cold clump, 
with some substructure visible, for example, in the PACS data. 
Based on the estimated virial masses, the clump is not gravitationally bound, at least within the used aperture. However, the potential pre-stellar nature of this object needs to be investigated in
more detail.

The mass estimates from the Robitaille models for sources B-1, B-2, and B-3 (1.71, 3.07, and 0.13 $M_{\sun}$, respectively, see Table~\ref{tab:Robitaille_fits_range})
divided by the cloud mass of 72.1 $M_{\sun}$ (see Table~\ref{tab:result_areas}) gives a star formation efficiency of $\sim$~7~\% for
L1642.

\section{Conclusions} \label{sect:conclusions}

We have studied the properties of L1642, one of the two high galactic latitude ($|b| > 30^{\circ}$) clouds with active star-formation.
The high latitude ($-$36.4$^{\circ}$), indicating no foreground contamination, the small distance ($\sim$140 pc), the modest column density ($A_V < 12^{\rm m}$), and the spatially distinct sequence of star-forming cores make L1642 one of the best targets for studying the initial stages of low-mass star formation.
We presented in detail high-resolution far-infrared and submillimetre observations with \emph{Herschel} and AKARI satellites and millimetre observations with AzTEC/ASTE telescope and combined them with archive data from NIR (2MASS) and MIR (WISE) to millimetre (Planck). We studied both larger regions (A1, A2, B, C) and compact sources (both previously known binary T Tauri stars and potential new objects) within the main cloud. Our main conclusions are the
following:

\begin{itemize}

\item The high-resolution \emph{Herschel} observations, combined with other data, show an evolutionary sequence from a cold clump (B-4) to young stellar objects of different spectral classes: B-1 (L1642-1, Class II), B-2 (L1642-2, Flat spectrum), and B-3 (Class III).

\item Based on \emph{Herschel} FIR--submm data, the point source B-3 (2MASS J04351455-1414468) appears to be a YSO forming inside the L1642 cloud, instead of a foreground brown dwarf, contrary to previous classification.

\item \emph{Herschel} data reveal striation in the diffuse dust around the cloud L1642, especially in the northern part. Region A shows striation extending towards the NE and has a correspondingly steeper column density gradient on its southern side. Region B has the appearance of being compressed from the west and has a filamentary tail extending eastward. The differences suggest that these may be spatially distinct structures, aligned only in projection.

\item Both \emph{Herschel} FIR--submm and AzTEC/ASTE mm data show an elongated structure north-east of the binary star B-2 (L1642-2), which was not visible in earlier observations. The structure is aligned with the general striation pattern, but it has a significantly higher column density.

\item We derived values of the dust emission cross-section per H nucleon of $\sigma_e(250) = $ 0.56, 0.47, 1.45, and 1.04 $\times10^{-25} {\rm cm}^2/{\rm H}$ for the different regions.

\item Modified black-body fits to the SED of \emph{Herschel} and Planck data give emissivity spectral index $\beta$ values 1.8--2.0 for the different regions of L1642.

\item Markov chain Monte Carlo fits of the SEDs show a strong anticorrelation between $\beta$ and $T$ errors and the importance of Planck data in constraining $\beta$ and $T$ estimates.

\item The compact sources show a $T$-$\beta$ anticorrelation, but on larger scale the data show no clear correlation or anticorrelation.

\item Radiative transfer modelling suggests that low beta values towards the sources might be caused to a large extent by the LOS temperature variations. On the other hand, the optical depth estimates are expected to be quite accurate (to within $\sim$ 10 \%) away from the sources.

\end{itemize}

\acknowledgements
We thank the anonymous referee for the detailed comments which improved the paper. We thank Kimmo Lehtinen and Delphine Russeil for providing us data from their earlier studies. JMa and MJ acknowledge the support of the Academy of Finland Grants No. 250741 and 127015. 
The development of Planck has been supported by: ESA; CNES and CNRS/INSU-IN2P3-INP (France); ASI, CNR, and INAF (Italy); NASA and DoE (USA); STFC and UKSA (UK); CSIC, MICINN and JA (Spain); Tekes, AoF and CSC (Finland); DLR and MPG (Germany); CSA (Canada); DTU Space (Denmark); SER/SSO (Switzerland); RCN (Norway); SFI (Ireland); FCT/MCTES (Portugal); and The development of Planck has been supported by: ESA; CNES and CNRS/INSU-IN2P3-INP (France); ASI, CNR, and INAF (Italy); NASA and DoE (USA); STFC and UKSA (UK); CSIC, MICINN and JA (Spain); Tekes, AoF and CSC (Finland); DLR and MPG (Germany); CSA (Canada); DTU Space (Denmark); SER/SSO (Switzerland); RCN (Norway); SFI (Ireland); FCT/MCTES (Portugal); and PRACE (EU).
This research is based on observations with AKARI, a JAXA project with the participation of ESA.
We also acknowledge the ASTE staffs both for operating ASTE and for helping us with the data reduction. Observations with ASTE were (in part) carried out remotely from Japan by using
NTT's GEMnet2 and its partner R\&E (Research and Education) networks, which are based on the AccessNova collaboration of the University of Chile, NTT Laboratories, and the National
Astronomical Observatory of Japan.
This publication makes use of data products from the Wide-field Infrared Survey Explorer, which is a joint project of the University of California, Los Angeles, and the Jet Propulsion Laboratory/California Institute of Technology, funded by the National Aeronautics and Space Administration.
This research has made use of the NASA/IPAC Infrared Science Archive, which is operated by the Jet Propulsion Laboratory, California Institute of Technology, under contract with the National Aeronautics and Space Administration.
This research has made use of the NASA/IPAC Extragalactic Database (NED) which is operated by the Jet Propulsion Laboratory, California Institute of Technology, under contract with the National Aeronautics and Space Administration. 

\bibliographystyle{aa}
\bibliography{biblio_j3}

\begin{appendix}

\section{Additional maps of L1642}

\begin{figure*}
\centering
\includegraphics[width=8.5cm]{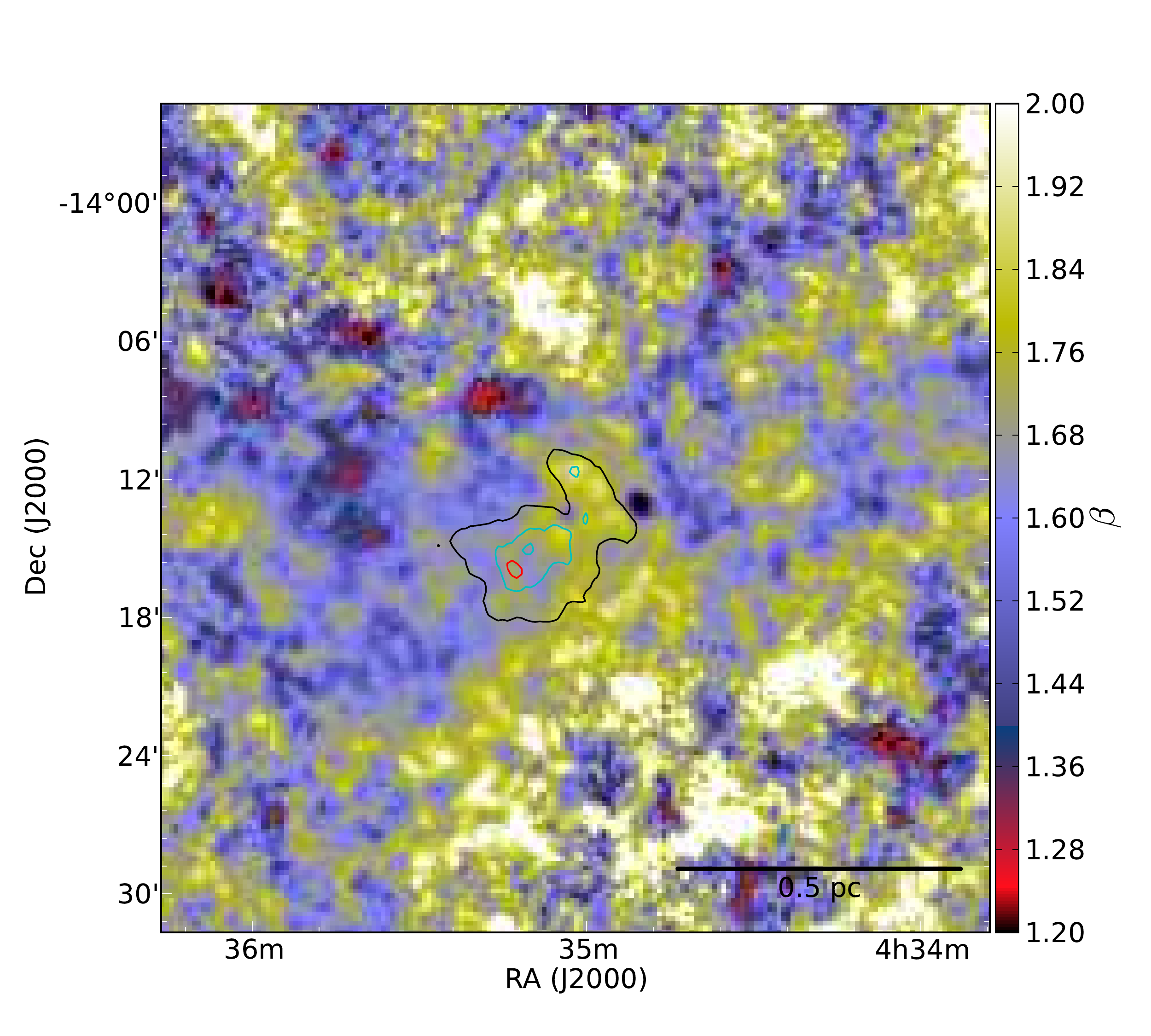}
\includegraphics[width=8.5cm]{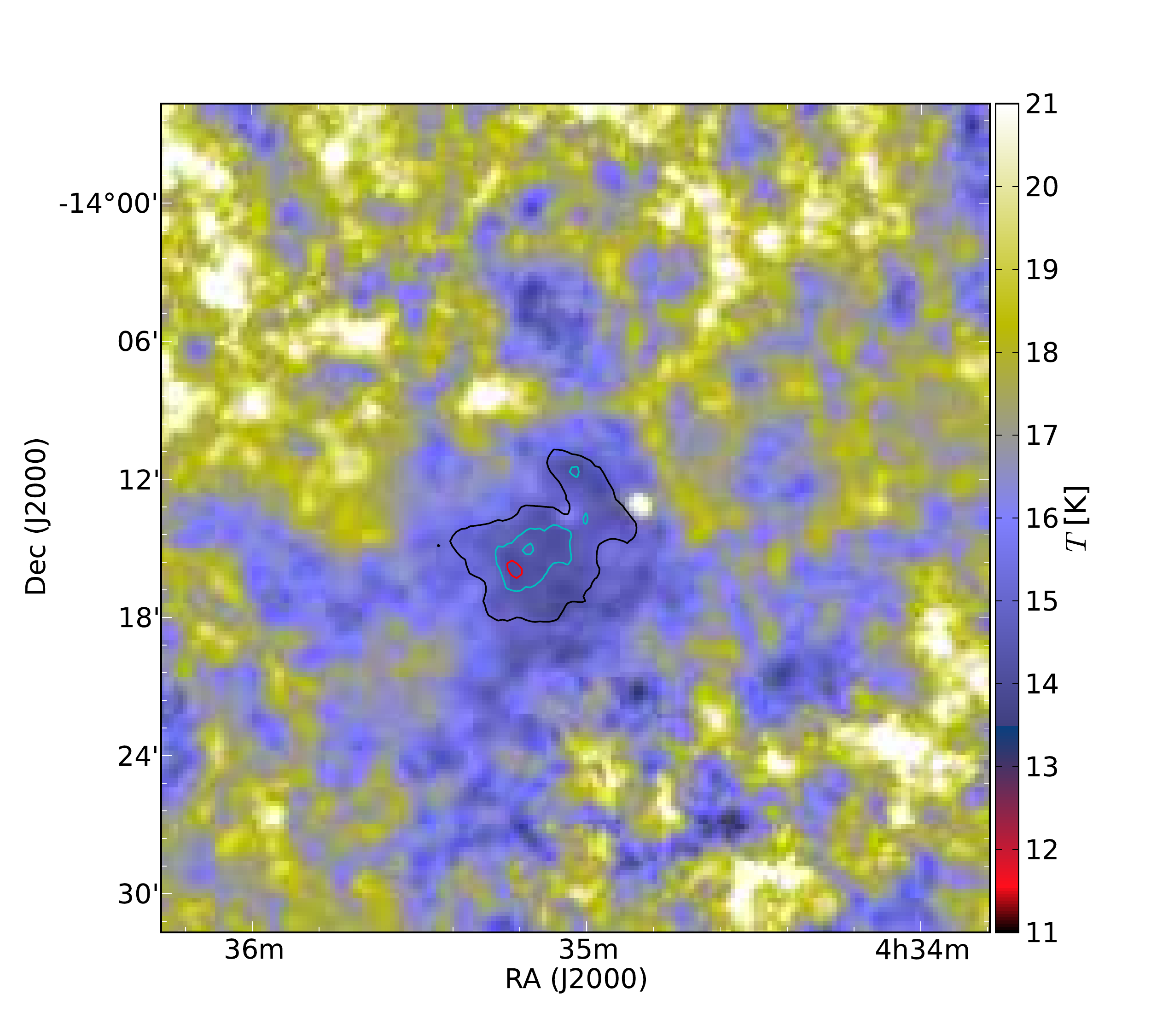}
\caption{Maps of $\beta$ (left) and $T$ (right) derived by MCMC fitting with free $\beta$, using \emph{Herschel} 160--500 $\mu$m maps. Background is not subtracted. Contours of the $\tau_{250}$ map are drawn on both maps at levels 0.0035, 0.0029, and 0.0019.}
\label{fig:tau250_T_maps_freeb}
\end{figure*}

\begin{figure*}
\centering
\includegraphics[width=8.5cm]{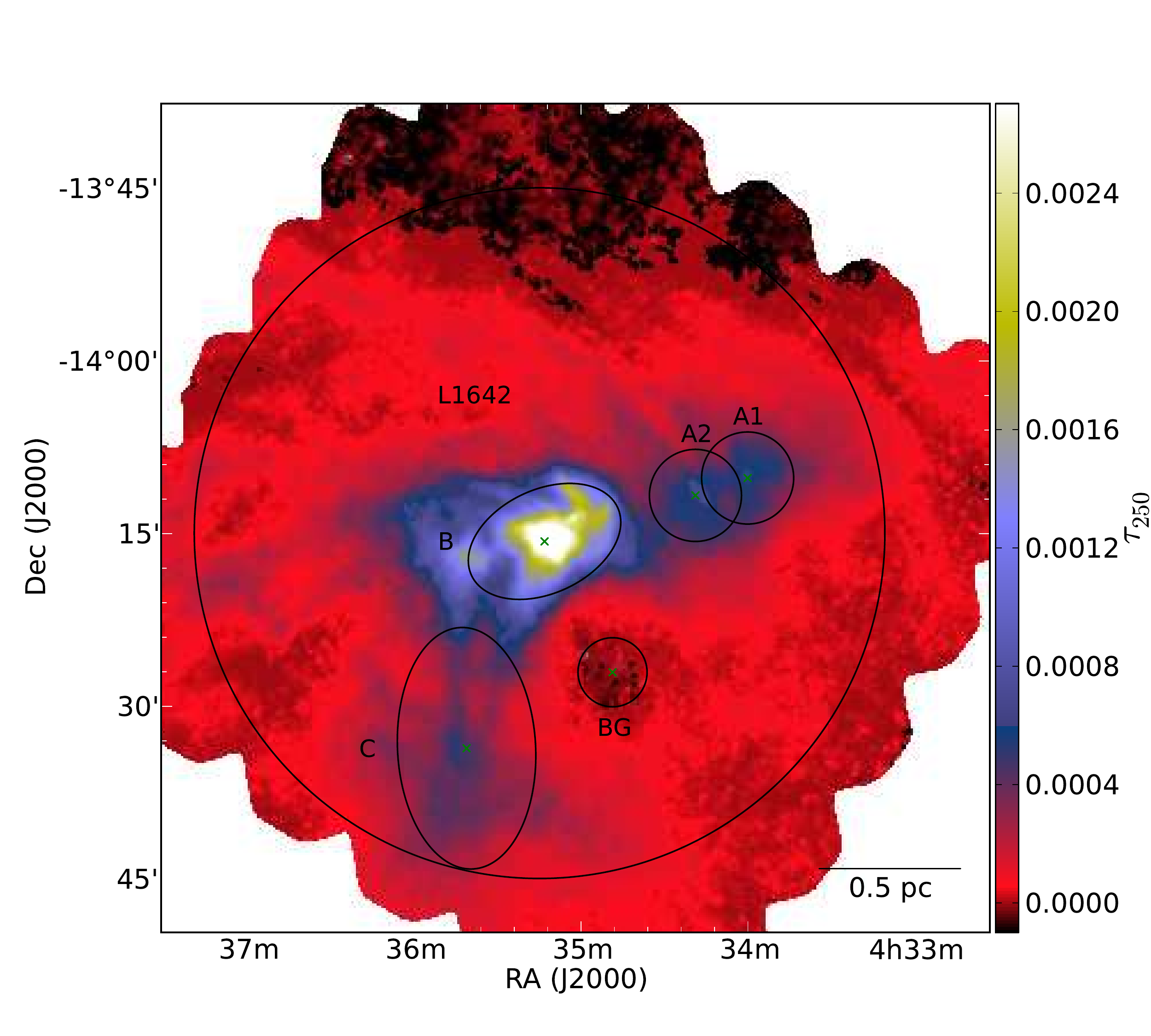}
\includegraphics[width=8.5cm]{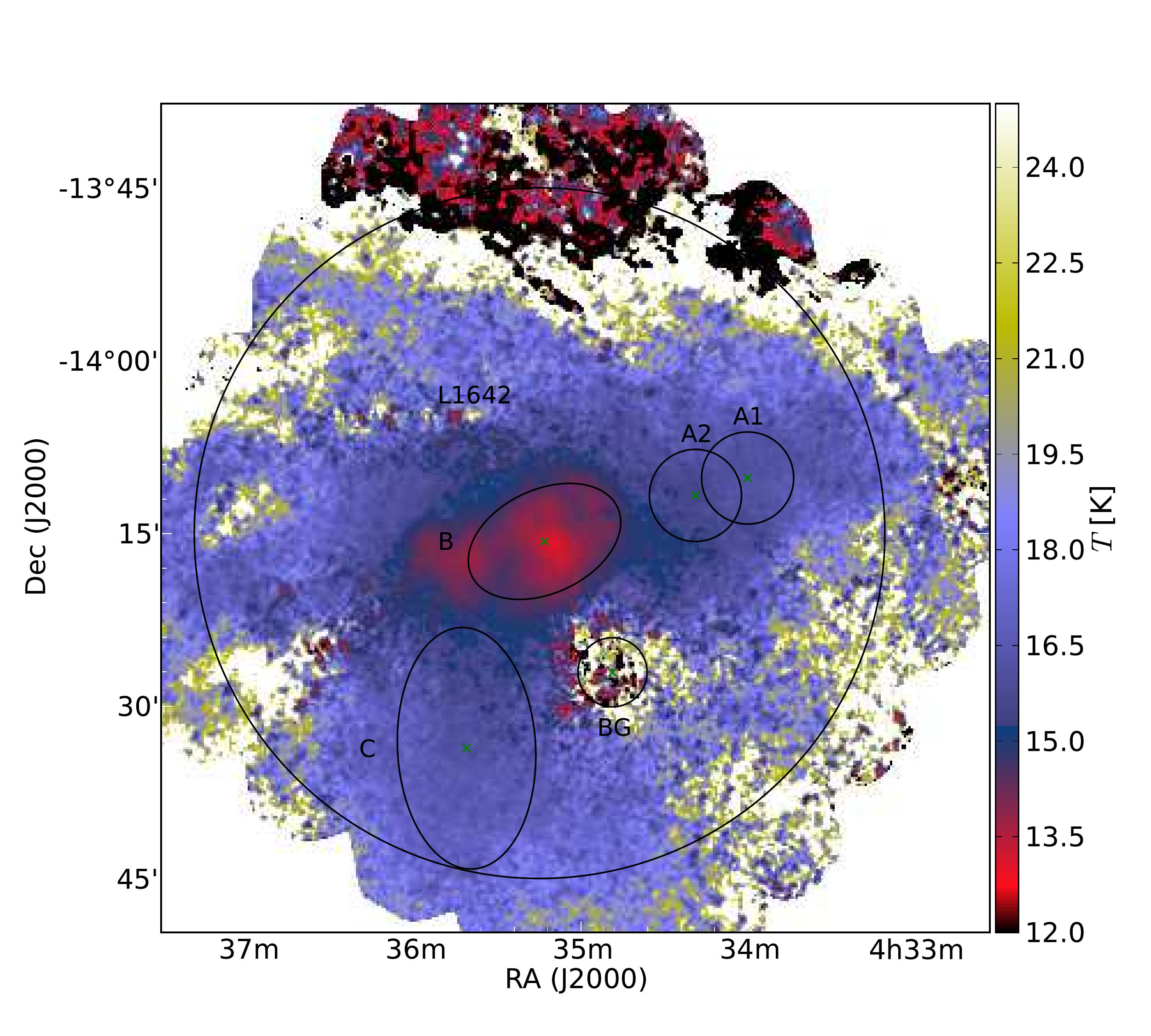}
\caption{Maps derived by MCMC fitting with a constant $\beta = 1.8$, using \emph{Herschel} 250--500 $\mu$m maps at 40$\arcsec$ resolution.
The maps are based on background-subtracted intensity maps, using the area marked BG as background. Areas of the clumps A1, A2, B, and C, and the main part of cloud (marked L1642) used in the analysis. The maps are $\tau_{250}$ (left) and $T$ (right).
}
\label{fig:tau250_T_maps_app}
\end{figure*}

\begin{figure*}
\centering
\includegraphics[width=8.5cm]{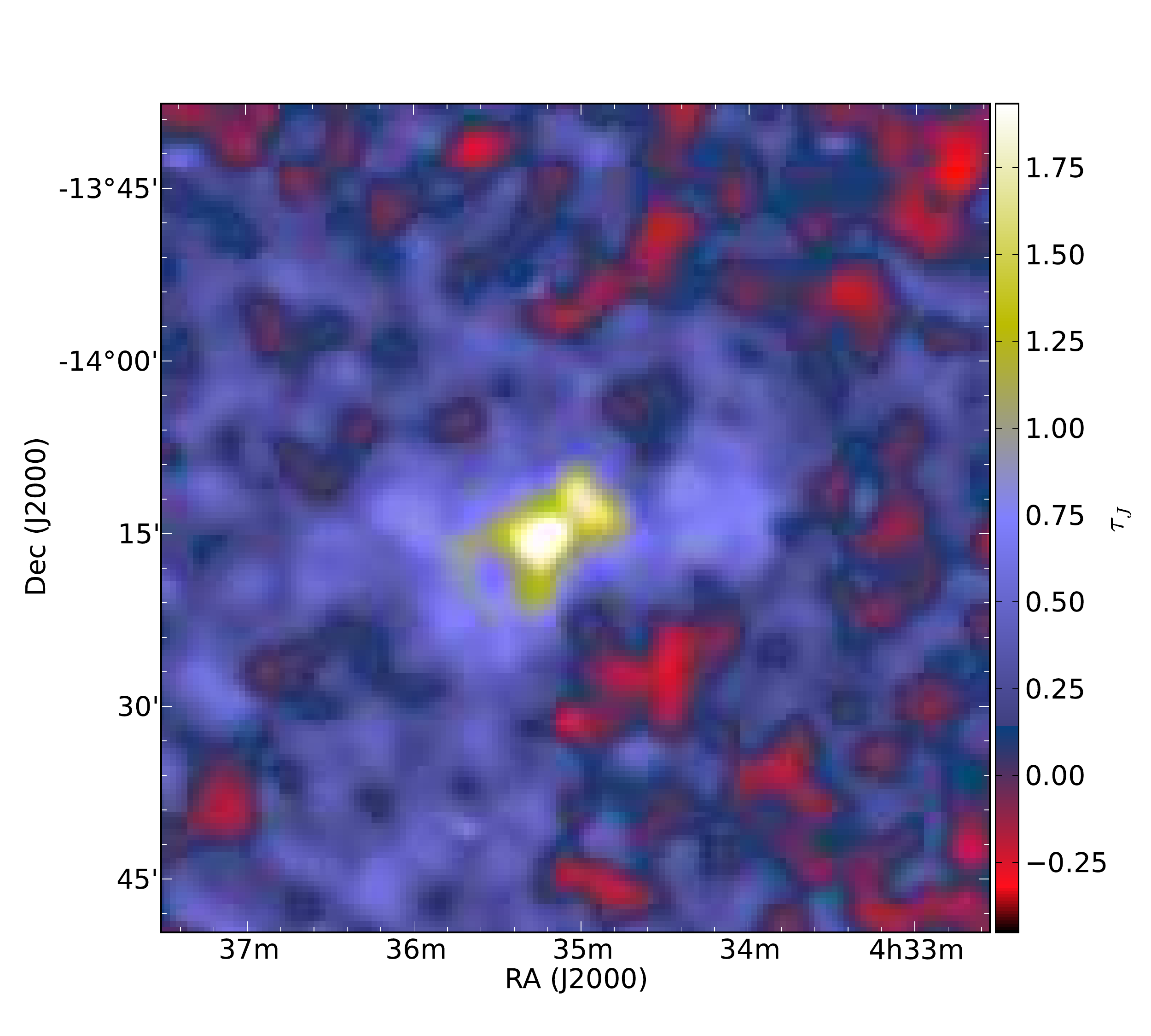}
\includegraphics[width=8.5cm]{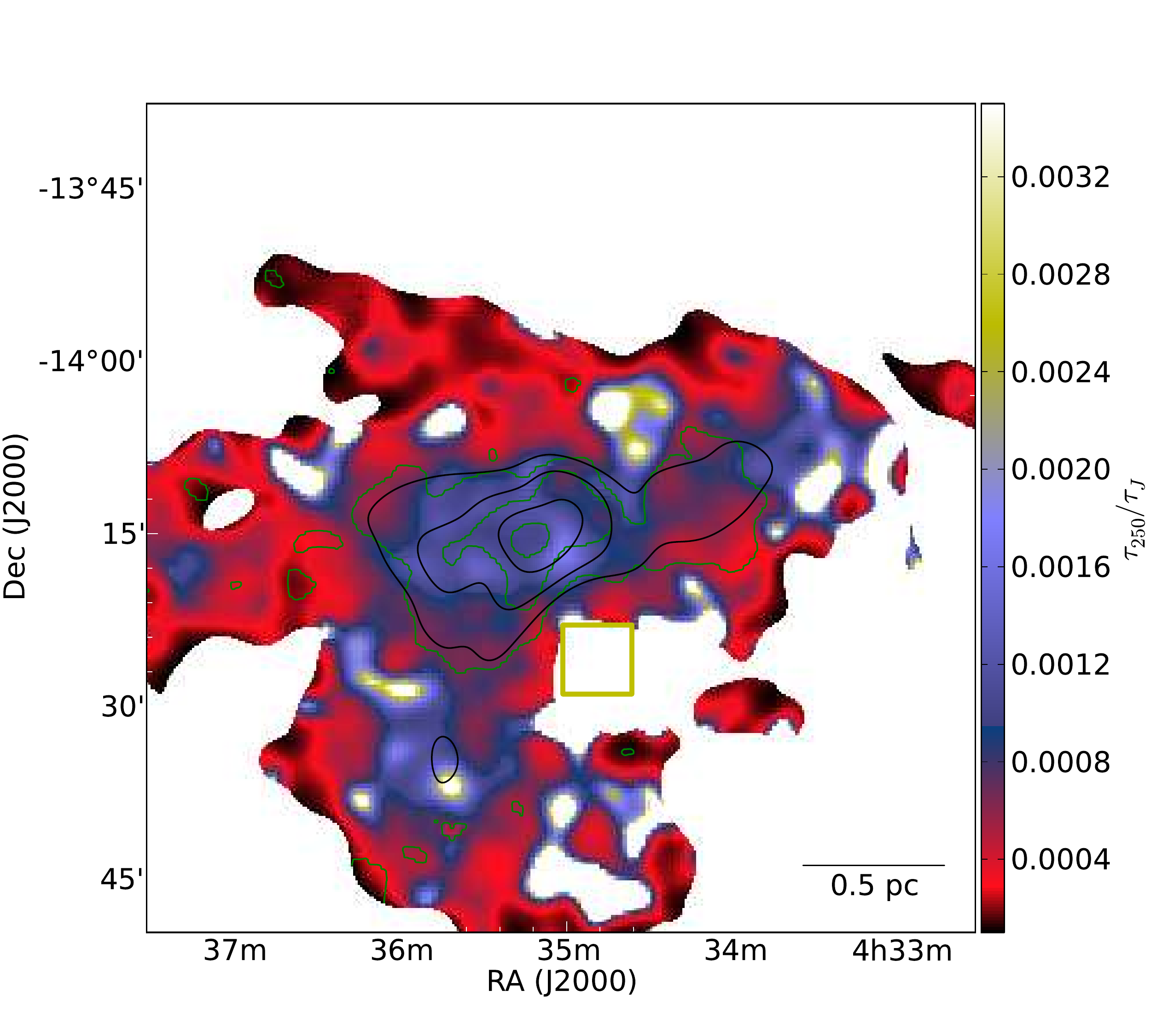}
\caption{(Left) Optical depth $\tau_J$ map derived from 2MASS data. The resolution of the map is 180$\arcsec$. (Right) $\tau_{250}/\tau_J$ map with resolution 180$\arcsec$. The black contours are drawn on $\tau_{250}$ levels 0.0016, 0.0008, and 0.0004. The green contours are drawn on $\tau_J$ levels 2, 1, and 0.5. Areas where $\tau_J < 0.0001$ or $\tau_{250} < 0.00004$ are masked. The area marked with a yellow rectangle was used as a background when subtracting the background from the two optical depth maps.
}
\label{fig:2mass_tauJ}
\end{figure*}

MCMC fitting with free $\beta$ (see Sect.~\ref{sect:MC_fitting}) produces the $\beta$ and $T$ maps shown in Fig.~\ref{fig:tau250_T_maps_freeb}. 
MCMC fitting with a constant value $\beta = 1.8$, using background-subtracted intensity maps, produces the $\tau_{250}$ and $T$ maps in Fig.~\ref{fig:tau250_T_maps_app}. These maps were used in the analysis of the different regions marked in the figures. An
extinction map and a $\tau_{250}/\tau_J$ map (described in Sect.~\ref{sect:correlations}) are shown in Fig.~\ref{fig:2mass_tauJ}.

\section{Other possible YSOs}   \label{sect:other_YSOs}

We used a subset of the WISE catalogue to search for other YSO candidates. This subset contained sources with A photometric quality (S/N $>$ 10) in at least three of the four bands and at least B quality (S/N $>$ 3) in one of the bands. We searched previously classified sources in the SIMBAD and VizieR catalogues with a radius of 5$\arcsec$. We used the quadratic discriminant analysis~\citep[QDA,][]{McLachlan1992}
to separate the sources with known object types (galaxies, evolved stars, single stars, ISM-related objects, and YSOs) and to determine
the boundaries in the multidimensional parameter space. These boundaries were then applied to the sources with unknown object types. The sources that were similar to the known YSOs were selected as YSO candidates. The multidimensional parameter space included the following colours: $J$--$H$, $H$--$K$, $K$--W1, W1--W2, W2--W3, W3--W4, and also the W1 brightness. $J$, $H$, and $K$ data are from 2MASS catalogue.

This method for YSO selection has been previously used in several studies~\citep{Toth2013, Marton2013, Cambresy2013} and will be discussed in more detail in a forthcoming paper (Marton et al., in prep.). \citet{Toth2013} found that the reliability of the classification is higher than 90 \% compared with known YSOs. Marton et al. (in prep.) have found
a quite good reliability using the current WISE catalogue. In their study, only 0.33 \% of all the other known sources (including galaxies, stars, etc.), 0.77 \% of galaxies and 73.5 \% of the known YSOs were classified as YSO candidates. Based on the classification in the SDSS catalogue, 6.79 \% of the sources classified as YSO candidates are galaxies. However, biases in the WISE catalogue might affect these results.

\begin{figure}
\centering
\includegraphics[width=8.5cm]{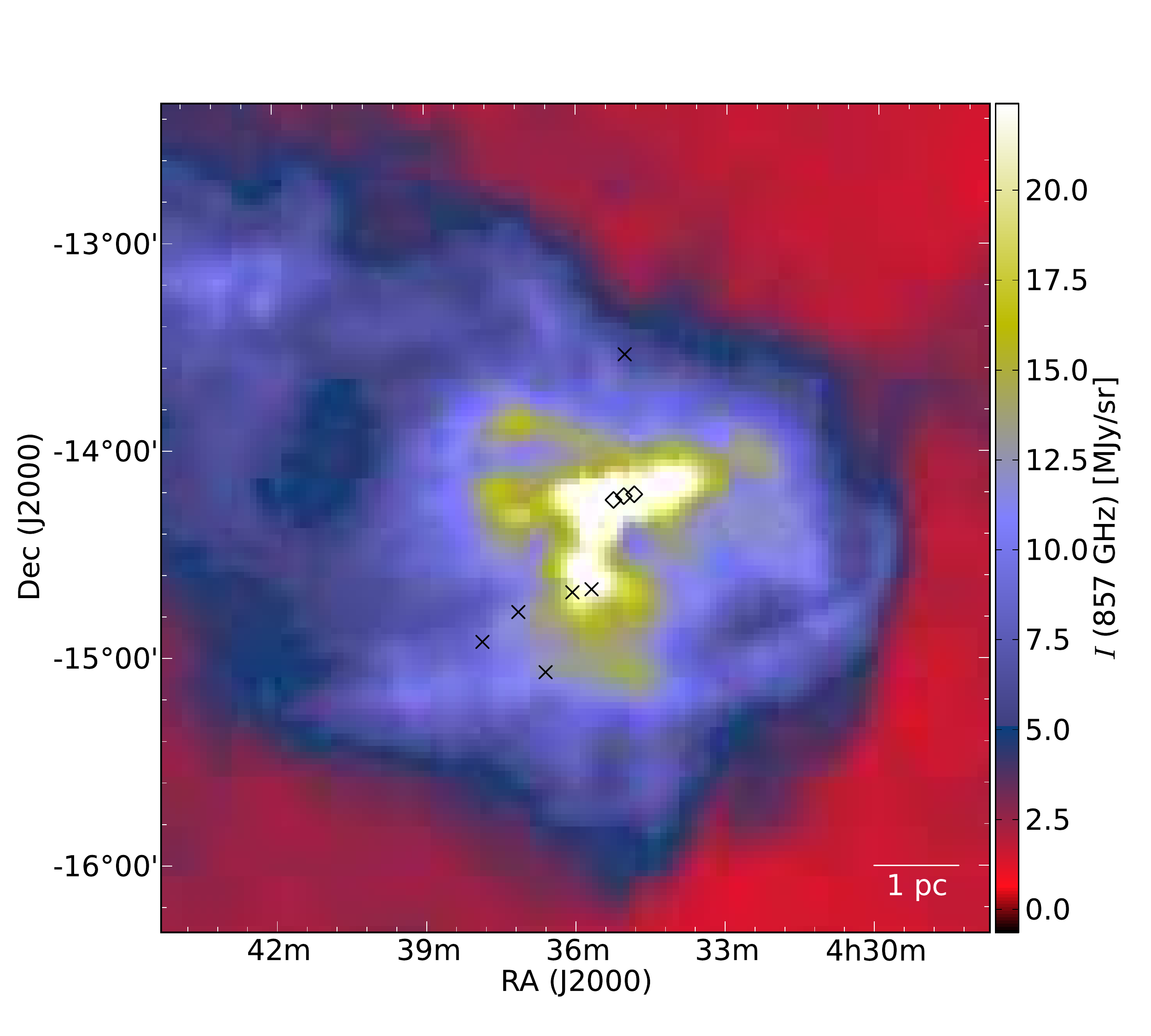}
\caption{Point sources B-1, B-2, and B-3 (diamonds) and other YSO candidates (crosses) marked on the Planck 857 GHz intensity map.}
\label{fig:ysos}
\end{figure}

We located six other YSO candidates in addition to B-1, B-2, and B-3 within a 1$^\circ$ radius of the centre of L1642, based on WISE data. 
These objects are shown in Fig.~\ref{fig:ysos}. Two of the objects are within or at the edge of the densest part of the cloud on the sky, in region C. These two are plotted in our \emph{Herschel} maps and can be seen in the 100 $\mu$m map as faint objects.
The other four are projected on a more diffuse area, but still within the larger structure of L1642 cloud. We fitted these objects with the SED models of~\citet{Robitaille2007} using optical, NIR (2MASS), and MIR (WISE) data.
The most probable distances provided by the SED model fits for the objects are 
mostly between 350\,pc and 800\,pc, clearly farther away than the estimated distance of L1642. 
There is no evidence that any of these sources are inside the L1642 cloud, but confirming this would require more study.

\section{Comparing SEST and Planck CO maps} \label{sect:Planck_SEST_CO}

We used the SEST CO data of~\citet{Russeil2003} to correct the
Planck 217 and 353 GHz maps for the contribution of CO line emission.
We compared the SEST data with the maps of integrated CO line area that
have been derived from Planck data and are available in the Planck
Legacy Archive\footnote{http://www.sciops.esa.int/wikiSI/planckpla/index.php?\\title=CMB\_and\_astrophysical\_component\_maps\&\\instance=Planck\_Public\_PLA\#CO\_emission\_maps}.
The Planck archive includes three separate estimates of the CO emission.
Type 1 products are based on the bandpass measurements of individual
Planck channels, include separate estimates of the $J$=1--0, $J$=2--1,
and $J$=3--2 line intensities, but have relatively low SN. Type 2 maps
include estimates of the CO $J$=1--0 and $J$=2--1 emission that were
obtained by a joint analysis of multiple Planck frequency channels and
have a higher S/N. Type 3 maps additionally use prior information of the
CO line ratios resulting in a combined CO emission map with a
good S/N.

\begin{figure*}
\centering
\includegraphics[width=6cm]{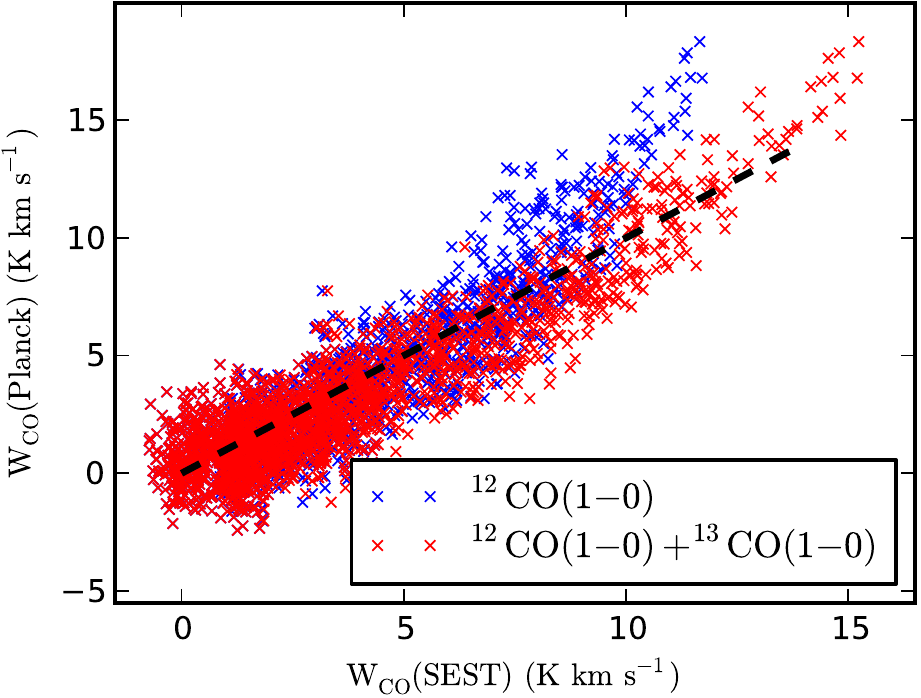}
\includegraphics[width=6cm]{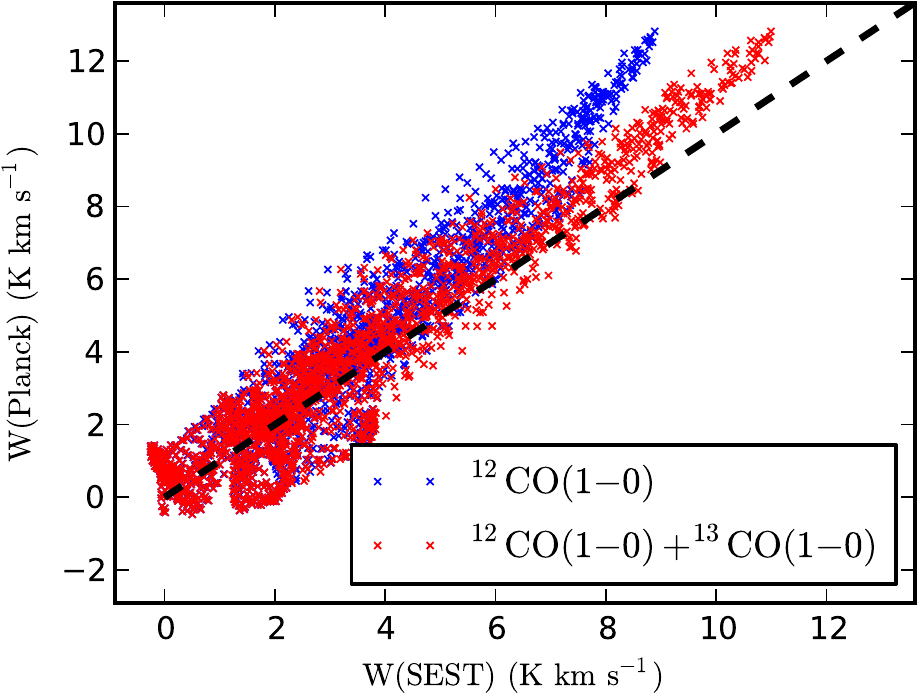}
\includegraphics[width=6cm]{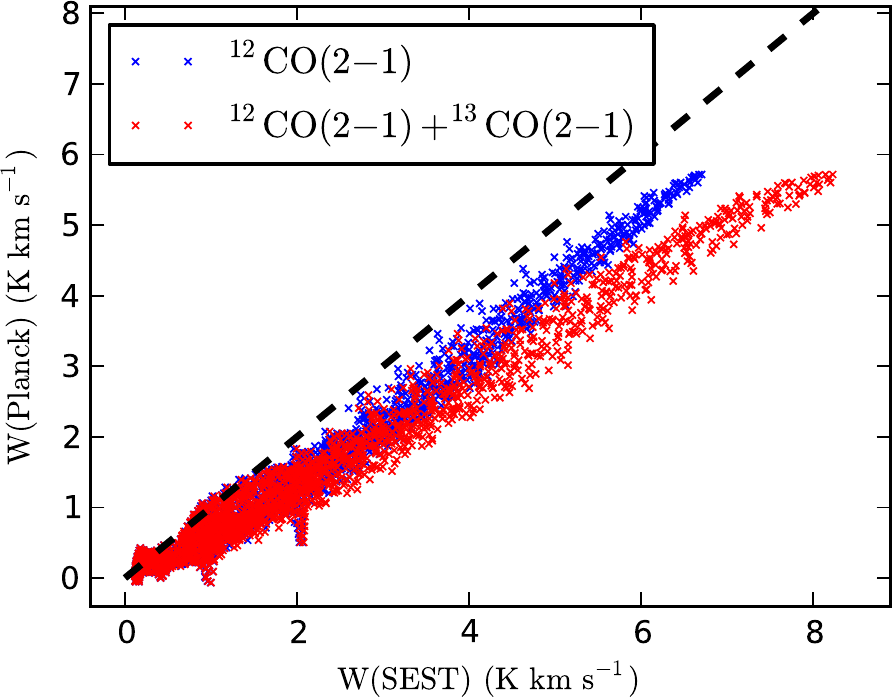}
\caption{
Relation between the Planck Legacy Archive CO maps and the SEST CO
data of~\citet{Russeil2003} using Planck Type 3 $J$=1--0 data (left
frame, 6$\arcmin$ resolution) and using Type 2 data (15$\arcmin$
resolution) for $J$=1--0 (middle frame) and $J$=2--1 (right frame).
The blue crosses show
the relation between Planck CO and the SEST $^{12}$CO line area, the
red crosses the relation between Planck CO and the sum of SEST
$^{12}$CO and $^{13}$CO line areas.
}
\label{fig:Planck_CO}
\end{figure*}

The comparison between SEST and the Type 2 and Type 3 data is shown
in Fig.~\ref{fig:Planck_CO}. The correlation is very good between Type III
$J$=1--0 line areas and the sum of SEST $^{12}$CO(1--0) and
$^{13}$CO(1--0) data (left frame). Type 2 (1--0) maps give slightly
higher line areas than Type 3 maps, and Type 2 (2--1) maps give lower
values than SEST. However, because the calibration errors of the SEST data
can be $\sim$ 10 \% , the correlations are still quite good. Note that
the SEST values are antenna temperatures corrected using the Moon
efficiency.

\section{Radiative transfer model of L1642} \label{sect:models}

We carried out radiative transfer modelling of L1642 to qualitatively
estimate the effects that LOS temperature variations have
on estimates of the optical depth and the emissivity spectral index.
The initial model is based on the column density map derived from
observations, the \citet{Mathis1983} model of the interstellar
radiation field (ISRF), and the dust properties corresponding to
observations of diffuse interstellar medium \citep{Draine2003b}.
However, to be consistent with the previous analysis, the dust model
was adjusted so that the spectral index was exactly $\beta=1.8$ for
all wavelengths $\lambda>100 \mu$m.

The density distribution was defined using a 132$^3$ Cartesian grid with a cell size corresponding to 6$\arcsec$. Thus the model covers an area of 13.2$\arcmin \times 13.2\arcmin$ that includes the sources B-1, B-2, B-3, and B-4. The initial column density distribution was obtained from the analysis of \emph{Herschel} observations. The LOS density distribution was assumed to be Gaussian with a FWHM corresponding to the value of $\sim 6\arcmin$ visible in the plane of the sky. The external radiation field was assumed to be isotropic. Corresponding to the sources B-1, B-2, and B-3, we added three radiation sources inside the cloud. In the plane of the sky these are the positions of the local peaks measured in the 250\,$\mu$m surface brightness maps, and along the LOS they are in the middle of the cloud. We treated B-4 as a clump heated from the outside, and did not add an internal source there. A non-LTE radiative transfer program was used to estimate the dust temperature distributions and to calculate synthetic surface brightness maps at the \emph{Herschel} wavelengths \citep{Juvela2003}.

The initial model was optimised by comparing the observed and modelled surface brightness maps. We used observations that were background subtracted using a reference area around the position RA=4${\rm h}$34$^{\rm m}$52$^{\rm s}$, DEC=$-$14$\degr$25$\arcmin$15$\arcsec$ (J2000.0). For each LOS, the densities of the model cloud were scaled so that one reproduces the observed 250\,$\mu$m surface brightness data at the observed resolution. The ISRF was scaled to reproduce the observed surface brightness ratio 160\,$\mu$m/500\,$\mu$m that was estimated from the central area of $4.4\arcmin \times 4.4\arcmin$. The embedded sources B-1, B-2, and B-3 were assumed to be black-bodies at a temperature of 8000\,K. Their luminosities were adjusted so that the surface brightness ratios between 160\,$\mu$m and 250\,$\mu$m bands had the observed values when measured towards the sources, using the resolution of the observed maps (12.0$\arcsec$ and 18.3$\arcsec$ for 160\,$\mu$m and 250\,$\mu$m, respectively). The calculations were iterated until the 250\,$\mu$m surface brightness, the average colour of the central regions, and the colours towards the three sources all matched to a few percent. This is only a crude model and, in particular, the effect of the sources depends sensitively on their position along the LOS, their temperature, and the density structures in their immediate vicinity.

\begin{figure*}
\centering
\includegraphics[width=16cm]{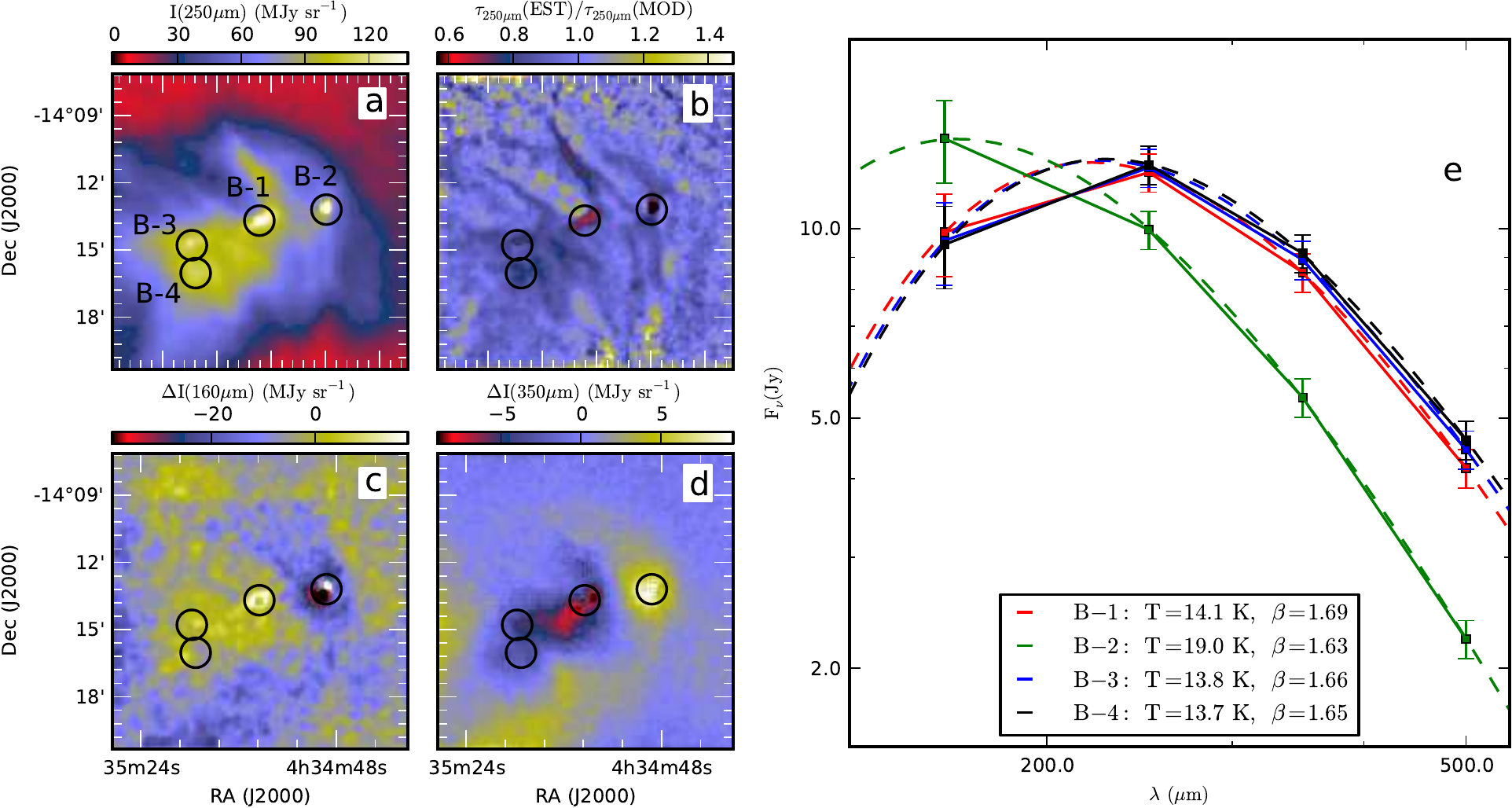}
\caption{
Analysis of model surface brightness and optical depth. 
Frame a shows the 250\,$\mu$m surface brightness and the apertures used for flux measurements of sources B-1, B-2, B-3, and B-4.
Frame b shows the ratio of optical depths that are estimated from
synthetic surface brightness data and the actual 250\,$\mu$m optical depth of the model cloud.
Frames c and d show the residuals between the actual \emph{Herschel} observations and the model predictions at 160\,$\mu$m and 350\,$\mu$m, respectively.
Frame e shows the SEDs derived for the sources using the model-predicted intensities and quotes the results of the MBB fits.
}
\label{fig:model_figures}%
\end{figure*}

Figure~\ref{fig:model_figures} shows the final 250\,$\mu$m surface
brightness map (within $\sim$ 1 \% of the observed values) as well as the
residuals of the 160\,$\mu$m and 350\,$\mu$m surface brightness data. 
The residuals are low at the very positions of the three sources.
However, the heating effect of source B-2 is too extended. Because of
the constraint on the average colour, the model is correspondingly too
cold around the sources B-1 and B-3, leaving positive
residuals in the 160\,$\mu$m map.
In the immediate vicinity of B-2 the high residuals are caused by the
fact that the source is associated with some extended emission and the
peak position changes with wavelength. At 160\,$\mu$m the relative
error is up to $\sim$ 20 \% around B-2 and rises to $\sim$ 30 \% around
the sources B-1 and B-3. Compared with 160\,$\mu$m, the relative errors
are of similar magnitude at 350\,$\mu$m and 500\,$\mu$m, but are, of
course, in the opposite direction. Because of these inaccuracies, we
used the model mainly for qualitative analysis. 

Figure~\ref{fig:model_figures}e shows the spectra calculated for the
four sources using 80$\arcsec$ apertures. Because the effect of point
sources appears to be more extended than in \emph{Herschel} data, we used flux
densities in the apertures without further subtracting the surrounding
annulus. Thus, the values do not exactly correspond to those derived from
\emph{Herschel} data. The most noticeable fact is that the estimated spectral
indices are lower than the opacity spectral index $\beta=1.8$ of the dust
grains. The strongest source B-2 has a value of $\beta=1.63$ that is 0.17 units
lower than the intrinsic beta value. In the observations, the value was
even lower, $\beta=$1.25 (see the right frame of Fig.~\ref{fig:PSareas_SED}). Nevertheless, the model suggests that most if
not all of the difference to diffuse regions might be caused by
temperature variations within the beam. 

Figure~\ref{fig:model_figures}b compares the optical depth map obtained directly from the model density cube with the optical depth $\tau(250\mu$m) derived from the synthetic surface brightness maps at a resolution of 40$\arcsec$. 
The optical depth was estimated as described in Sect.~\ref{sect:MC_fitting}, using a fixed value of $\beta=1.8$ when performing least-squares fits of MBB spectra.
Because the result depends on the assumed dust opacity, we normalised the results so that the mean optical depth ratio is one in the area where the \emph{Herschel} column density is between $1\times 10^{21}$\,cm$^{-2}$ and $2\times 10^{21}$\,cm$^{-2}$. The result shows that at the moderate column densities of L1642, the temperature variations (colour temperature exceeding the mass-averaged temperature) generally leads to no more than an underestimation of $\sim$ 10 \% in optical depth. However, the errors increase to $\sim 30$ \% towards the sources B-1 and B-3 and the largest error towards B-2 is larger by a factor of $\sim$ 3.

\end{appendix}
\end{document}